\documentclass[12pt]{article}
\usepackage{multirow}
\usepackage{booktabs}
\usepackage{mathrsfs}
\usepackage{fullpage}
\usepackage{amsmath}
\usepackage{graphicx}%
\usepackage{amsfonts}%
\usepackage{amssymb}
\usepackage{afterpage}
\usepackage{url}
\usepackage{fancyhdr}
\usepackage{isomath}
\usepackage{amsmath}
\usepackage{amsbsy}
\usepackage{amssymb}
\usepackage{amscd}
\usepackage{amsfonts}
\usepackage{graphicx}

\usepackage{verbatim}
\usepackage{euscript}
\usepackage{alltt}
\usepackage{stmaryrd}
\usepackage{float}
\usepackage[titletoc,title]{appendix}
\usepackage{caption}
\usepackage[english]{babel}
\usepackage{setspace} 
\usepackage{enumitem}

\usepackage{subcaption} 
\usepackage{layout}
\usepackage{longtable}
\usepackage{tikz}

\input{temp-definitions.inp}

\usepackage[margin=1in]{geometry}
\newcommand{\myhline}{\hline \\[-1.0em]}

\usepackage{titlesec}

\titleformat{\paragraph}
{\normalfont\normalsize\bfseries}{\theparagraph}{1em}{}
\titlespacing*{\paragraph}
{0pt}{3.25ex plus 1ex minus .2ex}{1.5ex plus .2ex}

\PassOptionsToPackage{normalem}{ulem}
\usepackage{ulem}

\newcommand{\added}[1]{#1}

\usepackage[symbol]{footmisc}

\begin{document}

\pagenumbering{roman}

\title{Finite Element Approximation of Finite Deformation Dislocation Mechanics}
\author{Rajat Arora\thanks{Dept.~of Civil \& Environmental Engineering, Carnegie Mellon University, Pittsburgh, PA, 15213.  Currently: R\&D Engg. II,  Ansys, Inc. rajat.arora9464@gmail.com.}  \qquad Xiaohan Zhang\thanks{Senior Data Scientist, Salesforce.com, Sunnyvale, CA, 94086. xiaohanzhang.cmu@gmail.com.}   \qquad Amit Acharya\thanks{Dept.~of Civil \& Environmental Engineering, and Center for Nonlinear Analysis, Carnegie Mellon University,
Pittsburgh, PA, 15213. acharyaamit@cmu.edu.}
}
\date{}

\maketitle

\begin{abstract} 
\noindent  We develop and demonstrate the first general computational tool for finite deformation static and dynamic dislocation mechanics. A finite element formulation of finite deformation (Mesoscale) Field Dislocation Mechanics  theory  is presented. The model is a minimal enhancement of classical crystal/$J_2$ plasticity that fundamentally accounts for  polar/excess dislocations at the mesoscale.  It  has the ability to compute the static and dynamic finite deformation stress fields of arbitrary (evolving) dislocation distributions in finite bodies of arbitrary shape  and elastic anisotropy under general boundary conditions. This capability is used to present a comparison of the  static stress fields, at finite and small deformations, for screw and edge dislocations, revealing heretofore unexpected differences. The computational framework is verified against the sharply contrasting predictions of geometrically linear and nonlinear theories for the stress field of a spatially homogeneous dislocation distribution in the body, as well as against other exact results of the theory.  Verification tests of the time-dependent numerics are also presented. Size effects in crystal and isotropic versions of the theory  are shown to be a natural consequence of the model and are validated against available experimental data. With inertial effects incorporated, the development of an (asymmetric) propagating Mach cone is demonstrated in  the finite deformation theory when a dislocation moves at speeds greater than the linear elastic shear wave speed of the material.






	

\end{abstract}

\pagenumbering{arabic}


\section{Introduction}


 Before the advent of the finite element (FE) method, solving boundary value problems of elasticity theory in any generality, especially for practical purposes of engineering design, was an essentially impossible task. That state of affairs has significantly changed today, with the approximation of solutions to complex problems of industrial design having become routine, with even robust commercial FE software being available for such tasks. A similar situation exists today in materials science related to a large class of important problems. Many  materials   physics problems in  structural and electronic materials demand the comparison of stress and energy density fields of two (or more) specific defect distributions in an elastically anisotropic, possibly inhomogeneous, body of geometrically complex shape, in order to determine which may be energetically more favorable and therefore physically observable (as a first estimate). For example, a first guess at whether dislocation nucleation is possible or not in a nanostructure may be addressed by computing the total energy content of the loaded structure without defects at finite deformation, and comparing this energy content with energy content of the body containing the putative, expected defect configuration to be nucleated - the Matthews-Blakeslee critical thickness criterion for strained epitaxial heterostructures is very much in this spirit. These are   problems that involve large elastic strains,   elastic anisotropy and often, inhomogeneity, and finite bodies - and, today, there exists no general purpose capability to address such questions of design (and theory), taking the burden of creative, ad-hoc, case-by-case approximation off of the analyst and transferring it to a robust  computational capability based on fundamental principles. As part of this paper, we present such a framework. The presented development can also perform several other important tasks relevant to the materials science of hard solids (as well as soft, e.g.~liquid crystal elastomers) where defects play an important role.

Conventional elasto-plastic theories model plastic flow through constituitive assumptions without explicitly recognizing dislocation motion. Owing to the lack of an inherent length scale, these theories also fail to capture any size dependence in the elasto-plastic response of (homogeneous) materials. However, it is now well-established through a vast literature, e.g.  \cite{fleck1994strain,liu2012size, stelmashenko1993microindentations, ebeling1966dispersion, ma1995size, stolken1998microbend} that metals exhibit size effects at micron to submicron length scales. For example, there is a strong size effect in the measured indentation hardness of single crystals when the indenter size is below $10$ microns  \cite{ma1995size, stelmashenko1993microindentations}.  For a given volume fraction of elastic particles in a ductile metal, the strengthening is greater for smaller particle size \cite{ebeling1966dispersion, kelly_precipitation}.  Another example is that under torsional loading of copper wires,  the scaled shear strength, as a function of average shear strain, has been shown to increase by a factor of around $3$ as the wire diameter is reduced from $100$ microns to $10$ microns \cite{fleck1994strain, liu2012size}, with negligible size effect observed under tensile load. Direct support for the notion that GNDs,  so-called geometrically-necessary dislocations or polar/excess dislocation density at the mesoscale, lead to enhanced hardening comes from the experiments performed in \cite{russell1970slip}. The  prediction of size-dependent behavior requires the presence of an inherent length scale in the theory  based on dimensional grounds. 




Beyond  hardening in the material,  dislocations have also been observed to develop intricate microstructures under the action of their mutual interactions and applied loads  such as dislocation cells \cite{mughrabi1976asymmetry,mughrabi1979persistent,mughrabi1981cyclic, hughes2000microstructure}  and labyrinths \cite{jin1984dislocation}, often with dipolar dislocation walls, and mosaics \cite{theyssier1995mosaic}. The presence of  such  dislocation microstructures, in particular their `cell size' and orientations can have a strong influence on the macroscopic response of materials \cite{humphreys2012recrystallization, reed_2006}.

To our knowledge, there is no continuum formulation that takes into account the stress field of signed dislocation density and its transport at finite deformation and can predict realistic microstructure development. Toupin's couple-stress elasticity theory \cite{toupin1962elastic} is computationally implemented within the Isogeometric Analysis method in \cite{wang2016three} to compute the stress fields of static dislocations at finite strains, representing defects by force dipole distributions. As shown there, the force-dipole representation is `non-local' w.r.t representing a dislocation line/loop and therefore can become onerous for computing the static field of a complex network of dislocations as well as its stress-coupled evolution. Computational implementations of gradient plasticity models at finite deformation of various flavors  \cite{acharya2000grain,arsenlis2004evolution,
tang2004effects,evers2004non,kuroda2008finite,
ma2006dislocation, niordson2004size,niordson2005instabilities,
lynggaard2019finite,niordson2019homogenized,
erdle2017gradient,kaiser2019dislocation,
ling2018reduced}, including inertia \cite{kuroda2019nonuniform}, have been developed with the goal of predicting length-scale effects, with some accounting for some version of dislocation transport; none of these models, however, can compute the stress field of a specified dislocation distribution. The use of Discrete Dislocation Dynamics (DD) and Molecular Dynamics to model elastic-plastic material response and microstructure development at realistic time scales and mesoscopic length scales is currently  an active area of research, as is the development of  continuum scale models that can overcome the limitations of conventional theories -- modeling size effects, calculating finite deformation stress fields of signed dislocation density, and predicting (realistic) microstructure in the material.  All DD models take the closed-form stress fields of individual dislocations as input. Current versions of DD accounting for some features of finite deformation have been reviewed in \cite{arora_acharya_ijss} - these models are not capable of computing the finite deformation stress and energy density fields of dislocation distributions. This  paper reports  the development of one such model by developing a mechanics based, novel parallel computational tool to make finite element method based computational predictions of finite deformation dislocation plasticity. 

 The current work presents a numerical framework based on Field Dislocation Mechanics (FDM) theory \cite{acharya2001model, acharya2003driving, acharya2004constitutive},  an extension of conventional plasticity that exactly accounts for the finite deformation  stress fields of dislocations and their spatio-temporal evolution, for solving  a large class  of  initial boundary value problems in finite strain dislocation mechanics at microscopic scales.  A model for mesoscale plasticity, Mesoscale Field Dislocation Mechanics (MFDM) \cite{acharya2006size, acharya2011microcanonical}, was developed from FDM by elementary averaging techniques utilized in the study of multiphase flows (see e.g.~\cite{babic1997average}). This averaging procedure does not provide closure equations, and the resulting structure can be interpreted, for operational purposes, as the equations of FDM augmented by an extra term.   This term describes the plastic strain rate due to unresolved dislocations and is  phenomenologically specified in MFDM. Despite the phenomenology, the averaging procedure and the fine-scale theory involved impart to the coarse model a rich structure that enables a gamut of relevant predictions, with only two extra material parameters over and above conventional macroscopic continuum plasticity.

The finite-element formulation for  finite deformation MFDM, presented subsequently,  uses an  updated Lagrangian description. The formulation generalizes the FEM implementation of small deformation MFDM theory developed in \cite{roy2005finite, roy2006size} to finite deformation. Since the equations of MFDM are identical to those of FDM except for an additional term in the plastic strain rate, denoted by $\bfL^p$, the algorithm and the computational framework for FDM and MFDM are similar. It adopts the static finite elasticity framework developed in \cite{puri2009modeling}. The implementation for the evolution problem utilizes an additional equation of incremental equilibrium that enables a staggered formulation akin to the small-deformation formulations \cite{roy2005finite, roy2006size}, thus overcoming the limitations of \cite{puri2009modeling} wherein the numerical formulation for finite deformation MFDM was first attempted. The formulation presented in this paper consists of the governing  balance of linear momentum equation (and its rate form    for quasi-static and equilibrium problems), a $div$-$curl$ system for the elastic incompatibility, an evolution equation for the compatible part of the elastic distortion tensor, and a first-order wave propagation equation for the evolution of the (spatially averaged)  dislocation density, singularly perturbed by a second order parabolic term.  The potential and generality of the model (both FDM and MFDM) are demonstrated through several illustrative examples.



This paper is organized as follows:  after introducing notation and terminology immediately below, Sec.~\ref{sec:theoritical_framework} presents an introduction to the governing equations of finite deformation MFDM. The details of the finite element discretization of the equations of finite deformation (M)FDM are then presented in Sec.~\ref{sec:fem}. The staggered computational algorithm for the problems within the quasi-static and dynamic (with inertia) settings,  including   time-stepping criteria, are discussed in Sec.~\ref{sec:algorithm}.   Sec.~\ref{sec:results} presents the results that verify and validate the computational framework.


\section*{Notation and terminology}
\label{sec:notation}
Vectors and tensors  are represented by bold face lower and upper-case letters, respectively. The action of a second order tensor $\bfA$ on a vector $\bfb$ is denoted by $\bfA\bfb$. The inner product of two vectors is denoted by $\bfa\cdot\bfb$ and the inner product of two second order tensors is denoted by $\bfA:\bfB$. A superposed dot denotes a material time derivative. A rectangular Cartesian coordinate system is invoked for ambient space and all (vector) tensor components are expressed with respect to the basis of this coordinate system. $(\cdot)_{,i}$ denotes the partial derivative of the quantity $(\cdot)$ w.r.t.~the $x_i$ coordinate direction of this coordinate system. $\bfe_{i}$ denotes the unit vector in the $x_i$ direction. Einstein's summation convention is always implied unless mentioned otherwise. All indices span the range $1$-$3$ unless stated otherwise. The condition that any quantity (scalar, vector, or tensor) $a$ is defined to be $b$ is indicated by the statement $a := b$ (or $b =: a$).  $tr(\bfA)$ and $det(\bfA)$ denote the  trace and the determinant of the second order tensor $\bfA$, respectively. The symbol $|(\cdot)|$ represents the magnitude of the quantity $(\cdot)$.  The symbol $a\,en$  in figures  denotes $a \times 10^n$.

The current configuration and its external boundary is denoted by $\mOmega$ and $\partial \mOmega$, respectively. $\bfn$ denotes the unit outward normal field on $\partial \mOmega$. The symbols $grad$, $div$, and $curl$ denote the gradient, divergence, and curl on the current configuration. For a second order tensor $\bfA$, vectors $\bfv$, $\bfa$, and $\bfc$, and a spatially constant vector field $\bfb$, the operations of $div$, $curl$, and cross product of a tensor ($\times$) with a vector are defined as follows:
\begin{align*}
(div\bfA)\cdot\bfb &= div(\bfA^T \bfb), ~~~~~~~~~ \forall ~ \bfb \\
\bfb\cdot(curl\bfA)\bfc &=  \left[curl(\bfA^T \bfb)\right]\cdot \bfc, ~~~ \forall ~ \bfb, \bfc \\
\bfc\cdot(\bfA\times\bfv)\bfa &= \left[(\bfA^T \bfc)\times \bfv \right]\cdot\bfa ~~~~\forall ~ \bfa, \bfc.
\end{align*}
In rectangular Cartesian coordinates, these are denoted by
\begin{align*}
(div\bfA)_i =  A_{ij,j},\\
(curl\bfA)_{ri} =  \varepsilon_{ijk}A_{rk,j}, \\
(\bfA \times \bfv)_{ri} = \varepsilon_{ijk}A_{rj}v_k,
\end{align*} where $\varepsilon_{ijk}$ are the components of the third order alternating tensor $\bfX$.  The corresponding operations on the reference configuration are denoted by the symbols $Grad$, $Div$, and $Curl$. $\bfI$ is the second order Identity tensor whose components w.r.t.~any orthonormal basis are denoted by $\delta_{ij}$. The vector $\bfX(\bfA\bfB)$ is defined by
\begin{align*}
\left[\bfX(\bfA\bfB)\right]_i =  \varepsilon_{ijk}A_{jr}B_{rk}.
\end{align*}

The following list describes some of the mathematical symbols we use in this paper.\newline
$\bbC$ : Constant fourth order elasticity tensor assumed to be positive definite on the space of second order symmetric tensors\\
$E$ : Young's modulus\\
$\mu $: Shear modulus\\
$\nu$ : Poisson's ratio\\
$\bfC^e$ : Right Cauchy-Green deformation tensor\\
$I_1(\bfC^e)$ : First invariant of $\bfC^e$\\
 $\phi$ : Elastic energy density of the material\\
$\rho$ : Mass density of the current configuration\\
$\rho^*$ : Mass density of the pure, unstreched lattice\\
$(\cdot)_{sym}$ : Symmetric part of $(\cdot)$\\
 $m$: Material rate sensitivity\\
  $\hat{\gamma}_0$ : Reference strain rate\\
  $\hat{\gamma}$ : Magnitude of SD slipping rate for the $J_2$ plasticity model\\
   $\hat{\gamma}^k$ : Magnitude of SD slipping rate on the $k^{th}$ slip system for the crystal plasticity model\\
   $n_{sl}$ : Number of slip systems\\ 
   $sgn(\tau^k)$ : Sign of the scalar $\tau^k$\\
    $\tau^k$ : Resolved shear stress on $k^{th}$ slip system\\
      $\bfm^{k}$, $\bfn^k$ : Slip direction and the slip plane normal for the $k^{th}$ slip system in the current configuration\\
      $\bfm^k_0$, $\bfn^k_0$ : Slip direction and the slip plane normal for the $k^{th}$ slip system in the pure, unstretched lattice\\
            $g_0$ : Initial material strength\\ 
            $g_s$ : Saturation material strength\\
            $g$ : Material strength \\
            $\mTheta_0$ : Stage $2$ hardening rate\\
            $k_0$ and $\eta$ : Material constants\\
            $\epsilon$ : Material constant with dimensions of $stress \times length^2$\\
			$b$: Burgers vector magnitude of a full dislocation in the crystalline material\\
            $h$: Length of the smallest edge of an element in the finite element mesh under consideration.

\section{Theory}
\label{sec:theoritical_framework}

This section presents the governing equations, constitutive assumptions, and initial and boundary conditions of finite deformation Mesoscale Field Dislocation Mechanics, the model that is computationally implemented and evaluated in this paper. The development of the relevant field equations is detailed in Appendix \ref{app:MFDM}; here we summarize briefly:
\begin{subequations}
	\begin{align}
	&~\mathring{\bfalpha}\equiv (div\,\bfv)\bfalpha+\dot{\bfalpha}-\bfalpha\bfL^T = -curl\left(\bfalpha\times \bfV + \bfL^p\right)\label{eq:mfdm_a}\\[1mm]
	&~\bfW = \bfchi+grad\bff \nonumber\\[1.25mm]
	&\left.\begin{aligned}
	&curl{{\bfW}} = curl{\bfchi} = -\bfalpha\\
	&div{\bfchi} = \bf0
	\label{eq:mfdm_chi} 
	\end{aligned}\right\}\\[1.25mm]
	&~div\left(grad\dot{\bff}\right) = div\left(\bfalpha\times \bfV + \bfL^p - \dot{\bfchi}-\bfchi\bfL\right)\label{eq:mfdm_fevol}\\
	&~div\,[\bfT(\bfW)]  = \begin{cases}
	\bf 0 & \text{quasistatic} \\
	\rho\,\dot{\bfv} & \text{dynamic}.\\
	\end{cases}
	\label{eq:mfdm_f}
	\end{align}
	\label{eq:mfdm}
\end{subequations}
The upshot of the development in Appendix \ref{app:MFDM} is that if $\bfL^p = \bfzero$ then the system \eqref{eq:mfdm} refers to the governing equations of FDM theory; otherwise, it represents the MFDM model. FDM applies to understanding the mechanics of small collections of dislocations, resolved at the scale of individual dislocations. MFDM is a model for  mesoscale plasticity with clear connections to  microscopic FDM.  The fields involved in the MFDM model are space-time averaged counterparts of the fields of FDM \eqref{eq:fdm}, with $\bfL^p$ being an emergent additional mesoscale field. In \eqref{eq:mfdm}, $\bfW$ is the inverse-elastic distortion tensor, ${\bfchi}$ is the incompatible part of $\bfW$, $\bff$ is the plastic position vector, $grad\bff$ represents the compatible part of $\bfW$, $\bfalpha$ is the dislocation density tensor, $\bfv$ represents the material velocity field, $\bfL=grad\bfv$ is the velocity gradient, $\bfT$ is the (symmetric) Cauchy stress tensor, and $\bfV$ is the dislocation velocity field.

\subsection{Constitutive equations for $\bfT$, $\bfL^p$, and $\bfV$} 
\label{sec:dissipation}
MFDM requires constitutive statements for the stress $\bfT$, the plastic distortion rate $\bfL^p$, and the dislocation velocity $\bfV$. The details of the thermodynamically consistent constitutive formulations are presented in \cite[Sec.~3.1]{arora_acharya_ijss}.  This constitutive structure is summarized below.
	
		{\renewcommand{\arraystretch}{1.2}
\begin{table}
	\centering
\begin{tabular}{|p{3.5cm}|p{6.5cm}|p{5cm}|}
\hline
~~\newline
	Saint-Venant-Kirchhoff Material
		& \begin{align*}\phi(\bfW) = \dfrac{1}{2\rho^*} \bfE^e:\bbC:\bfE^e\end{align*} & 		\begin{align}	\bfT = \bfF^e\left[\bbC:\bfE^e\right]\bfF^{eT}\label{eq:stress_svk}\end{align}\\ \hline
		
		~~\newline
		Neo-Hookean Material & \begin{align*}\phi(\bfW) = \dfrac{\mu}{2 \rho^*}\left(I_1(\bfC^e) - \ln\left(det(\bfC^e)\right)\right)\end{align*}  & \begin{align}\bfT = \mu (\bfF^{e}\bfF^{eT} - \bfI)\label{eq:stress_nh}\end{align} \\		
\hline
Core energy density & \multicolumn{2}{|c|}{ $~~~~~~~\mUpsilon(\bfalpha) :=  \dfrac{1}{2\rho^*}{\epsilon} \,\bfalpha:\bfalpha$} \\
		  \hline
		
	\end{tabular}
	\caption{Constitutive choices for elastic energy density, Cauchy stress, and core energy density.}
	\label{tab:const_T}
\end{table}
}


Table \ref{tab:const_T} presents the Cauchy stress expressions for the Saint-Venant-Kirchhoff and a compressible Neo-Hookean material. It also presents the assumed constitutive form of the  mesoscopic core energy density (per unit mass) for the material.

{\renewcommand{\arraystretch}{1.2}
\begin{table}
	\centering
	\begin{tabular}{|p{4cm}|p{11.5cm}|}
\hline
		~ & {\begin{align}
			\hat{\bfL}^p&= \bfW\, \left(\sum_k^{n_{sl}} \hat{\gamma}^k \, \bfm^k\otimes\bfn^k \right)_{sym} \\
			\bfL^p &= \hat{\bfL}^p + \quad  \left( \dfrac{l^2}{n_{sl}} \sum_k^{n_{sl}} |\hat\gamma^k| \right) \,curl \bfalpha \label{eq:Lp_crystal}\\
			\hat{\gamma}^k &= sgn(\tau^k)\, \hat{\gamma_0}^k\left(\frac{|\tau^k|}{g} \right)^{\frac{1}{m}}\end{align}}\\
		\vspace{-3cm} Crystal plasticity & \vspace{-1cm}{\begin{align*}
			\tau^k = \bfm^k\cdot\bfT\bfn^ k  ;~~
			\bfm^k = {\bfF^e\bfm_0^k}  ; ~~
			\bfn^k = {{\bfF^e}^{-T}\bfn_0^k} 
			\end{align*}}\\[-2em] \myhline 
		\vspace{1.5cm}			$J_2$ plasticity &  {\begin{align}\hat{\bfL}^p &=\, \hat{\gamma}\bfW\frac{\bfT^{'}}{|\bfT^{'}|}  ;  ~~ \hat{\gamma} = \hat{\gamma_0} \left(\dfrac{|\bfT'|}{\sqrt{2} \,g}\right)^{\frac{1}{m}}\nonumber\\[.1cm]
			\bfL^p &=\hat{\bfL}^p +  l^2  \hat\gamma \, curl\bfalpha \label{eq:Lp_j2}  
			\end{align}} \\ \hline
	\end{tabular}
	\caption{Constitutive choices for plastic strain rate due to SDs $\bfL^p$.}
	\label{tab:const_Lp}
\end{table}
}

\begin{table}
	\centering
	\begin{tabular}{|p{15cm} |}
		\hline 
		{\begin{align}
			T'_{ij} = T_{ij} - \dfrac{T_{mm}}{3}\delta_{ij}; ~~~~ a_i &:= \dfrac{1}{3}T_{mm} \varepsilon_{ijk}{F^e}_{jp}\alpha_{pk}; ~~~~			c_i := \varepsilon_{ijk}T'_{jr}{F^e}_{rp}\alpha_{pk}\nonumber\\
			\bfd = \bfc - \left(\bfc\cdot \frac{\bfa}{|\bfa|} \right) \frac{\bfa}{|\bfa|}; ~~&~~			\hat{\gamma}_{avg} =
			\begin{cases} 
			\hat{\gamma} & J_2~\textrm{plasticity} \\
			\dfrac{1}{n_{sl}}\sum_k^{\,n_{sl}} |\hat{\gamma}^k| & \textrm{Crystal plasticity}.
			\end{cases}\nonumber\\
			\bfV = \zeta & \frac{\bfd}{|\bfd|} ~~ ; ~~\zeta =  \left(\dfrac{\mu}{g}\right)^2\,\eta^2\, b  \, \hat{\gamma}_{avg}\label{eq:V_lpj2} 
			\end{align}}\\
		\hline
	\end{tabular}
	\caption{Constitutive choices for dislocation velocity $\bfV$.}
	\label{tab:const_V}
\end{table}

Table \ref{tab:const_Lp} presents the constitutive assumptions for $\bfL^p$ for Crystal and $J_2$ plasticity models.  Table \ref{tab:const_V} presents the constitutive assumptions for $\bfV$ for Crystal and $J_2$ plasticity models.  Table \ref{tab:const_g} presents the governing equation for the evolution of material strength $g$ for the two models. \added{The use of $\hat{\gamma}_{sd}$ in \eqref{eq:softening} stems from the fact that isotropic (or Taylor) hardening is used for the  evolution of strength on every slip system with equal initial values, i.e.,
\begin{align}\nonumber
\hat{\gamma}^k &= sgn(\tau^k)\, \hat{\gamma_0}^k\left(\frac{|\tau^k|}{g^k} \right)^{\frac{1}{m}},\qquad k = 1,\ldots,n_{sl} \nonumber\\
\dot{g}_{kj} &= h(\bfalpha,g)\left(\left|\bfF^e\bfalpha\times\bfV\right| + \sum_{j = 1}^{n_{sl}}  [q + (1-q)\delta_{kj}]\left| \hat{\gamma}^j \right| \right), \qquad 1 \leq q \leq 1.4, \qquad k,j = 1,\ldots,n_{sl},
 \label{eqn:latent_hardening}
\end{align}
where the function $h$ is defined in \eqref{eq:softening} and \eqref{eqn:latent_hardening} is a simple modification of standard latent hardening phenomenology assumed in classical crystal plasticity (see, e.g., \cite{peirce1983material}). Isotropic hardening is not a necessary condition for the formulation.} 

\begin{table}
	\centering
	\begin{tabular}{|p{15cm} |}
		\hline 
		{\begin{align}
			\hat{\gamma}_{sd} &=
			\begin{cases} 
			\hat{\gamma} & J_2~\textrm{plasticity} \\
			 \sum_k^{\,n_{sl}} |\hat{\gamma}^k| & \textrm{Crystal plasticity}.
			\end{cases}
			\label{eq:gaama_hat_ssd}
			\end{align}}
		{
		\begin{align}
		\dot{g} = h(\bfalpha, g) \left(\left|\bfF^e\bfalpha\times\bfV\right|+ \hat{\gamma}_{sd} \right); \qquad h(\bfalpha, g) = \frac{\mu^2\eta^2b}{2(g-g_0)}k_0 \left|\bfalpha\right|+ \mTheta_0 \left(\frac{g_s-g}{g_s-g_0}\right)
		\label{eq:softening}
		\end{align} 
	}\\
		\hline
	\end{tabular}
	\caption{Evolution equation for material strength $g$.}
	\label{tab:const_g}
\end{table}

All material parameters, except $k_0$ and $l$, are part of the constitutive structure of well-accepted models of classical plasticity theory. Our model requires these  two extra material parameters beyond the requirements of classical theory. $l$ (with physical dimensions of length) sets the length scale for the mesoscopic core energy to be effective,  and $k_0$ (non-dimensional) characterizes the plastic flow resistance due to ED.

We mention here that the length scale $l$, introduced in Eq. \eqref{eq:Lp_crystal} or \eqref{eq:Lp_j2}  as a dimensional consequence of including the core energy, is not responsible for producing enhanced size effects and microstructure in MFDM. Rather, the `smaller is harder' size effect  decreases with increasing magnitude of $l$ since its presence reduces the magnitude of the $\bfalpha$ field and consequently reduces hardening \eqref{eq:softening}.

\subsection{Boundary Conditions}
\label{sec:boundary_conditions}
The $\bfalpha$ evolution equation \eqref{eq:mfdm_a},  the incompatibility equation for $\bfchi$ \eqref{eq:mfdm_chi},  the $\bff$ evolution equation \eqref{eq:mfdm_fevol}, and the equilibrium equation \eqref{eq:mfdm_f} require specification of boundary conditions at all times.

The $\bfalpha$ evolution equation  \eqref{eq:mfdm_a} admits a `convective' boundary condition of the form $(\bfalpha \times \bfV + \bfL^p) \times \bfn=\bfPhi$ where $\bfPhi$ is a second order tensor valued function of time and position on the boundary characterizing the flux of dislocations at the surface, satisfying the constraint $\bfPhi\bfn=\bf0$.  The boundary condition is specified in one of following two ways:
\begin{itemize}
\item \textit{Constrained case}: It is modeled by taking $\bfPhi$ to be identically zero on the boundary at all times i.e.~$\bfPhi(\bfx, t) = \bf0$.  This makes the body plastically constrained on the boundaries which means the dislocations cannot exit the body while only being allowed to move in the tangential direction at the external boundary. It is also referred to as the no-slip or plastically rigid boundary condition.

\item \textit{Unconstrained case}: A less restrictive boundary condition where $\hat\bfL^p \times \bfn$ is specified on the boundary, along with the specification of dislocation flux $\bfalpha (\bfV\cdot\bfn)$ on the inflow part of the boundary (where $\bfV\cdot\bfn < 0$) can also be used. In addition to this, for non-zero $l$, specification of $l^2\hat{\gamma}_{sd} (curl\bfalpha \times \bfn)$ on the boundary is also required, where $\hat{\gamma}_{sd}$  is defined in Eq.\eqref{eq:gaama_hat_ssd}. 
\end{itemize}


 Incompatibility equation \eqref{eq:mfdm_chi} admits a boundary condition of the form 
\begin{align*}
\bfchi \bfn = \bf0 
\end{align*}
on the external boundary $\partial\mOmega$ of the domain. Such a boundary condition along with the system \eqref{eq:mfdm_chi} ensures vanishing of $\bfchi$ in the absence of any dislocation density $\bfalpha$. $\bff$ evolution equation \eqref{eq:mfdm_fevol} requires a Neumann boundary condition of the form
\begin{align*}
(grad\dot{\bff})\bfn  = (\bfalpha\times \bfV + \bfL^p - \dot{\bfchi} - \bfchi\bfL)\bfn
\end{align*}
on the external boundary of the domain. The equilibrium equation \eqref{eq:mfdm_f} requires specification of standard displacement/velocity and/or statically admissible tractions on complementary parts of the boundary of the domain. 


\subsection{Initial Conditions}
\label{sec:initial_conditions}
The evolution equations for $\bfalpha$  and  $\bff$  (Eqs.~\eqref{eq:mfdm_a} and \eqref{eq:mfdm_fevol}, respectively) require specification of initial condition on the domain.

For $\bfalpha$ equation, an initial condition of the form $\bfalpha(\bfx, t = 0) = \bfalpha_0(\bfx)$ is required.  To determine the initial the initial condition on $\bff$, the problem can be more generally posed as follows: determine the $\bff$ and $\bfT$ fields on a given configuration with a known  dislocation density $\bfalpha$. This problem can be solved by solving for $\bfchi$ from the incompatibility equation and then $\bff$ from the equilibrium equation as described by the system
\begin{eqnarray}
\left.\begin{aligned}
curl\bfchi  &= -\bfalpha \\
div\bfchi &= \bf0\\
div\left[\bfT(\bff, \bfchi)\right] &= \bf0\\
\end{aligned}\quad\right\} \text{on } \mOmega  \label{eq:ECDD}
\\
\left.\begin{aligned}
\bfchi\bfn &= \bf0\\
\bfT\bfn &= \bft\\
\end{aligned}\qquad \right\} \text{on }\partial\mOmega
\label{eq:ECDD_bc}
\end{eqnarray}
where $\bft$ denotes the statically admissible traction field on the boundary. This determination of $\bfchi$, $\bff$, and $\bfT$ for a given dislocation density $\bfalpha$ on  any known configuration will be referred to as the ECDD solve on that configuration.  Hence, we do the ECDD solve on the `as-received' configuration, i.e.~the current configuration at $t = 0$,  to determine the initial value of $\bff$ which also determines the stress $\bfT$ distribution at $t = 0$. For the dynamic case, an initial condition on material velocity field $\bfv(\bfx, t = 0)$ is required.

The model admits an arbitrary specification of $\dot{\bff}$ at a point to uniquely evolve $\bff$ from  Eq.~\eqref{eq:mfdm_fevol}  in time and we prescribe it to be $\dot{\bff} = \bf0$.


\section{Variational formulations} 
\label{sec:fem}
This section presents the weak form of the governing equations of MFDM at finite deformation for the quasistatic and dynamic cases. The  algorithms summarizing the implementation are then presented in Sec.~\ref{sec:algorithm}. 

Modeling material behavior through the use of MFDM requires the concurrent solution to a coupled nonlinear system of pdes given by \eqref{eq:mfdm}. To efficiently solve the system for the quasistatic case within a staggered scheme in each time increment as in \cite{roy2005finite,roy2006size} in the small deformation case, we augment the system \eqref{eq:mfdm} with the rate (or incremental) form of the equilibrium equation.   This rate form is solved to get the material velocity field $\bfv$ on the domain which can be used to (discretely) update the geometry of the body. In the absence of body forces and inertia,  the statement of local force balance (on the current configuration) w.r.t.~any choice of reference configuration can be expressed by 
\begin{equation}
\label{eq:eqb_piola}
Div\bfP = \bf0,
\end{equation} where $\bfP$ represents the first Piola-Kirchhoff stress w.r.t.~that reference. This implies
\begin{align*}\label{eq:eqb_rate_form}
\dot{\overline{Div\, \bfP}} = \bf0 ~~;~~ \mathnormal{Div \left[J\, tr(\bfL) \,\bfT \bfF^{-T} + J \dot{\bfT}\bfF^{-T} + J \bfT \dot{\overline{\bfF^{-T}}} \right] = \bf0},
\end{align*}
 where $J = det(\bfF)$ and choosing the reference configuration to be the current one i.e.~$\bfF = \bfI$, one obtains  \cite{mcmeeking1975finite}  
\begin{equation}
div\left[ tr(\bfL) \, \bfT + \dot{\bfT} - \bfT\bfL^T \right] = \bf0.
\label{eq:rate_form}
\end{equation} The system \eqref{eq:mfdm} is then augmented with Eq.~\eqref{eq:rate_form}  for the quasistatic case. For the dynamic case, the balance of linear momentum can be solved directly to give the material velocity field on the domain.

The discretization methods  for solving the equations for the finite deformation MFDM \eqref{eq:mfdm} are similar to the small deformation case as presented earlier in \cite{roy2005finite, roy2006size}.  The following numerical schemes are used:  the Galerkin FEM for the  equilibrium equation \eqref{eq:mfdm_f} and its rate form \eqref{eq:rate_form}, and evolution equation \eqref{eq:mfdm_fevol} for the compatible part of inverse of elastic distortion; the Least-squares FEM \cite{jiang2013least} for the incompatibility equation \eqref{eq:mfdm_chi}; and the  Galerkin-Least-Squares FEM \cite{hughes1989new} for the dislocation evolution equation \eqref{eq:mfdm_a}.  Apart from the changes due to finite deformation, primary changes include the protocols needed to carefully integrate incremental reaction force rates, corresponding to imposed velocity-Dirichlet b.c.s, to obtain consistent traction b.cs.~(in the weak form)  for balance of linear momentum in the quasistatic case.  The (finite element) mesh motion is also taken into account.

\added{The FEM based computational framework for MFDM results in a total of $10$ and $24$ degrees of freedom (DOFs) per node for a simulation in $2$-d and $3$-d, respectively. In $2$-d, this includes $2$ unknowns in $\bfalpha$ ($\alpha_{13}$ and $\alpha_{23}$), $4$ in $\bfchi$ ($\chi_{11}, \chi_{12}, \chi_{21}, \chi_{22}$), and $2$ each in $\bfv$ and $\bff$. However, given the staggered nature of the algorithm, the largest linear system to be solved consists of $4$ and $9$ dofs per node in $2$-d and $3$-d, respectively.} 

The FEM formulation and algorithm presented here are independent of the constitutive choices made for $\bfL^p$, $\bfV$, and $\bfT$. We now discuss the numerical schemes to solve the governing equations. A typical time increment between times $t^n$ to $t^{n+1}$ is considered. $(\cdot)^n$ and $(\cdot)^{n+1}$ denote the quantity $(\cdot)$ at time $t^n$ and $t^{n+1}$, respectively. $\Delta t^n$ is defined as $\Delta t^n = t^{n+1} - t^n$.
\subsection{Weak form for $\bfv$}
The material velocity field $\bfv$ is required to update the geometry discretely by moving the finite element mesh. For the quasistatic case, $\bfv$ is obtained by solving the rate form of equilibrium equation \eqref{eq:rate_form} while for the dynamic case, it is obtained by solving the balance of linear momentum equation \eqref{eq:mfdm_f}$_2$.
\subsubsection{Quasistatic case}
\label{sec:fem_v} 
We solve the rate form of the equilibrium equation to obtain the material velocity field $\bfv$ on the current configuration $\mOmega$ following the assumed strain formulation of \cite{nagtegaal1974numerically}. 
We define $\delta \bar{\bfL}$ to be
\begin{align*}
\delta \bar{\bfL}(\bfx) := grad \delta \bfv(\bfx) - \frac{1}{3}  div\delta \bfv (\bfx) \bfI + \frac{1}{3\,|B(\bfx)|} \int_{B(\bfx)} \bfI\,div\delta \bfv \,  dV,
\end{align*}
where $B(\bfx)$ represents the element (in mesh) containing the point $\bfx$ and $|B(\bfx)|$ is the volume of the element $B$. The weak form is then written as
\begin{align*}
\int_\mOmega \delta \bar{\bfL} : \left[ div\bfv \,\bfT + \dot{\bfT} - \bfT\bar{\bfL}^T\right] \, dV = \int_{\partial \mOmega} \delta 
{\bfv} \cdot \dot{\bft} \, dA.
\end{align*}
where 
\begin{align*}
\bar{\bfL}(\bfx) := \bfL(\bfx) - \frac{1}{3}  div\,\bfv(\bfx) \bfI + \frac{1}{3\,V(\bfx)} \int_{B(\bfx)} \bfI\,div\,\bfv(\bfx) \,  dV
\end{align*}
and $\dot{\bft}$ is the specified Neumann boundary condition of nominal traction rate based on the current configuration as the reference.  Using the evolution equation \eqref{eqn:HJ}$_1$ for $\bfW$ and the identity $\bfF^e\bfW = \bfI$, the material time derivative of $\bfT$ is calculated as
\begin{align*}
\dot{\bfT} = \left(\dfrac{\partial \bfT}{\partial \bfF^e }\right) : \left(\bfL-(\bfF^e \bfalpha)\times \bfV - \bfF^e \bfL^p\right)\bfF^e .
\end{align*}

The calculation of $\parderiv{\bfT}{\bfF^e}$  for the Saint-Venant-Kirchhoff and the Neo-Hookean materials are shown in Appendices \ref{app:dTdFe_svk} and \ref{app:dTdFe_nh}, respectively.
Using the expression for $\dot{\bfT}$, the weak form of Eq.~\eqref{eq:rate_form} is  expressed  as
\begin{equation}
\begin{split}
\int_{\mOmega} \delta \, \bar{\bfL} : \left[tr(\bar\bfL) \bfT-\bfT\bar{\bfL}^T + \frac{\partial \bfT}{\partial \bfF^e}:(\bar{\bfL}\cdot \bfF^e)\right] \, dV &= \mathunderline{blue}{\int_{\mOmega} \delta\, \bar{\bfL} : \left[ \frac{\partial \bfT}{\partial \bfF^e }:\left( (\bfF^e\bfalpha)\times \bfV \right)\bfF^e\right] dV}\\  
\qquad & + \mathunderline{blue}{\int_{\mOmega} \delta\, \bar{\bfL} : \left[ \frac{\partial \bfT}{\partial \bfF^e }: (\bfF^e\bfL^p\bfF^e) \right] dV}\\  
\qquad & + \int_{\partial\mOmega} \delta\bar{\bfv}\cdot \dot{\bf\bft}\, dA.
\end{split}
\label{eq:fem_v}
\end{equation}

The underlined terms denote the contribution of  the plastic strain rate to the Cauchy stress rate in the body and Eq.~\eqref{eq:fem_v} shows their effect as a forcing term in the determination of material velocity field on the body. For a given state of the system $(\bfT^n, \bfF^{e,n}, \bfL^{p,n}, \bfalpha^n, \bfx^n$, and $\bfV^n)$ at time $t^n$, the weak form generates a system of linear equations which is then solved to get the velocity field $\bfv^{n}$ on the configuration $\mOmega^{n}$. 

On part of the boundary where Dirichlet conditions on the velocity are applied, the nodal reaction force rates at time $t^n$ are calculated after solving \eqref{eq:fem_v} on the configuration $\mOmega^n$. For a finite element mesh node $A$  on the velocity-Dirichlet part of the boundary, the nodal reaction force rate  corresponding to the degree of freedom pair $\{(A,a)\}$ is expressed as 
\begin{align}
\hat{T}^A_{a} = \int_{\mOmega}  \left(\parderiv{N^A}{x_j}\mathscr{P}_{aj} - \dfrac{\mathscr{P}_{ii}}{3}  \parderiv{N^A}{x_a} + \dfrac{\mathscr{P}_{ii}}{3 |B(\bfx)|} \int_{B(\bfx)}\parderiv{N^A}{x_a}\,dV\right)  \, dV,
\label{eq:traction_rate}
\end{align}
where $\pmb{\mathscr{P}}$ denotes the first Piola-Kirchhoff traction rate (evaluated on the assumed strain velocity gradient) given by
\begin{align*}
\begin{split}
\pmb{\mathscr{P}} = tr(\bar\bfL) \bfT-\bfT\bar{\bfL}^T +  \frac{\partial \bfT}{\partial \bfF^e }:\left[\bar{\bfL}\cdot \bfF^e - \left( \bfF^e\bfalpha\times \bfV \right)\bfF^e -  \bfF^e\bfL^p\bfF^e \right].
\end{split}
\end{align*}

  For each node on this part of the boundary, this reaction force rate physically corresponds to the spatial integration of the nominal/First Piola-Kirchhoff traction rate, based on the configuration $\mOmega^n$ as reference, over the area patch (on the same configuration) that contributes to the node in question. Since such a nodal force rate, viewed as a discrete function of time, corresponds to the evolving current configuration of the body (recall the definition of the First Piola-Kirchhoff stress tensor), we simply (discretely) integrate it in time and accumulate the result on the known nodal force at time $t^n$ to obtain the nodal force (on the velocity-Dirichlet-part of the boundary) at time $t^{n+1}$. On the part of the boundary where Cauchy tractions are specified (including  null), nothing needs to be done.

\subsubsection{Dynamic case}
\label{sec:vsolve_dyna}
For the dynamic case, the balance of linear momentum equation is directly solved to obtain the velocity on the given configuration.  Assuming the stresses and material velocity on the current configuration $\mOmega^n$ are given, we solve for $\bfv^{n+1}$ using the Forward Euler method as follows:

\begin{align}
\label{eq:fem_vdyna}
\int_{\mOmega^n} \rho\, v_i^{n+1} \delta v_i \,dV = \int_{\mOmega^n} \rho \,v_i^{n} \delta v_i \, dV + \Delta t^n \left( \int_{\partial{\mOmega}^n} t_i \delta v_{i} \, dA  - \int_{\mOmega^n} T_{ij}\delta v_{i,j} \,dV \right).
\end{align}


%

\subsection{Weak form for $\bfchi$}
\label{sec:fem_chi}
For a given dislocation density $\bfalpha$  and a configuration of the body $\mOmega$, $\bfchi$ is evaluated by solving the system \eqref{eq:mfdm_chi} along with the Dirichlet boundary conditions mentioned in Sec.~\ref{sec:boundary_conditions}.  We use the  Least-Squares finite element method to solve for $\bfchi$ from the $div$-$curl$ system \eqref{eq:mfdm_chi}. The objective  functional $J$ for this system is written as
\begin{equation}
J = \frac{1}{2}\int_{\mOmega} \left( curl\bfchi + \bfalpha \right ) : \left( curl\bfchi + \bfalpha \right )\,dV + \frac{1}{2} \int_\mOmega div \bfchi \cdot div \bfchi \,dV, \nonumber
\end{equation} resulting in the weak form
\begin{align}
\int_\mOmega e_{ijk}\delta\chi_{rk,j} \left( e_{imn}\chi_{rn,m} + \alpha_{ri} \right) \, dV + \int_\mOmega \delta \chi_{ij,j} \chi_{im,m}\, dV = 0.
\label{eq:fem_chi}
\end{align}

The above system of linear equations can be easily solved to obtain $\bfchi$ on a given  configuration $\mOmega$ for a given dislocation density.





\subsection{Weak form for  $\bfalpha$}
\label{sec:fem_alpha}
{The transport equation for $\bfalpha$ \eqref{eq:mfdm_a} exhibits nonlinear wave type solutions. In the presence of a non-zero core energy $\mUpsilon({\bfalpha})$, the dislocation evolution equation is singularly perturbed by a second order parabolic term which behaves as a small diffusive term leading to a convection\textendash diffusion equation.} Following \cite{roy2005finite, roy2006size}, we adopt the Galerkin-Least-Squares FEM approach as described in \cite{hughes1989new} wherein the Galerkin residual is added to a non-negative (may be spatially varying) scalar multiple of the least squares residual. Writing $\bfL^p = \hat{\bfL}^p + \beta\, curl\bfalpha$, \eqref{eq:mfdm_a} can be rewritten as
\begin{align}
tr(\bfL)\bfalpha+\dot{\bfalpha}-\bfalpha\bfL^T = -curl\left(\bfalpha\times \bfV + \hat{\bfL}^p + \beta curl\bfalpha\right).
\label{eq:mfdm_fem_a_exp}
\end{align}
Using a linearly implicit scheme, the Galerkin-Least-Squares residual for Eq.~\eqref{eq:mfdm_fem_a_exp} can be written as
\begin{align}\label{eq:fem_a}
\begin{split}
R&= \mathunderline{blue}{\int_{\mOmega^n} \delta\alpha_{ij}\left(\Delta t^n\, L_{pp}\alpha_{ij} - \Delta t^n\,\alpha_{ip} L_{jp} \right) \,dV} + \int_{\mOmega^n} \delta\alpha_{ij}\left( \alpha_{ij}-\alpha^n_{ij} \right) \,dV \\ &\quad + \Delta t^n \, \int_{\mOmega^n} \varepsilon_{jqp} \, \varepsilon_{jab} \alpha_{ia} V_{b} \delta \alpha_{ip,q} \, dV \\
&\quad + \Delta t^n \, \int_{\mOmega^n} \hat{L}^p_{ij} \varepsilon_{jqp}\delta\alpha_{ip,q} \, dV + \Delta t^n \, \int_{\mOmega^n} \beta \varepsilon_{jab}\alpha_{ib,a} \varepsilon_{jqp}\delta\alpha_{ip,q} \, dV \\
&\quad + \Delta t^n \, \int_{\partial\mOmega_i^n} B_{ij}\delta\alpha_{ij} \, dA +  \Delta t^n \, \int_{\partial\mOmega_o^n} \alpha^n_{ij} V_p n_p \delta\alpha_{ij} \, dA- \Delta t^n \, \int_{\partial\mOmega^n} \alpha_{iq}V_jn_q\delta\alpha_{ij} \, dA \\
&\quad- \Delta t^n \, \int_{\partial\mOmega^n}  \varepsilon_{jpq} \hat{L}^p_{ip}n_q\delta\alpha_{ij} \, dA - \Delta t^n \, \int_{\partial\mOmega^n} \beta \varepsilon_{jpq}\varepsilon_{pba} \alpha_{ia,b}n_q \delta\alpha_{ij} \, dA \\
&\quad + c\, \biggl[ \int_{\mOmega_e^n} A_{ri}  \delta\alpha_{ri} \, dV + \mathunderline{blue}{\Delta t^n  \int_{\mOmega_e^n} L_{pp} A_{ri}  \delta\alpha_{ri} \, dV - \Delta t^n \int_{\mOmega_e^n} A_{ri} \delta\alpha_{rp} L_{ip} \, dV} \\
&\quad+  \Delta t^n \int_{\mOmega_e^n} A_{ri} \left( \delta\alpha_{ri,q}V_{q} - \delta\alpha_{rq,q}V_{i} + \delta\alpha_{ri}V_{q,q} - \delta\alpha_{rq}V_{i,q}\right)  \, dV\\
&\quad + \Delta t^n \int_{\mOmega_e^n} A_{ri} (\beta_{,p}  \delta\alpha_{rp,i} +  \beta \delta\alpha_{rp,ip} -\beta_{,p}  \delta\alpha_{ri,p} - \beta \delta\alpha_{ri,pp})\, dV \biggr],
\end{split}
\end{align}
where
\begin{align*}
A_{ri} &= \alpha_{ri} - \alpha^n_{ri} + \Delta t^n \left[ \mathunderline{blue}{\alpha^n_{ri}L_{pp} - \alpha^n_{rp} L_{ip}} +  \alpha^n_{ri,q}V_{q} - \alpha^n_{rq,q}V_{i} + \alpha^n_{ri}V_{q,q} - \alpha^n_{rq}V_{i,q} + \varepsilon_{ipq}\hat{L}^p_{rq,p} + \right.\\
&  \left. + \beta_{,p}  \alpha^n_{rp,i} +  \beta \alpha^n_{rp,ip} -  \beta_{,p}  \alpha^n_{ri,p} -  \beta \alpha^n_{ri,pp} \right].
\end{align*}  

%


 In \eqref{eq:fem_a}, no superscript on $\bfalpha$  refers to $\bfalpha^{n + 1}$. $\bfL$, $\bfV$,  and $\hat{\bfL}^p$ are treated as known data. $\partial \mOmega^n_{i}$ and $\partial \mOmega^n_{o}$ represent the inflow and outflow parts of the boundary $\partial \mOmega^n$. $\bfB$ is the input dislocation flux $\bfalpha (\bfV\cdot\bfn)$ on $\partial\mOmega^n$. $\mOmega^n_e$ denotes the element interiors.   The terms underlined in blue above are the additional terms that enter the discretization for the dislocation density evolution in the finite deformation setting.  We ignore the gradients of $\beta$ in in the Least-Squares stabilization Eq.~\eqref{eq:fem_a} as including these terms was found to degrade the computational approximation in our practical experience. $c$ is  the non-negative (possibly spatially varying) scalar that takes the value $1$ in the convection dominated regions and is equal to the grid P\'eclet number in diffusion dominated regions. Since we take $l$ (see Eqs.~\eqref{eq:Lp_crystal} and  \eqref{eq:Lp_j2}) to be very small, we choose $c = 1$ for MFDM calculations, unless stated otherwise.

%
%
\subsection{Weak form for  $\bff$}
For the dynamic and the quasistatic cases,  $\bff$ is determined in the domain at any time $t$ by evolving equation \eqref{eq:mfdm_fevol} in time. However, \added{since we solve the rate form of the equilibrium equation \eqref{eq:rate_form}  to generate the current configuration at discrete times, the discretely evolving $\bff$ and $\bfchi$ fields generate a stress field that may not satisfy discrete force balance on the current configuration. To correct for this, we periodically (see Table \ref{tab:quasi_algo}) solve the equilibrium equation \eqref{eq:mfdm_f}  to satisfy balance of forces which is now posed as a traction boundary value problem, with the boundary data implemented in the form of nodal reaction forces which are obtained by integrating the nodal reaction force rate as mentioned in Sec.~\ref{sec:fem_v} (see the discussion surrounding Eq.~\eqref{eq:traction_rate}) and minimal kinematic constraints to eliminate rigid deformation modes}. Solving the equilibrium equation on a given configuration  amounts to adjustment of the solution for $\bff$ obtained by solving Eq.~\eqref{eq:fem_fevol} as detailed in Sec.~\ref{sec:fem_staticfSolve} below. 

\subsubsection{Evolution of $\bff$}
\label{sec:fem_fevol}
The evolution equation  \eqref{eq:mfdm_fevol} for $\bff$ is solved on the current configuration at each time step with the natural b.c.s defined in Sec.~\ref{sec:boundary_conditions} imposed on the external boundary.  Letting $\bfY = (\bfalpha\times \bfV + \bfL^p - \dot{\bfchi} - \bfchi \bfL)$ and using a forward Euler scheme to update $\bff$, the weak form of \eqref{eq:mfdm_fevol} is  
\begin{align}
\int_{\mOmega^n} grad\,{\bff}^{n+1}: {grad\,\delta{\bff}} \, dV = \Delta t^n \int_{\mOmega^n} \bfY^n:{grad \,\delta{\bff}} \, dV + \int_{\mOmega^n} grad\,{\bff}^{n}: {grad\,\delta{\bff}} \, dV.
\label{eq:fem_fevol}
\end{align}
The  weak form implies the satisfaction of the natural boundary conditions as mentioned in Sec.~\ref{sec:boundary_conditions}.  We specify $\bff^{n+1}$ (equivalent to $\dot{\bff} = 0$) at an arbitrary point at all times to ensure a unique solution,  without loss of generality.

\subsubsection{Adjusting $\bff$ from equilibrium equation}

At small deformation, the equilibrium equation is linear in $\bff$ and can be solved in a single iteration. However, it is nonlinear in $\bff$ at finite deformation and therefore  Newton-Raphson technique is used to solve for $\bff$ at finite deformation. Here, we present the weak form for both the cases i) small deformation and ii) finite deformation.  After $\bff$ is determined, the stress for Saint-Venant-Kirchhoff and Neo-Hookean materials are given by Eqs. \eqref{eq:stress_nh} and \eqref{eq:stress_svk}, respectively.


\paragraph{Small deformation}
\label{sec:fem_z}
For the linear theory, $\bfW \approx \bfI - \bfU^e$ where   $\bfU^e = grad\bfz - \bfchi$ and $curl\bfW = -curl\bfU^e = -\bfalpha$. This implies
\begin{align}
\bfW &\approx \bfI - \bfU^e \nonumber\\
grad\bff +\bfchi &\approx  grad\bfx - grad\bfz + \bfchi \nonumber\\
\implies \bff &\approx \bfx - \bfz ~~~(\text{upto a constant})\label{eq:f_z_relation}
\end{align}
where $\bfx$ represents the points in the current configuration. At small deformation, the stress is a linear function of $grad\bfz$ which is related to $grad\bff$ as shown in Eq.~\eqref{eq:f_z_relation}.  Once $\bfchi$ is known, the residual for the equilibrium equation in the absence of body forces \cite{roy2005finite, zhang2018finite} is given as 
\begin{align}
\begin{split}
R(\bfz) &= \int_{\partial\mOmega} t_{i}\, \delta z_{i} \, dA - \int_\mOmega T_{ij} \delta z_{i,j} \, dV.
\end{split}
\label{eq:fem_fsmall}
\end{align}

The  Jacobian of the system is calculated by taking a variation of the residual \eqref{eq:fem_fsmall} in the direction $d\bfz$. For finite element mesh nodes $A$ and $B$, the discrete form of the Jacobian matrix  corresponding to the degree of freedom pair $\{(A,a),(B,b)\}$ is expressed as 
\begin{align*}
\begin{split}
J^{AB}_{ab} = - \int_\mOmega \parderiv{N^A}{x_j} \parderiv{T_{aj}}{(grad z)_{bc}}\parderiv{N^B}{x_c} dV.
\end{split}
\end{align*}
 The calculation of $\parderiv{\bfT}{(grad\bfz)}$ for the Saint-Venant-Kirchhoff and the Neo-Hookean materials is shown in Sections \ref{app:dTdH_svk} and \ref{app:dTdH_nh} respectively. After solving for $\bfz$, $\bff$ can be updated following the relation from Eq.~\eqref{eq:f_z_relation}.


\paragraph{Large deformation}
\label{sec:fem_staticfSolve}
We use the Newton-Raphson scheme to solve for $\bff$ at large deformation as the governing equation $div[\bfT(\bff,\bfchi)]= \bf0$ is nonlinear in $\bff$.  Following the scheme outlined in \cite{puri2009modeling}, we write the residual from the variational statement for  \eqref{eq:mfdm_f}$_1$ 
\begin{align}\label{eq:fem_flarge}
R(\bff) &=  \int_{\partial \mOmega} t_i \delta f_i \, dA - \int_{\mOmega} T_{ij}\delta f_{i,j} \, dV.
\end{align} The discrete form of the Jacobian matrix  corresponding to the degree of freedom pair $\{(A,a),(B,b)\}$ is expressed as 
\begin{align*}
J^{AB}_{ab} = - \int_{\mOmega} \parderiv{N^A}{x_j} \parderiv{T_{aj}}{F^e_{mn}} \parderiv{F^e_{mn}}{W_{bc}}  \parderiv{N^B}{x_c} \, dV.
\end{align*} 
The calculation of $\parderiv{\bfT}{\bfF^e}$  for the Saint-Venant-Kirchhoff and the Neo-Hookean materials is shown in Appendices  \ref{app:dTdFe_svk} and \ref{app:dTdFe_nh}, respectively.  The guess for this Newton-Raphson solve is crucial for success in solving for $\bff$. We denote this guess as $\bff_0$ and it is determined, following \cite{zhang2018finite}, as follows:
\begin{itemize}
	\item For ECDD solves ($t = 0$), $\bff_0$ is obtained  by solving the equilibrium equation on the current configuration by assuming small deformation as shown above in Sec.~\ref{sec:fem_z}.
	
	\item At any other time ($t > 0$), $\bff^{n+1}$ obtained by solving the evolution equation \eqref{eq:fem_fevol} serves as the guess $\bff_0$ for the Newton-Raphson based scheme.
\end{itemize}

The nonlinear system is then iteratively solved until the norm of the discrete residual $|R^A_a|$ is  less than a tolerance of $10^{-12}\,g_0\,h^{2}$ (in $2$-d), where $h$ denotes the length of the smallest edge of an element in the finite element mesh.

%


\section{Algorithms}
\label{sec:algorithm}
We choose a combination of explicit-implicit schemes to evolve the coupled system (Eqs.~\eqref{eq:fem_v}, \eqref{eq:fem_chi}, \eqref{eq:mfdm_fem_a_exp}, \eqref{eq:fem_fevol}) in time. An efficient time stepping criteria based on plastic relaxation, and purely elastic and ‘yield strain’ related physical model parameters has been developed. Furthermore, to ensure robust and stable evolution of state variables, an intricate cut-back algorithm is used that carefully controls the magnitude of  plastic strain in each increment.

The following notation is used for the description of the algorithm:
\begin{enumerate}
	\item $(\cdot)^n$ means a quantity at time $t^n$. $\bfx^n$ represents the coordinates of the finite element mesh on the configuration $\mOmega^n$.
	\item At any integration point $q$, the following state variables are stored at any given time $t^n$:
	material strength $g^n$, elastic distortion tensor ${\bfF^e}^n$, Cauchy stress ${\bfT}^n$, dislocation velocity ${\bfV}^n$, slip distortion rate ${\bfL^p}^n$. We will collectively refer to them as ${PH}^n$ (short for point history) of integration points.
	\item $\Delta t^n$ is defined as $t^{n+1}-t^n$. To evaluate $\Delta t^n$ we first calculate the following variables  at each time-step
	\begin{align*}
	\Delta t_1 &= \dfrac{\xi\, h}{\max(|\bfV^n|)}\\[.2cm]
	\Delta t_2 &= \dfrac{.002}{\max(|{\bfF^e}^n\bfalpha^n \times \bfV^n|)+ \max(\hat{\gamma}^n_{sd})}\\[.2cm]
	\Delta t_3 &= \dfrac{\xi\,g_0}{E\, \max(|\bfL^n|)}\\[.2cm]
	\Delta t_4 &= \dfrac{\xi \,h}{v_s}
	\end{align*}
	where $\max(\cdot)$ denotes the maximum of the quantity $(\cdot)$ over all integration points in the domain, $h$ denotes the length of the smallest edge of an element in the finite element mesh, $v_s$ is the shear wave speed of the material, and $\xi$ is a scalar currently chosen to be $0.1$.
	
$\Delta t_1$ and $\Delta t_4$ relate to the Courant conditions for numerical stability related to dislocation motion and elastic wave propagation (in the dynamic case), respectively; their specifications above enforce that the respective waves are allowed to propagate a fraction of $h$ in any given time step. $\Delta t_2$ ensures that the maximum plastic strain increment at any given point in a time step has an upper bound of $0.2\%$. $\Delta t_3$ puts a bound on the maximum strain increment that can be attained in a time step at any point in the domain.  The diffusive term in Eq.~\eqref{eq:fem_a} is treated implicitly and therefore it does not pose any restriction on the time step selection criteria. 	$\Delta t^n$ is then given as
	\begin{align}
	\Delta t^n = \begin{cases}
	\min \left(\Delta t_1, \Delta t_2, \Delta t_3 \right) & \text{Quasistatic case} \\
	\min \left(\Delta t_1, \Delta t_2, \Delta t_3, \Delta t_4 \right) & \text{Dynamic case} \\
	\end{cases} 
	\label{eq:delta_t}
	\end{align}
	
\end{enumerate}

The algorithms for the quasistatic and dynamic cases are shown in Tables \ref{tab:quasi_algo} and \ref{tab:dyna_algo}, respectively. 

\subsection{Quasistatic case}

\begin{singlespace}
	
	\begin{longtable}[t]{|p{0.97\textwidth}|}
	\myhline
		\textbf{Given}: material properties, initial conditions, boundary conditions, and applied loading conditions.\\
\myhline
		\textbf{Step 1}: Finding the initial stress field on the body in `as-received' configuration - ECDD solve.
		\begin{itemize}
			\item ECDD solve, mentioned in Sec.~\ref{sec:initial_conditions}, is done on the initial configuration, i.e.~current configuration at $t = 0$.
			\item This gives $\bff$, $\bfchi$, and $\bfT$ on the configuration of the body at $t=0$.
		\end{itemize}\\
\myhline
		\textbf{Step 2}: Evolution of the system:
		Assume that the state at time $t^n$ is known:
		$\bfx^n$, $\bfalpha^n$, $\bff^n$, $\bfchi^n$,  ${\dot{\bfchi}}^n$, $\bfV^n$, $\bfL^p$, $\Delta t^n$, $PH^n$ \newline
		To get the state at time step $t^{n+1}$ the following is done:
		\begin{itemize}
			\item The rate form of the equilibrium equation \eqref{eq:rate_form} is solved on $\mOmega^n$ to get the material velocity $\bfv^{n}$ using Eq.~\eqref{eq:fem_v}. 
			\item Weak form of $\bfalpha$ evolution equation \eqref{eq:fem_a} is solved on $\mOmega^n$ to obtain  $\bfalpha^{n+1}$  on $\mOmega^{n+1}$. 
			
			\item The configuration of the body is discretely updated i.e.~$\bfx^{n+1} = \bfx^{n} + \bfv^n \,\Delta t^n$.
			
			\item $\bfchi^{n+1}$ on $\mOmega^{n+1}$ is obtained by solving Eq.~\eqref{eq:fem_chi} on $\mOmega^{n+1}$. 
			
			\item $\bff^{n+1}$ on $\mOmega^{n+1}$ is obtained by doing one of the following:
			\begin{enumerate}
				\item Solve Eq.~\eqref{eq:fem_fevol} on  $\mOmega^{n}$ to obtain $\bff^{n+1}$. 
				
				\item Solve equilibrium equation \eqref{eq:mfdm_f} in alternate increments to adjust $\bff^{n+1}$  on $\mOmega^{n+1}$ as shown in Eq.~\eqref{eq:fem_flarge}. $\bff^{n+1}$ obtained by solving Eq.~\eqref{eq:fem_fevol} serves as the initial guess for the Newton Raphson scheme.
				
			\end{enumerate}
			
			\item $\dot{\bfchi}^{n+1}$ is calculated  as follows: $\dot{\bfchi}^{n+1} = \dfrac{\bfchi^{n+1} - \bfchi^n}{\Delta t^n}$.
			
			\item $PH^{n+1}$ is updated on the configuration $\mOmega^{n+1}$.
		\end{itemize}\\
		
\myhline
\textit{State acceptance criteria:}
 Let $PSR = ({\max(|\bfF^e\bfalpha \times \bfV|^{n+1})+ \max(\hat{\gamma}_{sd}^{n+1}))}$. If $PSR  \times \Delta t ^n \le 0.002$, the state is accepted. $\Delta t^{n+1}$, based on the new state, is calculated from  \eqref{eq:delta_t} and this algorithm is repeated to get state at increment $t^{n+2}$.
		If the condition is not satisfied:
		\begin{itemize}
			\item Go back to the state at time $t^n$.
			\item Use $\Delta t^{n, new} =  \min \left(\frac{0.002}{PSR}, 0.5\, \Delta t^n \right)$ and repeat the algorithm to obtain a new state at $t^{n+1}$.
		\end{itemize} \\
	\hline
		\caption{Quasi-static MFDM algorithm.}
		\label{tab:quasi_algo}
	\end{longtable}
	
\end{singlespace}

\subsection{Dynamic case}

\begin{singlespace}

	\begin{longtable}[t]{|p{0.95\textwidth}|}
	\myhline

		\textbf{Given}: material properties, initial conditions, boundary conditions, and applied loading conditions.\\
		\hline\\[-2pt]
		\textbf{Step 1}: ECDD solve as outlined in the Table \ref{tab:quasi_algo} is done on  the initial configuration, i.e.~current configuration at $t = 0$, to determine $\bff$, $\bfchi$, and $\bfT$ at $t=0$. \\
	\myhline

		\textbf{Step 2}: Evolution of the system:
		Assume that $\bfv^{n-1}$  and the state at time $t^n$ is known:
		$\bfx^n$, $\bfalpha^n$, $\bff^n$, $\bfchi^n$,  ${\dot{\bfchi}}^n$, $\bfV^n$, $\bfL^p$, $\Delta t^n$, $PH^n$ \\
		To get the state at time step $t^{n+1}$ the following is done:\\[-1.5em]
		\begin{itemize}
			\item The balance of linear momentum Eq.~\eqref{eq:rate_form} is solved on $\mOmega^n$ to get the material velocity $\bfv^{n}$ using  Eq.~\eqref{eq:fem_vdyna}.
			
			\item Weak form of $\bfalpha$ evolution equation \eqref{eq:fem_a} is solved on $\mOmega^n$ to obtain  $\bfalpha^{n+1}$  on $\mOmega^{n+1}$.
			
			\item  $\bff^{n+1}$ on $\mOmega^{n+1}$ is obtained by solving Eq.~\eqref{eq:fem_fevol}.
			
			\item The configuration of the body is discretely updated i.e.~$\bfx^{n+1} = \bfx^{n} + \bfv^n \,\Delta t^n$.

			\item $\bfchi^{n+1}$ on $\mOmega^{n+1}$ is obtained by solving Eq.~\eqref{eq:fem_chi} on $\mOmega^{n+1}$.
			\item $\dot{\bfchi}^{n+1}$ is calculated  as follows: $\dot{\bfchi}^{n+1} = \dfrac{\bfchi^{n+1} - \bfchi^n}{\Delta t^n}$.
			
			\item $PH^{n+1}$ is updated on the configuration $\mOmega^{n+1}$.
		\end{itemize}\\
		
\myhline
\textit{State acceptance criteria} is same as in the quasistatic case listed in Table \ref{tab:quasi_algo}.\\
		\hline
		\caption{Dynamic (with inertia) MFDM algorithm.}
		\label{tab:dyna_algo}
	\end{longtable}
\end{singlespace}

%
%

\subsection{Classical plasticity}
\label{sec:new_classical_algo}
 We solve problems of classical plasticity at finite deformation by considering the system
\begin{subequations}
	\begin{align}
	div\left[ tr(\bfL) \, \bfT + \dot{\bfT} - \bfT\bfL^T \right] = \bf0,\label{eq:inclusion_conv_v}\\
	\dot\bfW + \bfW\bfL = \bfL^p,\label{eq:inclusion_conv_wdot}\\
	div[\bfT(\bfW)]  = \bf0,\label{eq:inclusion_conv_t}
	\end{align} 
	\label{eq:inclusion_conv}
\end{subequations}
along with the evolution of the material strength $g$ \eqref{eq:softening} with $l$ and $k_0$ set to $0$. For a given state of the system ($\bfx^n, \bfW^n, \bfL^{p,n}$) at any time $t^n$, the solution to the system \eqref{eq:inclusion_conv} is obtained through  the following steps:

\begin{itemize}
	\item Solve Eq.~\eqref{eq:inclusion_conv_v} to obtain material velocity $\bfv^n$ using  Eq.~\eqref{eq:fem_v}.
	\item Evolve Eq.~\eqref{eq:inclusion_conv_wdot} locally at integration points to obtain $\tilde\bfW \equiv \bfW^{n+1}$
	\item The geometry is updated i.e. $\bfx^{n+1} = \bfx^x + \bfv^n\Delta t^n$.
	\item With $\bfW$ then written as $\bfW = \tilde{\bfW} + grad\bfw$,  \eqref{eq:fem_flarge} is then used to solve  \eqref{eq:inclusion_conv_t} for $\bfw$ on the updated configuration to maintain balance of forces which amounts to adjustment of $\bfW^{n+1}$. 
   \item $PH^{n+1}$ is updated on the new configuration.  
   \item \textit{State acceptance criteria} is same as in the quasistatic case  listed in Table \ref{tab:quasi_algo}.
  
\end{itemize}
The algorithm above is novel for solving classical plasticity problems at finite deformation.
 


\section{Results and Discussion}
\label{sec:results}

An MPI-accelerated finite element based computational framework for the full finite deformation MFDM is developed using C++.   The framework is based on the algorithms presented in Sec.~\ref{sec:algorithm} and uses comprehensive state-of-the-art libraries  Deal.ii \cite{dealII85}, P4est \cite{BursteddeWilcoxGhattas11}, MUMPS \cite{MUMPS:1}, and PetSc \cite{petsc-web-page}.  A  post-processing toolbox has been   developed in Python using its Scipy \cite{scipy}, Numpy \cite{numpy}, Pandas \cite{mckinney2010data, mckinney2011pandas}, Matplotlib \cite{matplotlib}, and Seaborn \cite{michael_waskom_2018_1313201} modules to produce publication-quality figures. The figures presented in this document are obtained using this developed toolbox.

 The finite element implementation is quite efficient. Table \ref{tab:comp_effeciency} presents average wall-clock times for $2$-d simulation for the quasistatic and dynamic evolution problems at finite deformation presented in Sections \ref{sec:results_mfdm_inc} and  \ref{sec:dyna_fdm_mach}, respectively.

\begin{table}[H]
	\centering
	\begin{tabular}{ c  c c  c}
		\hline
		Case &  Number of Nodes & Processors & Wall-clock time (Hours) \\ 		\hline\\[-.8em]
		Quasistatic & $47,241$ & $56$ & $3$\\     
		Dynamic &  $86,876$ & $84$ & $7$\\     \hline
	\end{tabular}
	\caption{Wall clock times for typical simulations.}
	\label{tab:comp_effeciency}
\end{table}

Here, we present results of some selected physically meaningful verification tests. To our knowledge, the results presented here are also the first fully nonlinear results (in kinematics, elasticity, and dissipation) involving dislocation mediated plasticity.  The organization of this section is as follows:


\begin{enumerate}
	
\item In Sec.~\ref{sec:NLE_ss}, we verify the framework by studying the problem of homogeneous elastic deformation, under simple shear and extensional loadings (both quasistatic), of  blocks of Saint-Venant-Kirchhoff and Neo-Hookean materials. When the velocity boundary conditions are applied for an assumed homogeneous purely elastic deformation history,  it is expected that the numerics should reproduce the homogeneous deformation with no numerically induced hysteresis upon unloading. However, given  the extensive use of the incremental equilibrium equation \eqref{eq:rate_form} in our scheme,  it is not a priori clear that no hysteresis is induced in the numerical approximation. This overall test also verifies the algorithm for the accumulation of reaction forces due to velocity-Dirichlet boundary conditions given in Sec.~\ref{sec:fem_v} (discussion surrounding  Eq.~\eqref{eq:traction_rate}).

\item In Sec.~\ref{sec:res_sesd}, we calculate the finite deformation stress fields of a screw dislocation in a body of finite extent that is assumed to behave  as a compressible  Neo-Hookean material, and verify it with the analytical solution for the same case. We then calculate the finite deformation stress field of an edge dislocation in a Saint-Venant-Kirchhoff material. The results demonstrate significant  deviations from small deformation  closed-form linear elastic solution for stress fields.

\item In Sec.~\ref{sec:res_bvc}, the framework is  verified against the elastic loading of a Saint-Venant-Kirchhoff material with defect evolution in the special case of no generation or motion of the  defect relative to the material. The dislocation density evolution solely takes place due to its coupling with the motion in the  transport equation \eqref{eq:mfdm_a} through its convected derivative. In the absence of any flux of dislocations, the Burgers vector content of any arbitrary area patch has to be conserved.  Sec.~\ref{sec:res_bvc} verifies these hypotheses under quasistatic simple shear and extensional loadings.

\item Section \ref{sec:ha} presents the stress field of a spatially homogeneous dislocation distribution in the domain. We show contrasting predictions of the stress field  by the linear and nonlinear FDM theory.

\item Section \ref{sec:results_mfdm_inc} studies the effect of inclusion size on the strength of  a model composite. We present these results for the Crystal and $J_2$ plasticity MFDM models and demonstrate that for a given volume fraction of inclusions, the material strength is enhanced for smaller inclusion sizes.

\item  Sec.~\ref{sec:dyna_fdm} presents results for   elastodynamics with finite deformations of moving dislocations with specified velocity. We show the evolution of the dislocation density and plastic deformation in the body. The result for the motion of a single dislocation core can be interpreted as the longitudinal propagation of a shear band.  Sec.~\ref{sec:dyna_fdm_mach} shows the formation of the Mach cone in the body when the dislocation moves at a speed higher than the (linear elastic) shear wave speed of the material.  The geometric nonlinearity has an effect that the observed Mach cone is unsymmetric under prestress.

\end{enumerate}

For all the results presented in this work, the input flux $\bfalpha (\bfV\cdot\bfn)$ and $curl\bfalpha \times \bfn$ are assumed to be  $\bf0$ on the boundary. Also, $\hat\bfL^p$ is directly evaluated at the boundary to calculate $\hat\bfL^p \times \bfn$. All fields are interpolated using element-wise bilinear/trilinear interpolation in $2$-d/$3$-d, unless otherwise stated.  The Burgers vector content of an area patch $A$ with normal $\bfn$ is given by
\begin{align}
\bfb_A = \int_A \bfalpha \bfn \, dA,
\label{eq:bv_area_patch}
\end{align} where $\bfalpha$ denotes the dislocation density field in the domain. When the dislocation distribution $\bfalpha$ is localized such that it is enclosed with the area patch $A$,  we denote its Burgers vector by $\bfb$. It must be noted that $\bfb$ is independent of the chosen area patch $A$.  We refer to the magnitude of Burgers vector, $|\bfb|$, as the strength of the dislocation. $b$ is a material constant which refers to the Burgers vector magnitude of a full dislocation in the crystalline material. $h$ denotes the length of the smallest edge of an element in the finite element mesh under consideration. We define  a dimensional measure of  magnitude of the dislocation density as  $\rho_g(\bfx, t) :=  \frac{|{\bfalpha}(\bfx,t)|}{b}$.

All algorithms in this paper have been verified to reproduce classical plasticity solutions for imposed homogeneous deformation histories by comparison with solutions obtained by integrating the evolution equation \eqref{eq:conventional_Fedot} for the elastic distortion tensor $\bfF^e$  to determine the Cauchy stress response for an imposed spatially homogeneous velocity gradient history, $\bfL$:
\begin{align}
\begin{split}
& \dot\bfF^e = \bfL\bfF^e - \bfF^e\bfL^p\bfF^e =: \tilde\bff(\bfF^e, g),\\
&\dot g = \tilde{g}(\bfF^e, g),
\end{split}
\label{eq:conventional_Fedot}
\end{align}
where $\bfL^p$ is defined from Eq.~\eqref{eq:Lp_j2} or \eqref{eq:Lp_crystal}  with $l = 0$, and $\tilde{g}$ is given by \eqref{eq:softening} with $k_0 = 0$.

A typical schematic of the basic geometry used in most problems (further details are mentioned as required) is shown in Fig.~\ref{fig:sch_homo_nle}. The stress-strain behavior of the body under shear loading is modeled by plotting the  averaged $T_{12}$ component of the stress tensor on the top surface, which is denoted by $\tau$. $\tau$ is calculated by  summing the tangential components of the nodal reaction force on the top surface and then dividing by the current area (line length) of the surface. The stress-strain behavior of the body under  extensional loading is modeled by plotting the averaged $T_{11}$ component of stress, on the right surface which is denoted by $\sigma$. $\sigma$ is calculated by  summing the normal components of the nodal reaction force on the right surface and then dividing by the current area (line length) of the surface. $\hat\mGamma$ represents the applied  strain rate. The shear and extensional strains are denoted by $\mGamma$ and $\epsilon$, respectively. These strains are engineering strains and are calculated as $\hat\mGamma t$  at any time $t$. 



\subsection{Nonlinear elasticity}
\label{sec:NLE_ss}
Due to our interest in calculating hyperelastic stress fields of dislocations, it is essential to make sure that the scheme accurately reproduces classical hyperelastic response. This includes the prediction of no hysteresis in an elastic loading-unloading cycle despite the extensive use of \eqref{eq:rate_form}.

A $2$-d plane strain problem is set up as follows: a  body of size ${(1\,mm)}^2$  (the size is immaterial) is set up for homogeneous extension and simple shear loadings  with details below. Upon reaching $100\%$ strain, the loading is reversed and the body is unloaded to its original configuration. The material constants $E$ and $\nu$ are chosen to be $62.78$  GPa and $0.3647$, respectively. A strain rate $\hat{\mGamma} = 1s^{-1}$ is used for both the loading cases.

\begin{figure}
	\centering
\begin{minipage}[b]{.650\linewidth}
	\centering
	{\includegraphics[width=\linewidth]{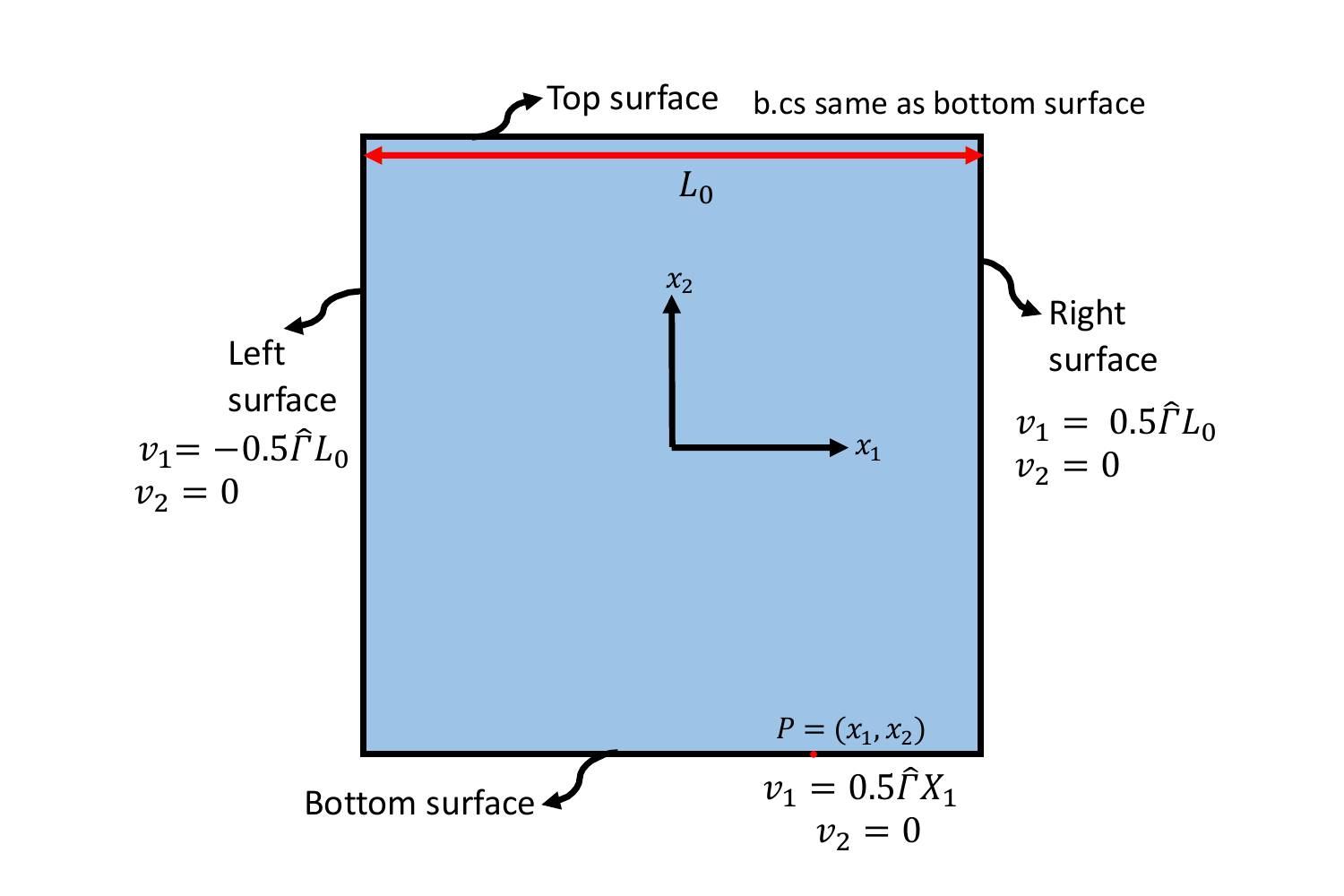}}
	\caption{Schematic of the geometry for extensional loading.}
	\label{fig:sch_homo_nle}
\end{minipage}
\end{figure}

To model a purely elastic process, the dislocation velocity and the plastic strain rate due to SDs are assumed to vanish i.e.~$\bfL^p = \bf0$ and $\bfV = \bf0$.   The  velocity boundary conditions for the simple shear loading are as follows:  at any point $P = (x_1, x_2)$ on the boundary in the current configuration, a velocity of the form $v_2 = 0$ and $v_1 = \hat\mGamma y(x_2)$ is imposed, where $y(x_2)$ is the height of the point $P$ from the bottom surface.  For the extension case, at the point $P = (x_1, x_2)$, the velocity boundary conditions of the form  $v_2 = 0$ and $v_1 = 0.5\hat\mGamma X_1$ are applied, where $\bfX$ denotes the coordinates of $\bfx$ in the configuration at $t = 0$ (reference configuration). The schematic of the set up for extensional loading is shown in Figure \ref{fig:sch_homo_nle}. 

\begin{figure}
	\centering
	\begin{subfigure}[b]{0.495\linewidth}
		\centering
		{\includegraphics[width=\linewidth]{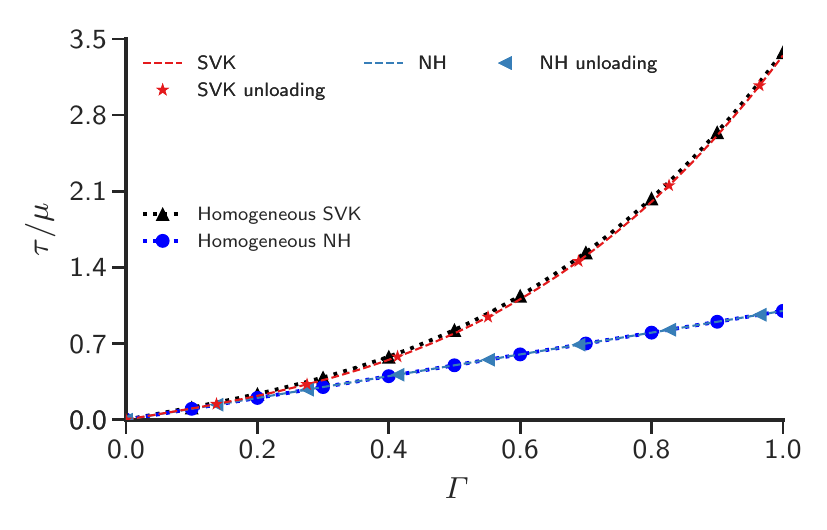}}
		\caption{}
		\label{fig:nle_ss}
	\end{subfigure}%
	\begin{subfigure}[b]{0.495\linewidth}
		\centering
		{\includegraphics[width=\linewidth]{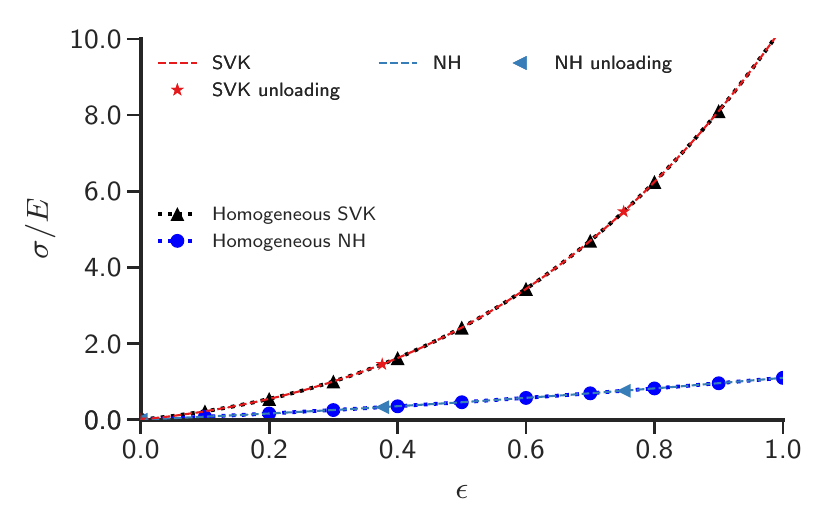}}
		\caption{}
		\label{fig:nle_st}
	\end{subfigure}%
	\caption{Stress-strain response for nonlinear elastic deformation a) Simple shear b)  Uniaxial  extension.}
\end{figure}

We plot the stress-strain response for the Saint-Venant-Kirchhoff (SVK) and the Neo-Hookean (NH) materials for the extension ($\sigma$ vs.~$\epsilon$) and shear ($\tau$ vs.~$\mGamma$) loadings in Figures \ref{fig:nle_ss} and \ref{fig:nle_st},  respectively.  The stress-strain plots overlap with the corresponding  homogeneous  deformation solutions obtained by simply evaluating the necessary tractions corresponding to the appropriate elastic stress-strain relationship \eqref{eq:stress_svk}-\eqref{eq:stress_nh} for the given imposed deformation history. The cyclic stress-strain curves also overlap each other, and the overall response does not show any hysteresis.

The effect of variation in stress with large homogeneous rigid rotations is studied next for the Saint-Venant-Kirchhoff material. We calculate the stress-strain response for a  a simple shear deformation superposed with rigid body motion given by
\begin{align}
\bfx^*(\bfX, t) = \bfQ(t)\bfF^{ss}(t)\bfX,
\label{eq:nle_xstar}
\end{align}
where $\bfX$ and $\bfx^*$ are the coordinates of the body in the reference and the current configurations, respectively. $\bfF^{ss}$ denotes the deformation gradient corresponding to a homogeneous simple shearing motion with $F^{ss}_{12} = \hat{\mGamma}\,t$ and $F^{ss}_{11} = F^{ss}_{22} = F^{ss}_{33} = 1$. $\bfQ(t)$ is the rotation tensor written as
\begin{align}
\bfQ(t) &= \begin{bmatrix}
cos(\theta(t)) & -sin(\theta(t)) & 0 \\
sin(\theta(t)) & cos(\theta(t)) & 0 \\
0 & 0 & 1
\label{eq:nle_Q}
\end{bmatrix},
\end{align}  where $\theta(t) = \omega t$ at any time $t$ and $\omega = 2\, rad.\,s^{-1}$ is a  (constant) angular  speed about the $x_3$ axis.   The velocity boundary conditions follow from 
\begin{align*}
\bfv^*(\bfX, t) = \dot{\bfQ}(t) \bfF^{ss}(t)\bfX + \bfQ(t) \dot{\bfF^{ss}}(t)\bfX,
\end{align*} evaluated on the boundary of the reference configuration.

Under the superposed rigid body motion defined by Eq.~\eqref{eq:nle_xstar}, the stress tensor for any frame-indifferent stress response function is given as  $\bfT^{*}(t) = \bfQ(t)\bfT^{ss}(t)\bfQ^T(t)$, where $\bfT^{ss}(t)$ denotes the stress field for the  simple shearing motion defined  by $\bfF^{ss}$. We compare the maximum error in the stress up to  $100\%$ strain defined by  $\displaystyle{\max_{(\bfX, t)}}$\,$ \frac{|\bfT(\bfX,t) - \bfT^{*}(\bfX,t)|}{|\bfT^*(\bfX,t)|}$, where $\bfT$ is the computed solution. The error accumulates at a very slow rate, leading to a maximum error of $\approx 1\%$ at $\mGamma = 1$.

We therefore conclude that the framework  is capable of dealing adequately with nonlinear elasticity, without any (numerically induced) hysteresis/dissipation. Moreover,  the protocol for accumulation of reaction tractions  (Sec.~\ref{sec:fem_v}, discussion surrounding  Eq.~\eqref{eq:traction_rate}) discretely in time is also  suf\mbox{}ficiently precise in dealing with large strains and rotations.

\subsection{Stress f\mbox{}ields of single dislocations}
\label{sec:res_sesd}
We calculate the finite deformation stress fields of single dislocations as finite element solutions of the ECDD system (Eqs.~\eqref{eq:ECDD} and \eqref{eq:ECDD_bc})  as explained in Sec.~\ref{sec:initial_conditions}. The material is assumed to be elastically isotropic, but this is not a restriction of the developed methodology \cite[Sec. 5.8.1]{zhang2018finite}. The material constants $E$ and $\nu$ are chosen to be $200$  GPa and $0.30$, respectively.  We define a difference measure $M(\bfT^*, \bfT)$ between two tensors (scalars, or tensor components) fields $\bfT^*$ and $\bfT$ as
	\begin{equation}
	M(\bfT^*, \bfT) = \dfrac{|\Delta \bfT|}{|\bfT|} = \frac{| \bfT^* - \bfT|}{|\bfT|}.
	\label{eq:ssd_M}
	\end{equation}

\begin{figure}
	\centering
		\centering
		{\includegraphics[width=.4\linewidth]{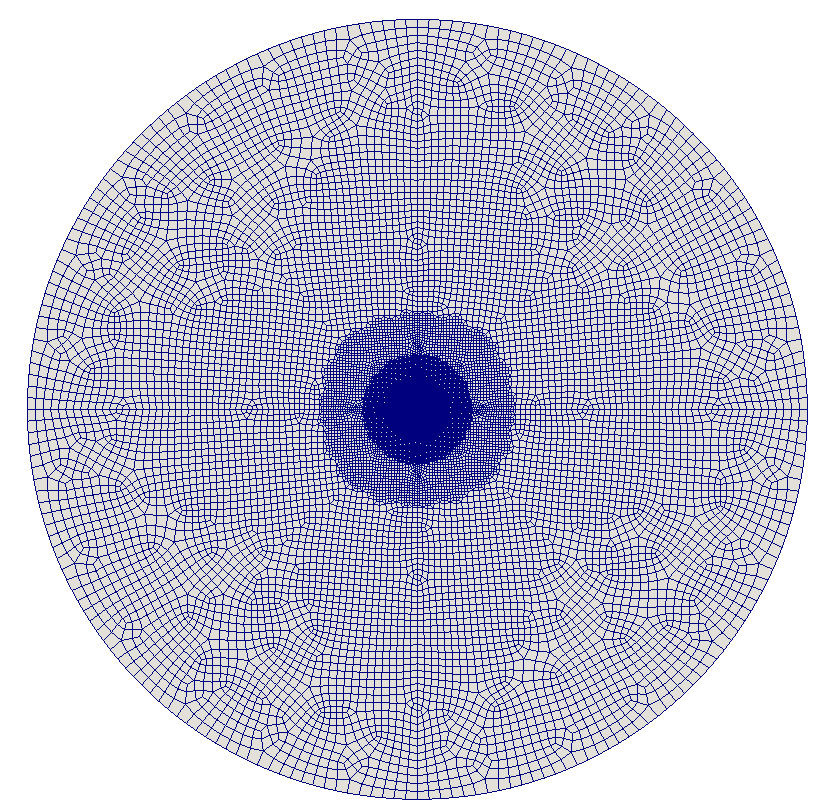}}
	\caption{Mesh on the $x_3 = 0$ plane for calculating the stress field of the screw dislocation.}
		\label{fig:ssd_mesh}
\end{figure}

\subsubsection{Screw dislocation}
A horizontal cylindrical plate, assumed to be thick in the $x_3$ direction and infinitely extended in the $x_1$-$x_2$ plane, has a screw dislocation embedded in it. The strength of the dislocation is assumed to be $b$ and its line direction is taken to be in the positive $x_3$ direction i.e.~$\bfb = b\bfe_3$. The body is discretized  with a non-uniform mesh of approximately $~125K$ elements which is refined near the core as shown in Fig.~\ref{fig:ssd_mesh}.  The dislocation  is modeled by specifying an $\bfalpha$ field of the form
\begin{equation}
\alpha_{33}(x_1, x_2, x_3) =  \begin{cases} 
\varphi_0 & r < r_0 \\
0 & r > r_0.
\end{cases}, ~~ \alpha_{ij} = 0 ~ \rm{if} ~ i \neq 3 ~\rm{and}~ j \neq 3, 
\label{eq:ssd_alpha33}
\end{equation}
where $r = \sqrt{x_1^2 + x_2^2}$ and $r_0$ is chosen to be $1.2b$. $\varphi_0$ is a constant chosen to make the dislocation Burgers vector equal to  $b\bfe_3$ by ensuring $\int_A \alpha_{33} \, dA = b$ on any cross section $A$ normal to $\bfe_3$ which encloses the dislocation  core, i.e. the disk $r\leq r_0$. This implies 
\begin{align*}
b = \int_0^{2\pi}\int_0^{r_0} \varphi(r) r \,dr\, d\theta ~; ~~ \varphi_0 = \dfrac{b}{\pi\,r_0^2}.
\end{align*}  We also assume that the front and back ends of the cylinder are capable of providing arbitrary tractions. An exact solution of the ECDD equations for this problem is developed in \cite{acharya2001model} for the incompressible Neo-Hookean material. That solution is easily adapted here to develop the same for the compressible Neo-Hookean material model whose elastic response is given by \eqref{eq:stress_nh}. We outline this exact solution first before using it for verification. The particular solution
$\bfW$ satisfying the ECDD equations \eqref{eq:ECDD} is given by $\bfI - H\bfalpha$, where the nonzero components of  $H\bfalpha$ are obtained as
\begin{align*}
H\alpha_{31}(x_1, x_2) = \frac{-x_2}{x_1^2 + x_2^2} \int_{0}^{r} \varphi(s) s\, ds ~~;~~
H\alpha_{32}(x_1, x_2) = \frac{x_1}{x_1^2 + x_2^2} \int_{0}^{r} \varphi(s) s\, ds,
\end{align*}
\begin{align*}
H\alpha_{31}(r) =  \begin{cases} 
-x_2 \dfrac{\varphi_0}{2} & r < r_0 \\[4mm]
-x_2 \dfrac{\varphi_0 r_0^2}{2r^2} & r > r_0
\end{cases} ~~~~;~~~~
H\alpha_{32}(r) =  \begin{cases} 
x_1 \dfrac{\varphi_0}{2} & r < r_0 \\[4mm]
x_1 \dfrac{\varphi_0 r_0^2}{2 r^2} & r > r_0.
\end{cases}
\end{align*}
The elastic distortion tensor $\bfF^{e}$ is then given by $\bfF^{e} = \bfW^{-1} = \bfI + H\bfalpha$ in this case. The exact stress field  $\bfT^*$, which satisfies equilibrium (without any further compensating fields) is  calculated from Eq.~\eqref{eq:stress_nh} as:
\begin{align}
\bfT^* &= \mu \begin{bmatrix}
0 & 0 & H\alpha_{31} \\
0 & 0 & H\alpha_{32} \\
H\alpha_{31} & H\alpha_{32} & (H\alpha_{31})^2+ (H\alpha_{32})^2 \label{eq:ssd_T_nh}
\end{bmatrix}.
\end{align}
To compute the finite element stress field, the problem is set  up in a full $3$-d setting as follows: we specify a dislocation density of the form given by Eq.~\eqref{eq:ssd_alpha33} in a cylinder of radius $50b$ extending from $x_3 = -25b$ to $x_3 = 25b$. To mimic the infinite domain size,  traction boundary conditions  corresponding to the analytical solution are imposed on the outer surface  including the front and the back of the cylinder of finite extent i.e.~$\bft = \bfT^*\bfn$ is used in Eq.~\eqref{eq:ECDD_bc}  where $r = 50b$ as well as $x_3 = \pm25b$ where $\bfT^*$ is  given by Eq.~\eqref{eq:ssd_T_nh}. 

\begin{figure}
	\centering
		\centering
		{\includegraphics[width=.65\linewidth]{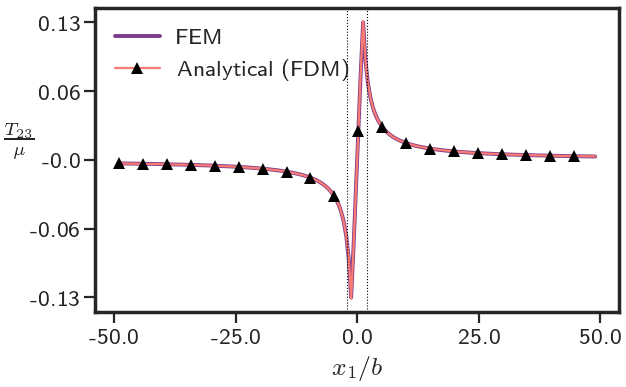}}

	\caption{Comparison of analytical solution of ECDD equations \eqref{eq:ECDD}  to its numerical solution for $T_{23}$ of the screw dislocation along $x_2 = 0$.}
			\label{fig:ssd_T23_x}
\end{figure}%

\begin{figure} 
	\centering
	\begin{subfigure}[b]{0.495\textwidth}
		\centering
		{\includegraphics[width=\linewidth]{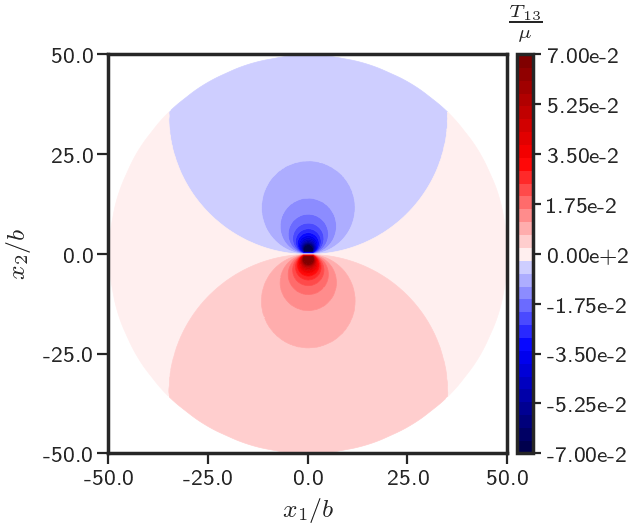}}
		\caption{}
		\label{fig:ssd_T13}
	\end{subfigure}%
	\begin{subfigure}[b]{0.495\textwidth}
		\centering
		{\includegraphics[width=\linewidth]{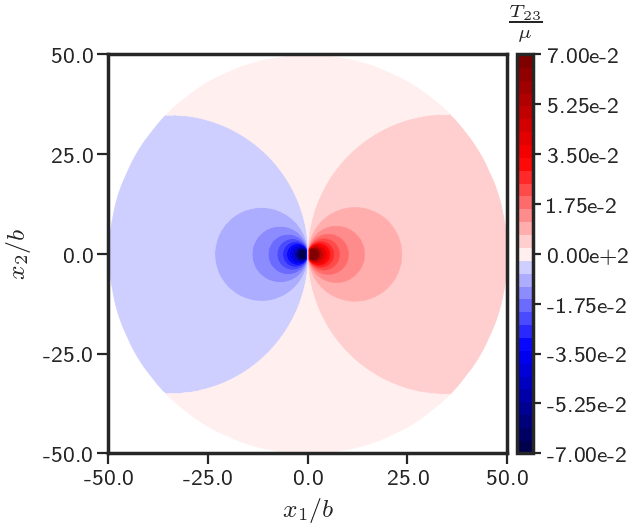}}
		\caption{}
		\label{fig:ssd_T23}
	\end{subfigure}
	\caption{Stress field of a screw dislocation embedded in the cylindrical domain behaving as a compressible Neo-Hookean material a) $\frac{T_{13}}{\mu}$ b) $\frac{T_{23}}{\mu}$.}
\end{figure}

Figure \ref{fig:ssd_T23_x} shows good  agreement of the numerically calculated stress field component $T_{23}$  plotted along $x_2 = 0$ on the $x_3 = 0$ plane  with the analytical result.  Figures \ref{fig:ssd_T13} and  \ref{fig:ssd_T23}  show finite deformation stress fields of the screw dislocation on the plane $x_3 = 0$ obtained from solving the ECDD system (Eqs.~\ref{eq:ECDD} and \ref{eq:ECDD_bc}) as shown in Sec.~\ref{sec:fem_staticfSolve}.


\begin{figure}
	\centering
	\begin{subfigure}[b]{0.495\textwidth}
		\centering
		{\includegraphics[width=\linewidth]{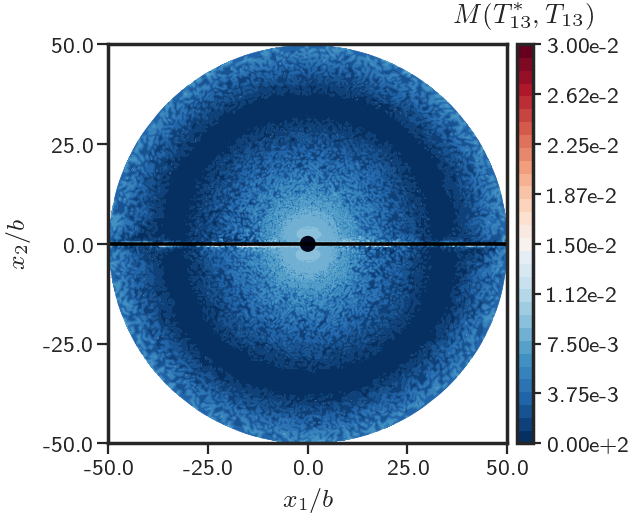}}
		\caption{}
		\label{fig:ssd_dT13}
	\end{subfigure}%
	\begin{subfigure}[b]{0.495\textwidth}
		\centering
		{\includegraphics[width=\linewidth]{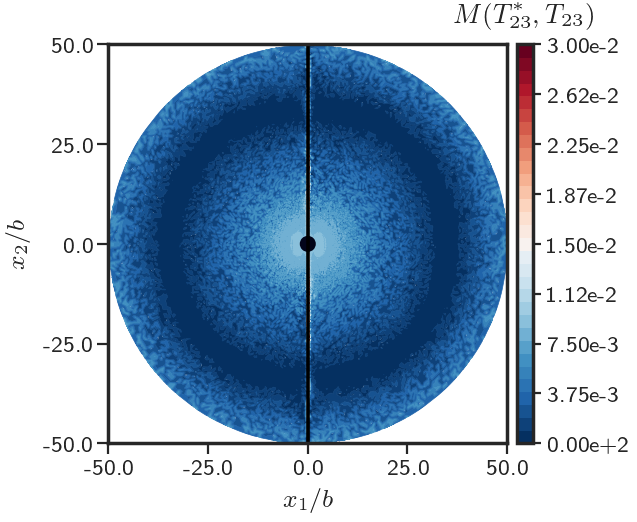}}
		\caption{}
		\label{fig:ssd_dT23}
	\end{subfigure}
	\caption{Comparison of the numerical and analytical solutions of stress field of screw dislocation in the cylindrical domain on the $x_3 = 0$ plane a) $M(T^*_{13}, T_{13})$ b) $M(T^*_{23}, T_{23})$.}
\end{figure}

Figures \ref{fig:ssd_dT13} and \ref{fig:ssd_dT23} show the relative difference between the numerical and analytical  stress fields.   The regions close to the core $(r \le 2b)$  as well as where the analytical stress components vanish have been marked by black lines. We can notice that  the error is less than $2\%$ everywhere.

\begin{figure}
	\centering
	{\includegraphics[width=.5\linewidth]{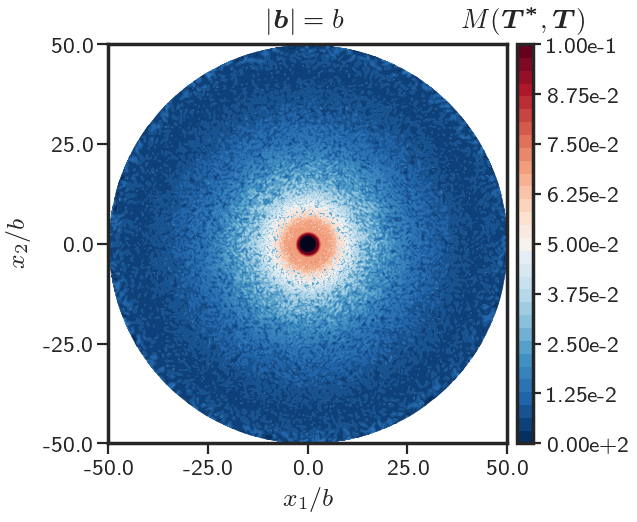}}
	\caption{Difference between stress field given by the finite deformation FDM theory and small deformation closed-form solution for a screw dislocation in a cylindrical domain for a Saint-Venant-Kirchhoff material.}
	\label{fig:ssd_fvs_sd_Tall}
\end{figure}

Next, for a Saint-Venant-Kirchhoff material, we compare the finite deformation stress field of a screw dislocation with the small deformation closed-form solution. The closed-form solution is given by \cite{hirth1982theory}
\begin{align}
\begin{split}
\label{eq:ssd_exact_sol_sd}
T^*_{13} &= -\frac{\mu b}{2\pi}  \cdot  \frac{x_2}{(x_1^2 + x_2^2)} \\
T^*_{23} &= \frac{\mu b}{2\pi}  \cdot \frac{x_1}{(x_1^2 + x_2^2)} \\
T^*_{11} &= T^*_{22} = T^*_{33} = T^*_{12} = 0.\\
\end{split}
\end{align}  The problem is set up for a Saint-Venant-Kirchhoff material in the same way as above except now the tractions imposed on the outer surface of the cylinder are determined by using $\bfT^*$ from Eq.~\eqref{eq:ssd_exact_sol_sd} in Eq.~\eqref{eq:ECDD_bc}.  The plot of $M(\bfT^*, \bfT)$ in the domain, shown in Figure \ref{fig:ssd_fvs_sd_Tall}, clearly displays that the stress fields differ in a   region around the core. Therefore, we establish that up to $\approx 6\%$ error arise as far as $10b$ from the core.



\subsubsection{Edge dislocation}
We calculate the finite deformation stress field of a single edge dislocation in a body and compare it with the closed-form classical (small deformation) linear elastic solution for the corresponding problem.  The small deformation closed-form solution for  stress field $\bfT^*$ for a single edge dislocation at the center of an infinite cylindrical solid is given by \cite{hirth1982theory}:
\begin{align}
\begin{split}
\label{eq:sed_exact_sol_ed}
T^*_{11} &= -D x_2  \left(- \frac{2x_2^2}{(x_1^2 + x_2^2)^2} +\frac{3}{(x_1^2 + x_2^2)} \right)\\ 
T^*_{22} &= -D x_2 \left(- \frac{2x_1^2}{(x_1^2 + x_2^2)^2} +\frac{1}{(x_1^2 + x_2^2)} \right)\\
T^*_{12} &= D x_1 \left(- \frac{2x_2^2}{(x_1^2 + x_2^2)^2} +\frac{1}{(x_1^2 + x_2^2)} \right) \\
T^*_{33} & = \nu (T_{11} + T_{22}), ~~~~ T^*_{13} = T^*_{23} = 0.
\end{split}
\end{align} where $D = \mu b  (2\pi (1-\nu))^{-1}$.  $x_1$ and  $x_2$ are the in-plane coordinates,  measured from the center of the dislocation. 
The computational problem is set up in a $2$-d plane strain setting for the Saint-Venant-Kirchhoff material as follows: an edge dislocation with a Burgers vector  $b\bfe_1$ and line direction $\bfe_3$ is assumed in a domain of dimensions $[-50b, 50b] \times [-50b, 50b]$. The edge dislocation is modeled by prescribing a dislocation density $\bfalpha$ at any $\bfx = (x_1, x_2)$ of the form
\begin{align}
\alpha_{13}(x_1, x_2)  &=  \begin{cases} 
\varphi_0 & |x_1| \le \frac{w}{2} \text{ and } |x_2| \le \frac{w}{2} \\
0 & \text{ otherwise},
\end{cases} ~~~ \alpha_{ij} = 0 \text{ if } i \ne 1 \text{ and } j \ne 3. 
\label{eq:sed_single_edge_core}
\end{align}
In this section, the core width $w$ is taken to be  $b$. The constant $\varphi_0$ is evaluated by making the Burgers vector of the dislocation equal to $b\bfe_1$, i.e. $\int_{A}\alpha_{13}\,dA = b$, where $A$ is any area patch in $\mOmega$ that encloses the dislocation core. Tractions $\bft$ on the external boundary  are applied such that $\bft = \bfT^* \bfn$  where $\bfT^*$ is given by Eq.~\eqref{eq:sed_exact_sol_ed}. The stress field $\bfT$ of the dislocation in the finite deformation setting is then calculated by solving the system \eqref{eq:ECDD} and \eqref{eq:ECDD_bc} in the rectangular domain along with the above-mentioned traction boundary conditions. \added{We use element-wise quadratic and linear interpolations for $\bff$ and $\bfchi$, respectively.}

\begin{figure}
	\centering
		{\includegraphics[width=.65\linewidth]{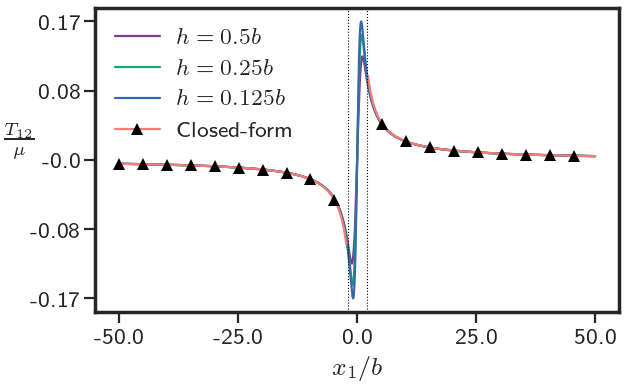}}
	\caption{$T_{12}$, calculated from finite deformation FDM for different element sizes compared against the small deformation closed-form solution along $x_2 = 0$.}%
 \label{fig:sed_T12_x}
\end{figure}

 Figure \ref{fig:sed_T12_x} compares $T_{12}$ obtained from the finite element solution and the closed-form solution $T^*_{12}$ along the line $x_2 = 0$ for three different  regularly spaced grids with element sizes: $h = 0.5b, 0.25b, 0.125b$. We see that the stress fields  are converged w.r.t.~the mesh sizes and an element size of $0.25b$ is adequate for stresses outside the core.  The two vertical lines in Fig.~\ref{fig:sed_T12_x} bound the small region $(|x_1| \le 2b)$ where the small deformation closed-form solution becomes large, as it  is singular at the origin.

\begin{figure}
	\centering
	\begin{subfigure}[b]{0.495\linewidth}
		\centering
		{\includegraphics[width=\linewidth]{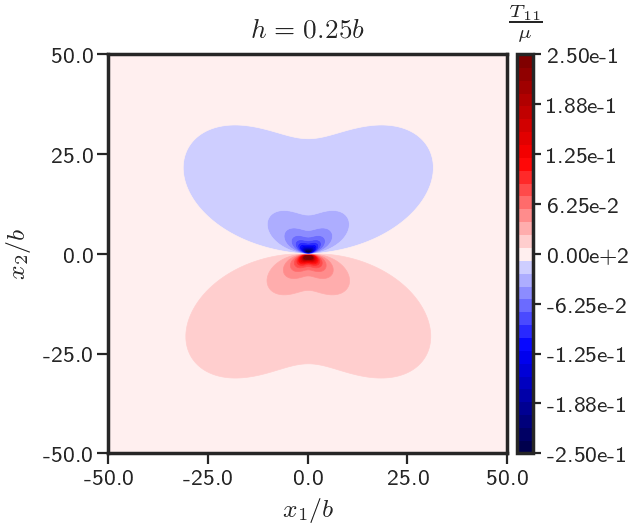}}
		\caption{}
		\label{fig:sed_T11}
	\end{subfigure}%
	\begin{subfigure}[b]{0.495\linewidth}
		\centering
		{\includegraphics[width=\linewidth]{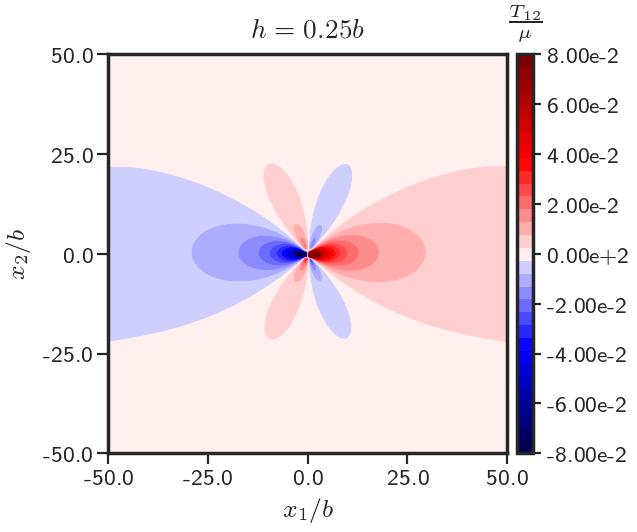}}
		\caption{}
		\label{fig:sed_T12}
	\end{subfigure}
	\caption{Finite deformation stress field of a single edge dislocation computed from FDM a) $T_{11}$  b) $T_{12}$.}
\end{figure}

\begin{figure}
	\centering

	{\includegraphics[width=.60\linewidth]{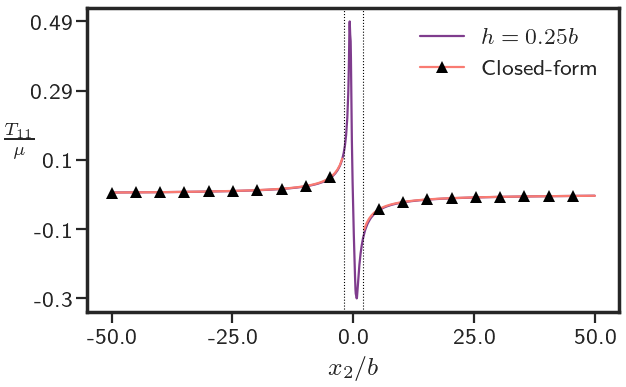}}
	
	\caption{$T_{11}$, calculated from finite deformation FDM compared against the small deformation closed-form solution along $x_1 = 0$.}%
	\label{fig:sed_T11_x}
\end{figure}

Figures \ref{fig:sed_T11} and \ref{fig:sed_T12} show the plots of finite element stress components $T_{11}$ and $T_{12}$ in the domain obtained by solving the ECDD system (Eqs.~\eqref{eq:ECDD} and \eqref{eq:ECDD_bc}) for the element size $h = 0.25b$. Figure \ref{fig:sed_T11_x} show the plot of $T_{11}$ along the line $x_1 = 0$. It may be noted that $T_{11}$ is not anti-symmetric about $x_2 = 0$  which is in contrast with the small deformation case. We conjecture that this is because at finite deformation the elastic modulus depends on $\bfF^e$ as $\bfT = \bfF^e (\mathbb{C}:\bfE^e) \bfF^{eT}$, and therefore the effective elastic moduli in a compressed region differ from those in a tensile region, states applicable to the dislocation above and below $x_2 = 0$.

\begin{figure}
	\centering
	\begin{subfigure}[b]{0.495\linewidth}
		\centering
		{\includegraphics[width=\linewidth]{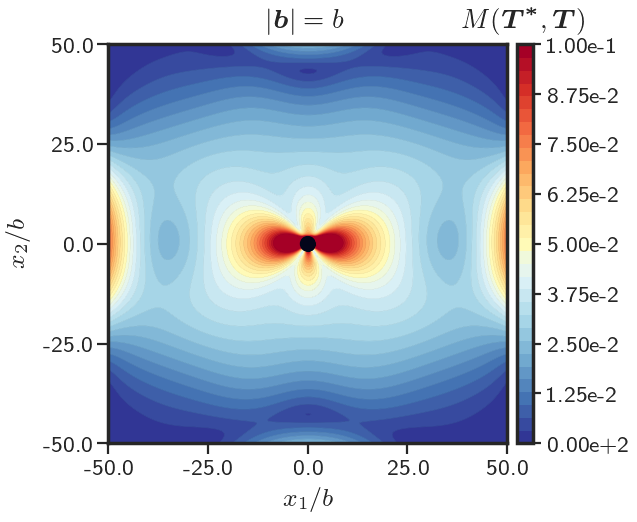}}
		\caption{}

		\label{fig:sed_M1}
	\end{subfigure}\hfill
	\begin{subfigure}[b]{0.495\linewidth}
		\centering
		{\includegraphics[width=\linewidth]{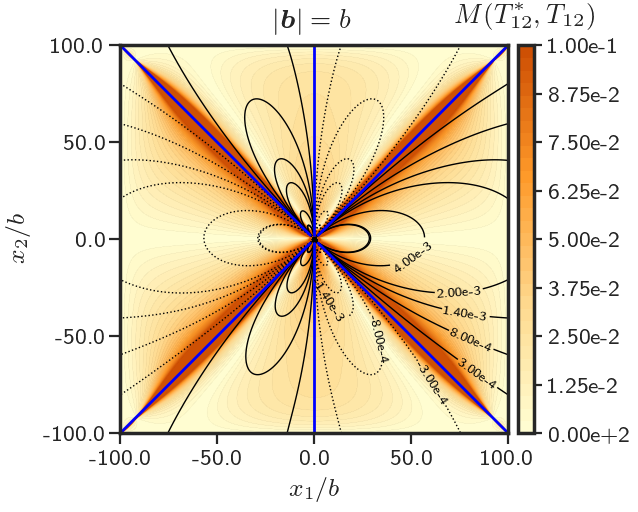}}	
		\caption{}
		\label{fig:sed_sbv_ed_12}
	\end{subfigure}%
	\caption{Difference in stress fields obtained from FDM and small deformation closed-form solution. a) $M(\bfT^*, \bfT)$  b) $M(T^*_{12}, T_{12})$, The solid and dashed lines represent the contours of positive and negative values of $\frac{T_{12}}{\mu}$, respectively.}
	
\end{figure}

We now compare the difference between the stress obtained from the finite element solution $\bfT$ and the closed-form solution for small deformation $\bfT^*$ by calculating the difference measures as defined in \eqref{eq:ssd_M}.  The plot of $M(\bfT^*, \bfT)$ in Figure \ref{fig:sed_M1} clearly displays that the stress fields differ   around the core with errors of approximately $10\%$ for up to $10b$ from the core. This is qualitatively consistent with DFT results of \cite{iyer2015electronic} wherein  it is shown  show that the energy contribution from the electronic-structure perturbations are significant up to a distance of $10b$  from the edge dislocation line where the strength of the dislocation is  $b$. \added{Given the very different nature of the two calculations, our result raises the intriguing question of how much of this specific aspect of the DFT calculations is a result simply of accommodating finite deformation elastic defect calculations or whether the observed correspondence with DFT results is entirely fortuitous for our model.}  

Figure \ref{fig:sed_sbv_ed_12} shows the plot of the difference measure $M(T^*_{12}, T_{12})$   for the same problem setup but solved in a comparatively larger domain ($200b \times 200b$),  and has been verified for convergence w.r.t mesh refinement. The figure is overlapped with contours of $T_{12}$ in the domain. The region where the small deformation closed-form solution vanishes has been marked in blue.   Two observations are noteworthy: a) The  $M(T^*_{12}, T_{12})$  along the $x_2 = 0$ line is negligible as compared to the other region in the domain.  Therefore, Figure \ref{fig:sed_T12_x} shows  no noticeable difference between the computed finite deformation $T_{23}$ and small deformation closed-form  $T_{12}$. b)  An  approximately $10\%$ normalized stress difference exists all along the diagonal of the extended body up to a distance of $\sim 120b$ from the center of the core. Moreover, the finite deformation $T_{12}$ stress component at the location $(67.78b, -56.76b)$, close to the diagonal, is $23$ MPa at a distance of $\sim 87 b$ from the center of the core. The latter stress magnitude, at a significant distance from the core, is not insignificant for affecting defect interactions in a plastically deforming body. These substantial differences between the small and finite deformation results over extended spatial regions, uncovered apparently for the first time here, warrant a careful examination of such conclusions against lattice statics calculations based on well-characterized interatomic potentials.



\begin{figure}
	\centering
	\begin{subfigure}[b]{0.495\linewidth}
		\centering
		{\includegraphics[width=\linewidth]{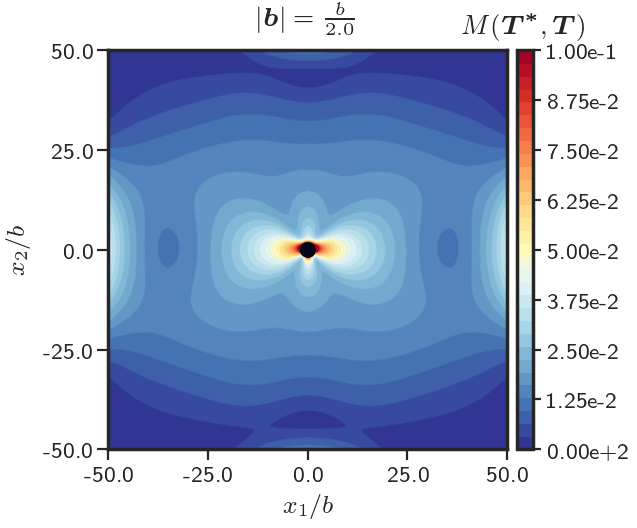}}
		\caption{}
		\label{fig:sed_sbv_2}
	\end{subfigure}%
	\begin{subfigure}[b]{0.495\linewidth}
		\centering
		{\includegraphics[width=\linewidth]{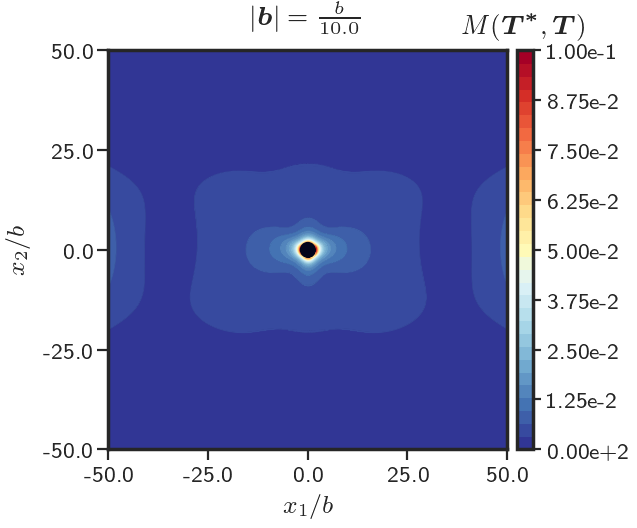}}
		\caption{}
		\label{fig:sed_sbv_10}
	\end{subfigure}\\
	\begin{subfigure}[b]{0.495\linewidth}
		\centering
		{\includegraphics[width=\linewidth]{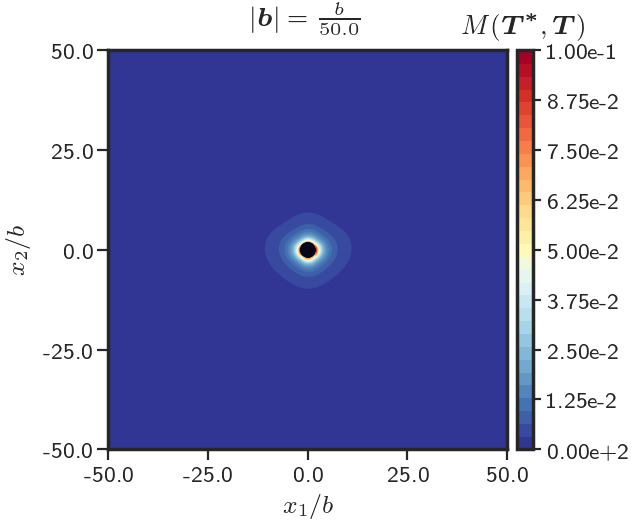}}
		\caption{}
		\label{fig:sed_sbv_50}
	\end{subfigure}
	\caption{Difference in magnitude of stress fields obtained from FDM and small deformation closed-form linear elastic solution for different strengths $|\bfb|$ of the dislocation  a) $\frac{b}{2}$ b) $\frac{b}{10}$ c) $\frac{b}{50}$.}
	\label{fig:sed_sbv_ed_all}
\end{figure}

Given the large differences  between $\bfT$ and $\bfT^*$ observed above,  we explore the domain of validity of classical dislocation fields by varying the strength of the dislocation. To do so,  we  again plot the difference measure $M(\bfT^*, \bfT)$ for different strengths $|\bfb|$ of the edge dislocation while keeping the domain size fixed and the problem set up the same as above. We show in Figure \ref{fig:sed_sbv_ed_all} that the  normalized difference between the finite deformation FDM stress field and small deformation closed-form solution becomes small as the strength $|\bfb|$ of the dislocation decreases, and the error is below $3\%$ in most of the domain when $|\bfb| = \frac{b}{50}$. The error goes to zero only in the limit $|\bfb| \to 0$. This exercise also serves as a verification of our code in that the correct limiting trends are produced for small forcing.

\subsection{Stress f\mbox{}ield of a spatially homogeneous dislocation density}
\label{sec:ha}

The dislocation density tensor $\bfalpha$ is related to the $curl$ of the inverse-elastic distortion tensor in the domain. Therefore, $\bfF^e$ being equal to a general (inhomogeneous) rotation field in the domain gives rise to a stress free configuration with a  non-zero dislocation density.  Similarly,  for the linear case, the stress free dislocation distributions belong to the class such that the linearised elastic distortion tensor $(\bfU^e \approx \bfF^e - \bfI)$ is skew-symmetric.  However, it can also be shown \cite{mura1989impotent, head1993equilibrium, acharya2018stress} that any uniform distribution of dislocation density in the domain is  stress free in the linear theory. This is because,  in the linear theory, a spatially constant dislocation density distribution has vanishing incompatibility ($\eta := (curl(\bfalpha^T))_{sym} $) and therefore the strain field is compatible and the body is stress free in the absence of any external forces. However, as recently shown in  \cite{acharya2018stress}, such distributions are not stress free when  geometric nonlinearity is taken into account in a two-dimensional setting.

We demonstrate this sharp contrast in the predictions for stress fields based on the linear and the nonlinear ECDD theories as a verification of our finite element scheme. The problem is set up in a $2$-d plane strain setting as follows: a spatially homogeneous distribution of edge dislocations $\alpha_{13}$ is specified in a domain with dimensions $[-50b, 50b] \times [-50b, 50b]$. The total Burgers vector of the dislocation distribution is assumed to be $\bfb = 100b\bfe_1 =  \int_\mOmega\alpha_{13}\,\bfe_1\,dA$.  The external boundary of the domain is considered to be traction free. A uniform  grid of $400 \times 400$ is used to mesh the domain. The materials constants $E$ and $\nu$ are chosen to be $200$  GPa and $0.30$, respectively. \added{We use element-wise quadratic and linear interpolations for $\bff$ and $\bfchi$, respectively.}

\begin{figure}
	\centering
	\begin{minipage}[b]{.48\linewidth}
		\centering
		{\includegraphics[width=\linewidth]{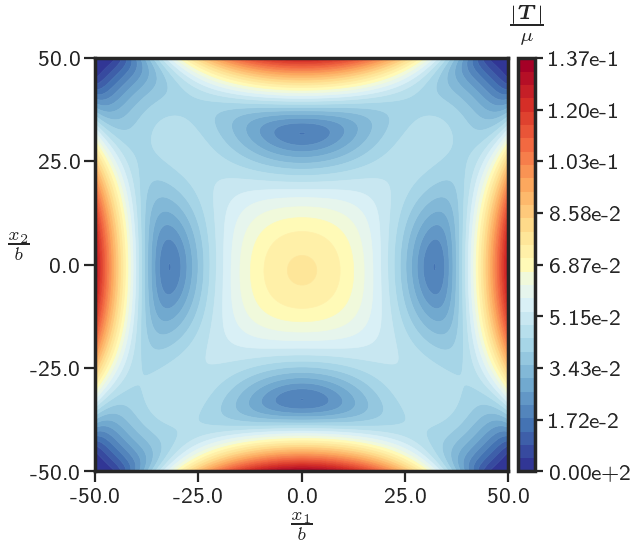}}
	\caption{Finite deformation stress field for spatially uniform distribution of $\alpha_{13}$ for  Saint-Venant-Kirchhoff material.}	
			\label{fig:ha_fd-svk}
	\end{minipage}\hfill
\begin{minipage}[b]{.48\linewidth}
			\centering
			{\includegraphics[width=\linewidth]{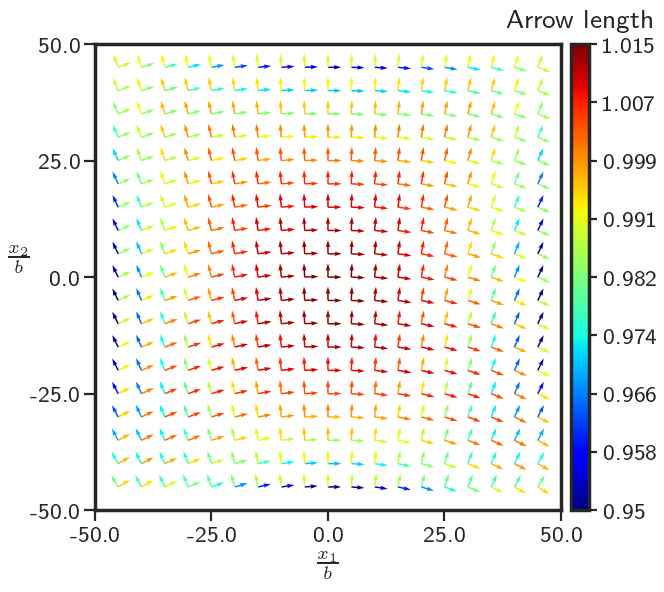}}
		\caption{Plot of $\bfe_1$ and $\bfe_2$ mapped by elastic distortion $\bfF^e$. The colorbar shows the length of the mapped vectors}
		\label{fig:ha_fd-svk_q}
\end{minipage}
\end{figure}

In the linear setting, the stress field is calculated by solving for $\bff$ (or $\bfz$) as described in Sec.~\ref{sec:fem_z}. The stress field in the nonlinear setting is obtained by solving  the ECDD system (Eqs.~\eqref{eq:ECDD} and \eqref{eq:ECDD_bc}) as shown in Sec.~\ref{sec:fem_staticfSolve}.  Figure \ref{fig:ha_fd-svk} shows the magnitude of stress field $|\bfT|$ obtained from the geometrically nonlinear FDM theory for the Saint-Venant-Kirchhoff material. \added{The $\frac{|\bfT|}{\mu}$ distribution for the Neo-Hookean material is similar to Figure \ref{fig:ha_fd-svk} except for smaller magnitude.} The linear  calculation predicts vanishing stress field for a spatially homogeneous dislocation distribution in the domain.

\added{Figure \ref{fig:ha_fd-svk_q} shows the plot of vectors $\widetilde{\bfe}_1$ and $\widetilde{\bfe}_2$ which are obtained by mapping the orthogonal unit vectors $\bfe_1$ and $\bfe_2$, respectively, by the elastic distortion tensor field $\bfF^e$ for the Saint-Venant-Kirchhoff material. The colorbar shows the length of the mapped vectors $\widetilde{\bfe}_1$ and $\widetilde{\bfe}_2$. The angle between the mapped vectors lies between $\mdeg{87.66}$ and $\mdeg{91.66}$ which corresponds to small shear strains. This, and the elastic stretches in $\bfe_1$ and $\bfe_2$ directions, result in the development of stresses inside the domain.}
	
\added{ The stretches in the $\bfe_1 (\bfe_2)$ directions are maximum near the center of the top and bottom (left and right) boundaries - which corresponds to the regions of large $|\bfT|$ in the body as observed in Fig.~\ref{fig:ha_fd-svk}. Interestingly, even with very small shear strains, the combination of rotation and stretch in the $\bfe_1$ and $\bfe_2$ directions generates Cauchy shear stresses due to the frame-indifferent, nonlinear elastic stress-strain relationship \eqref{eq:stress_svk}. Moreover, the variation in the direction of $\widetilde{\bfe}_1$ and $\tilde{\bfe}_2$ vectors in the domain shows the curvature of the deformed lattice (which is incompatible everywhere in this case).}

We remark here that although we demonstrated the contrasting predictions of linear and nonlinear theory for isotropic materials, the result holds true for any  possibly inhomogeneous and anisotropic  nonlinear elastic material with a single well energy density (in the elastic right Cauchy-Green deformation tensor). 

\subsection{Burgers vector constancy with dislocation density evolution in  nonlinear elastic motions}
\label{sec:res_bvc}

Several measures have been proposed to  define the dislocation density in a body as a function of the elastic or plastic distortion tensor  \cite{bilby1955continuous, eshelby1956continuum, fox1966continuum, willis1967second, acharya2000lattice, cermelli2001characterization}. Cermelli and Gurtin  \cite{cermelli2001characterization}   advocate  a single measure of GNDs based on `physically motivated requirements.' However, as is customary in continuum mechanics, relations should always exist between any two physically meaningful  measures of GNDs, and  it is these transformation rules that are physically significant rather than superficial differences in form  \cite{acharya2008counterpoint}.

The dislocation density tensor $\bfalpha$ in (M)FDM is a two point tensor that measures the local, undeformed Burgers vector of the dislocation distribution, per unit area of the current configuration.	For a given dislocation density $\bfalpha$ in the domain,  \eqref{eq:bv_area_patch} gives the Burgers vector $\bfb_A$ content of an area patch $A (t)$  at any time $t$. In the special case when there is no flux of dislocations into a  material area patch,  the dislocation density field  $\bfalpha$ has to evolve in such a way that the total Burgers vector of that material patch (given by \eqref{eq:bv_area_patch})  remains constant. Hence, under the conditions $\bfV = \bf0$ and $\bfL^p = \bf0$, the Burgers vector of any arbitrary area patch should not change in time regardless of the total deformation magnitude i.e.~at all times $t$,
\begin{align*}
\dot{\bfb}_{A}(t) = \deriv{}{t} \int_{A(t)} \bfalpha \bfn \, dA = \bf0, ~~ \forall ~~\  \mathnormal{A} \subset \mOmega \\
\implies \dot{\bfalpha} = \bfalpha\bfL^T - tr(\bfL) \, \bfalpha ~~~(\rm{from ~\eqref{eq:mfdm_a}}).
\end{align*}

 This constraint on the evolution of the dislocation density is verified under large extensional and simple shear loadings below. These are stringent tests of the numerics since the dislocation density evolution is coupled to the evolving deformation through the velocity gradient by an adapted convected derivative of a $2$-point tensor as shown in  Eq.~\eqref{eq:mfdm_a}. 

The problem is set up as follows: An edge dislocation is assumed to be present in a rectangular body of  dimensions $[-50b, 50b] \times [-50b, 50b]$.  The dislocation is modeled by prescribing an initial dislocation density $\bfalpha_{13}(\bfx, t = 0)$ of the form given by Eq.~\eqref{eq:sed_single_edge_core} at any point $\bfx = (x_1, x_2)$. The dislocation core width $w$ is taken as $2b$. A uniform grid of $100\times 100$ elements is used to mesh the domain. The materials constants $E$ and $\nu$ are chosen to be $200$  GPa and $0.30$, respectively. A strain rate of $\hat{\mGamma} = 1s^{-1}$ is used for both the loading cases.

%

\begin{figure}
		\centering
		{\includegraphics[width=.595\linewidth]{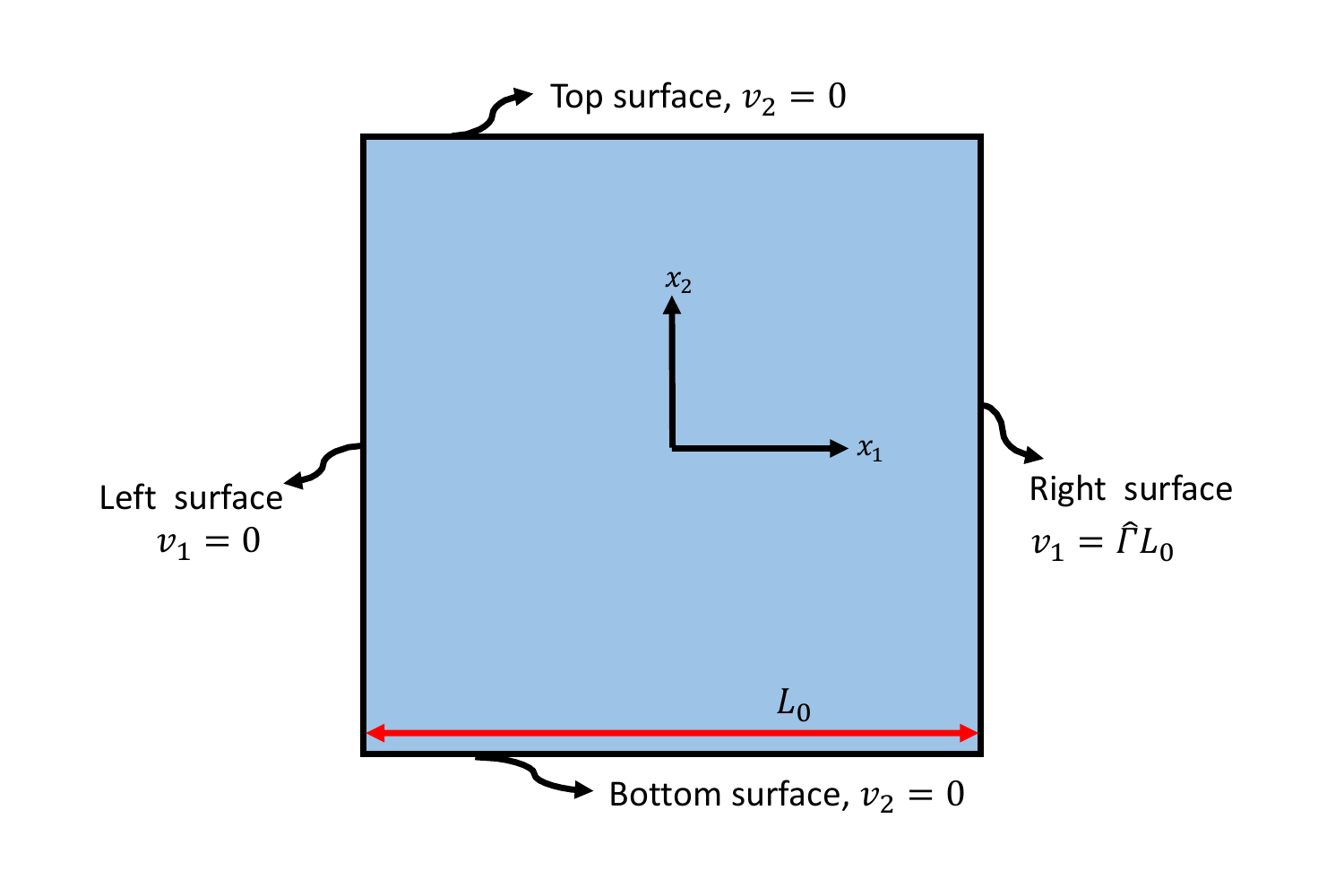}}
		\caption{Typical schematic of the geometry.}
		\label{fig:schematic_bvc_tensile}
\end{figure}

\subsubsection{Extensional loading}
The velocity boundary conditions on the left  and right faces are assumed to be $v_1 = 0$ and $v_1 = \hat\mGamma L_0$, respectively. $L_0 = 100b$ is the initial length of the body. On the  top and bottom surfaces,  $v_2 = 0$ is imposed. The schematic of the set up is shown in Fig.~\ref{fig:schematic_bvc_tensile}.



\begin{figure}
	\centering
	\begin{subfigure}[b]{0.395\textwidth}
		\centering
 		{\includegraphics[width=\linewidth]{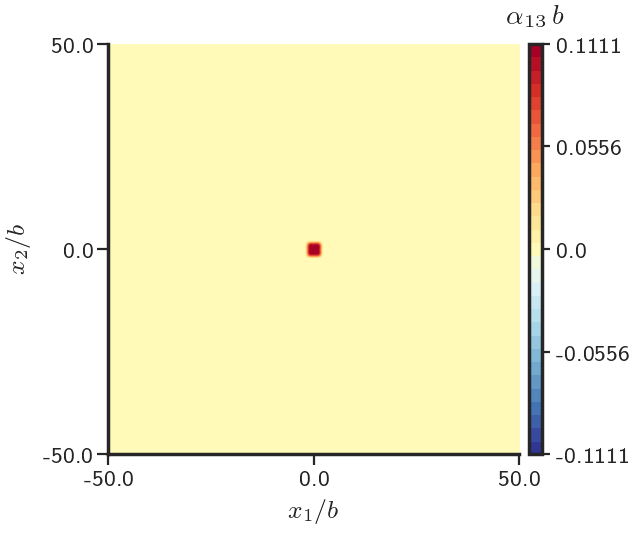}}
		\caption{}
		\label{fig:bvc_tensile_undeformed}
	\end{subfigure}%
	\begin{subfigure}[b]{0.595\textwidth}
		\centering
		{\includegraphics[width=\linewidth]{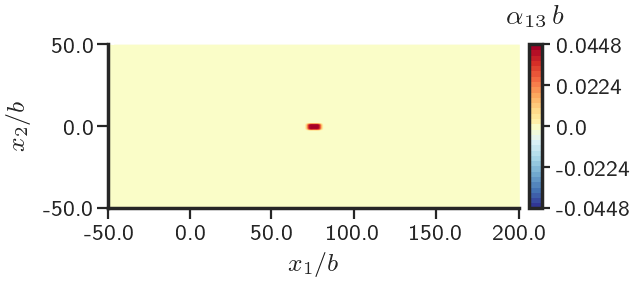}}
		\caption{}
		\label{fig:bvc_tensile_deformed}
	\end{subfigure}
	\caption{Configurations (not to same scale) during extensional loading a) undeformed b) deformed}
\end{figure}

\begin{figure}
	\centering
	\begin{subfigure}[b]{0.495\textwidth}
		\centering
		{\includegraphics[width=\linewidth]{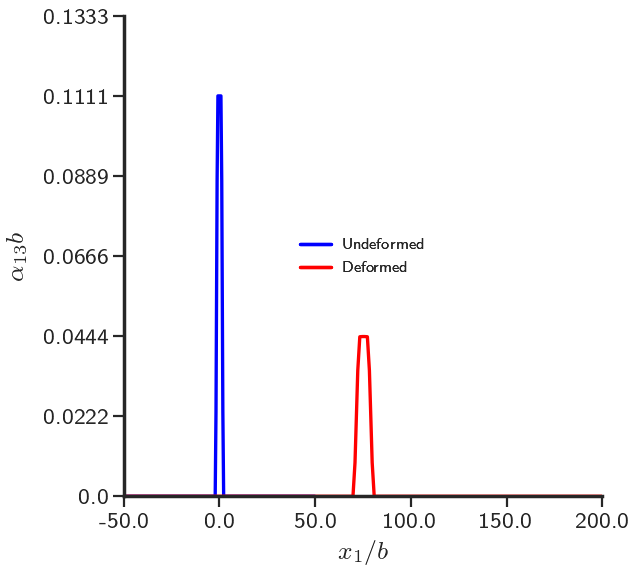}}
		\caption{}
		\label{fig:bvc_tensile_alpha}
	\end{subfigure}%
	\begin{subfigure}[b]{0.495\textwidth}
		\centering
		{\includegraphics[width=\linewidth]{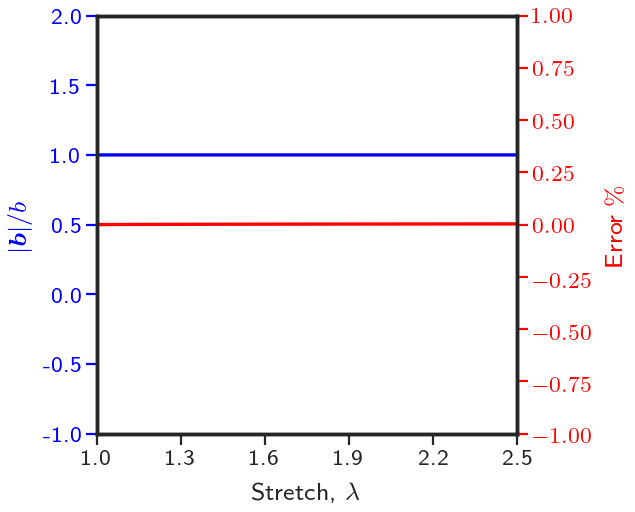}}
		\caption{}
		\label{fig:bvc_tensile_b}
	\end{subfigure}
	\caption{Variation of a) dislocation density  $\alpha_{13}$ along $x_2  = 0$ b) dislocation strength $|\bfb|$  with strain.}
\end{figure}

Figure \ref{fig:bvc_tensile_undeformed} shows the undeformed configuration of the body. The deformed configuration of the body at a stretch $\mLambda = 2.5$ under extensional loading is shown in Fig.~\ref{fig:bvc_tensile_deformed}.  Since the current area changes in the extensional loading case, the dislocation density evolves as well, as shown in Figure \ref{fig:bvc_tensile_alpha}. However, the change in dislocation density is such that the Burgers vector content, of the whole ($2$-d) body considered as the area patch, remains constant. Figure \ref{fig:bvc_tensile_b} shows the variation of the strength of dislocation with stretch. We observe that there is virtually no change  in the  strength of the dislocation even at large strains.

\subsubsection{Simple shear loading} 

\begin{figure}
	\centering
	\begin{subfigure}[b]{0.355\textwidth}
		\centering
		{\includegraphics[width=\linewidth]{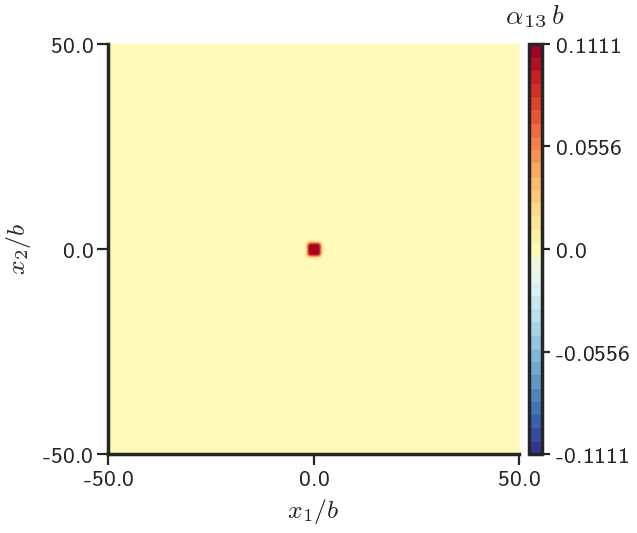}}
		\caption{}
		\label{fig:bvc_shear_undeformed}
	\end{subfigure}%
	\begin{subfigure}[b]{0.635\textwidth}
		\centering
		{\includegraphics[width=\linewidth]{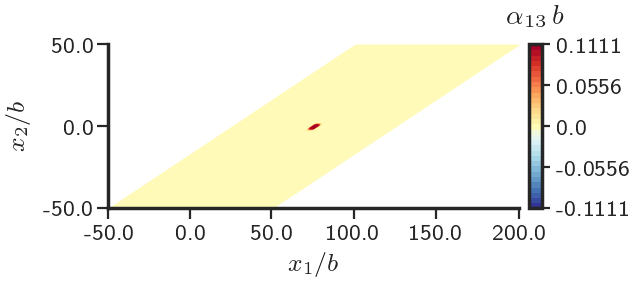}}
		\caption{}
		\label{fig:bvc_shear_deformed}
	\end{subfigure}
	\caption{Configurations (not to same scale) during shear loading a) undeformed b) deformed}
\end{figure}

\begin{figure}
	\centering
	\begin{subfigure}[b]{0.495\textwidth}
		\centering
		{\includegraphics[width=\linewidth]{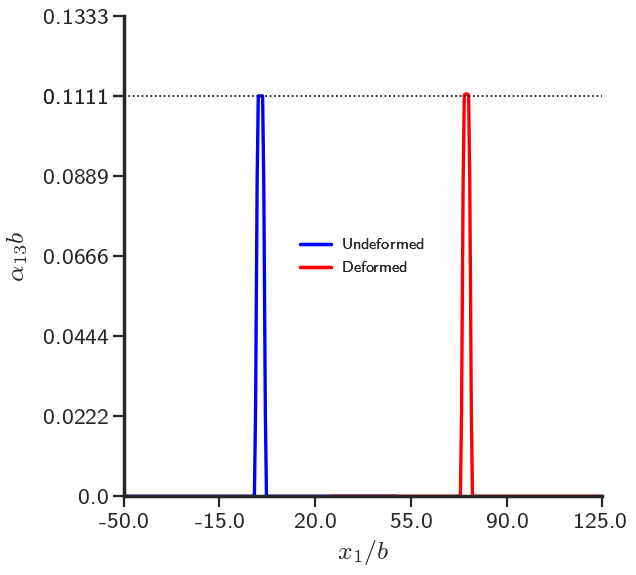}}
		\caption{}
		\label{fig:bvc_shear_alpha}
	\end{subfigure}%
	\begin{subfigure}[b]{0.495\textwidth}
		\centering
		{\includegraphics[width=\linewidth]{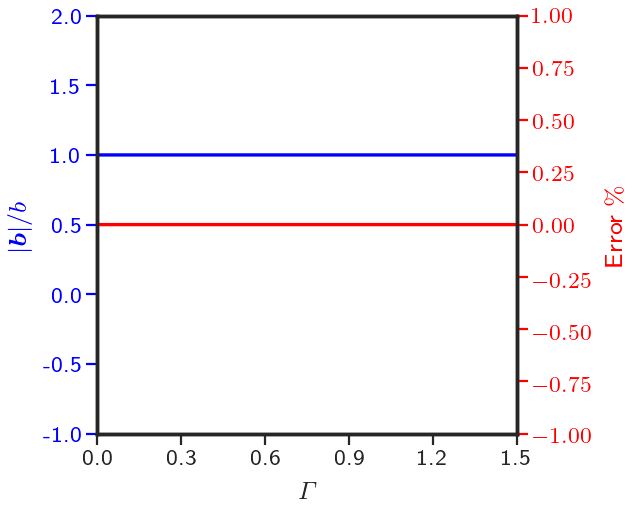}}
		\caption{}
		\label{fig:bvc_shear_b}
	\end{subfigure}
	\caption{Variation of a) dislocation density  $\alpha_{13}$ along $x_2  = 0$ b) dislocation strength $|\bfb|$  with strain.}
\end{figure}

Velocity boundary conditions corresponding to overall simple shear are imposed for a plane strain problem as described in Sec.~\ref{sec:NLE_ss}. 
Figure \ref{fig:bvc_shear_undeformed} shows the undeformed configuration of the body. The deformed configuration of the body at a strain $\mGamma = 1.5$ under simple shear loading is shown in \ref{fig:bvc_shear_deformed}.  In a $2$-d setting, dislocation density evolution in the absence of $\bfV$ can happen only if the material velocity field is such that $tr(\bfL) \ne \bf0$ (see~\eqref{eq:mfdm_a}). Unlike the small deformation case, the inhomogeneous stress field caused by the presence of a dislocation affects the  velocity solution through the rate form of the equilibrium equation \eqref{eq:fem_v}. Therefore, the velocity boundary conditions corresponding to the homogenous simple shear along with the  inhomogeneous stress field give rise to a non-zero $tr(\bfL)$ in the body. On the other hand, the perturbation, to the velocity field produced by solely the boundary condition (in this case isochoric), is mediated by $\frac{|\bfT|}{|\bbC|}$, the ratio of the `initial stress stiffness' to the `material stiffness' for the rate problem. We use our computational tool to  evaluate the magnitude of this perturbation through its effect on the  solution for the dislocation density \eqref{eq:fem_a}, a purely finite deformation effect.  The corresponding change in the dislocation strength  is shown in Figure \ref{fig:bvc_shear_b}. We observe that there is virtually no change in  the strength of the dislocation with progress of deformation over such large strains. The response has also been verified to be rate-independent w.r.t.~the rate of loading.

These examples demonstrate the robust performance of our code in stringent tests of finite deformation dislocation kinematics and also demonstrate the physical characteristics of our dislocation density measure \eqref{eq:mfdm_a} \cite{willis1967second,  acharya2008counterpoint}.


\subsection{Inf\mbox{}luence of inclusion size on material strength}
\label{sec:results_mfdm_inc}



 Materials can be hardened by distributing particles of another phase in them that either inhibit plastic flow in them or are elastically stiffer \cite{ebeling1966dispersion}. The experiments performed in Ebeling and Ashby  \cite{ebeling1966dispersion} used a  copper-silicon alloy for different particle sizes and volume fractions of SiO$_2$. The samples were then oriented for single slip and subjected to tensile loading. Their experimental results demonstrated that the strength of these alloys was greater than that of copper single crystals, and this enhancement in strength  (even though silica is elastically softer than copper) depended on the size, spacing, and volume fraction of the inclusions. Moreover, for a given volume fraction of inclusions, the strength was enhanced  for smaller inclusion sizes.

Several studies have qualitatively modeled the same phenomena under simple shearing motion  using different continuum models \cite{yefimov2004comparison, bassani2001plastic, richeton2011continuity}. Here, we use the crystal plasticity and $J_2$ plasticity models of MFDM to model this behavior under a simple shear loading. The inclusions are assumed to behave elastically and the matrix as a rate-dependent  elastic-plastic material. Table \ref{tab:inc_data} presents the values of the material constants used in this section. 

		{\renewcommand{\arraystretch}{1.2}
\begin{table}
	\begin{subtable}[h]{0.45\textwidth}
		\centering
		\begin{tabular}{ | c | c | }
			\hline
			Parameter &   Value  \\ \hline
			$E$ &  $70$ GPa \\     
			$\nu$ &  $.17$ \\ \hline 
		\end{tabular}
		\caption{}
		\label{tab:inc_data1}
	\end{subtable}
	\hfill
	\begin{subtable}[h]{0.45\textwidth}
		\centering
		\begin{tabular}{ | c | c | }
			\hline
			Parameter &   Value  \\ 	\hline
			$\hat\mGamma$ &  $1s^{-1}$\\     
			$\hat\gamma_0$ &  $1s^{-1}$\\     
			$m$ &  $0.03$\\     
			$\eta$ & $\frac{1}{3}$\\
			$b$ &  $2.5 \AA$\\  
			$g_0$ &  $.05$ GPa\\     
			$g_s$ &  $.21$ GPa\\     
			$\mTheta_0$ &  $.3925$ GPa\\     
			$k_0$ &  $20$\\     
			$l$ &  $\sqrt{3}\times 0.1\,\mu m$\\     
			$E$ &  $110$ GPa \\     
			$\nu$ &  $.34$ \\ \hline
			
		\end{tabular}
		\caption{}
		\label{tab:inc_data2}
	\end{subtable}
	\caption{Parameter values used to model inf\mbox{}luence of inclusion size on material strength: a) Inclusion b) Matrix.}
	\label{tab:inc_data}
\end{table}
}

\begin{figure}
	\centering
	\begin{subfigure}[b]{0.32\linewidth}
		\centering
		{\includegraphics[width=.90\linewidth]{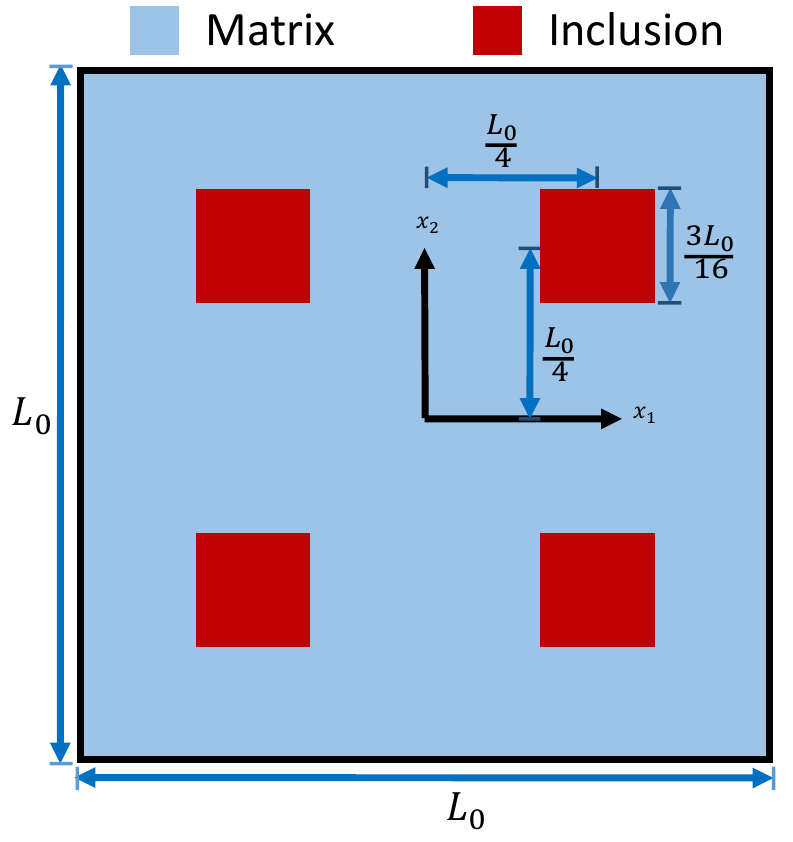}}
		\caption{}
		\label{fig:sch_4}
	\end{subfigure}%
	\begin{subfigure}[b]{0.32\linewidth}
		\centering
		{\includegraphics[width=.90\linewidth]{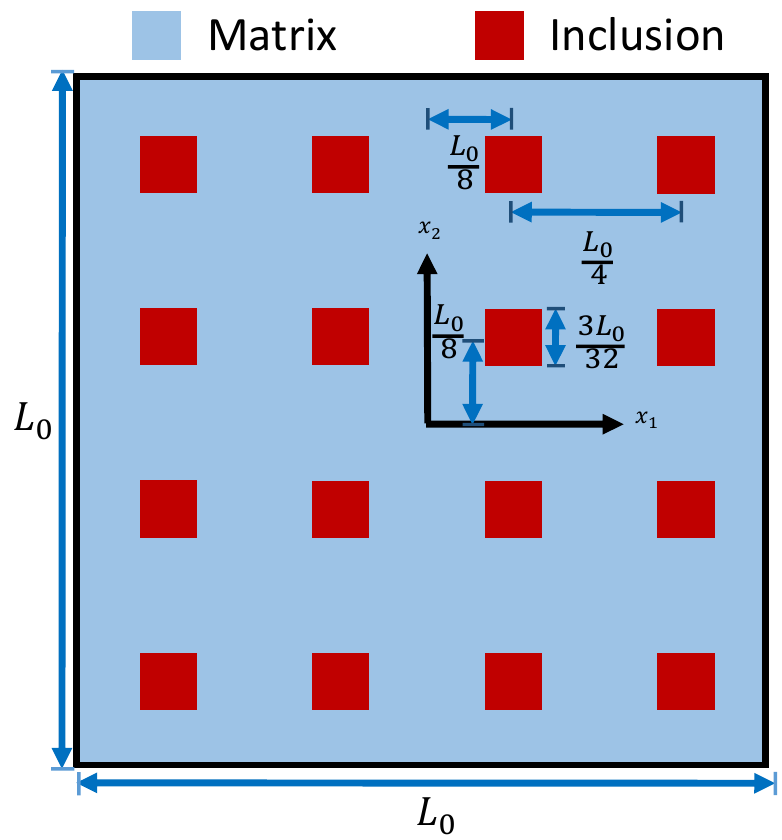}}
		\caption{}
		\label{fig:sch_16}
	\end{subfigure}
	\centering
\begin{subfigure}[b]{0.32\linewidth}
	\centering
	{\includegraphics[width=.86\linewidth]{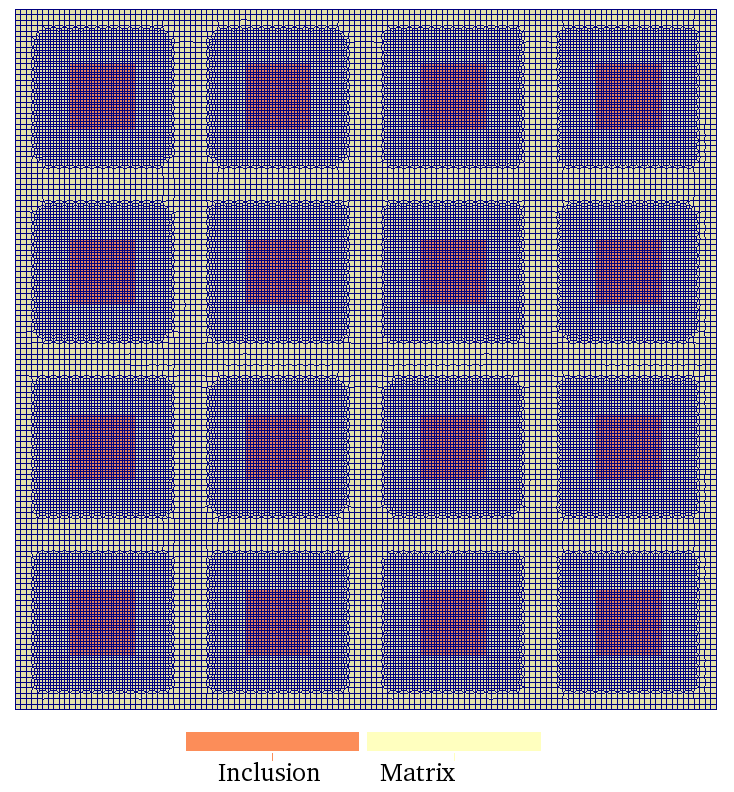}}
	\caption{}
	\label{fig:mesh_16}
\end{subfigure}
	\caption{Schematic layout of the geometry a) $4$ inclusions b) $16$ inclusions c) Finite element mesh with 16 inclusions. \added{The non-uniform \added{conforming} mesh uses element sizes of $\frac{L_0}{128}$ and $\frac{L_0}{256}$ in the regions far from and close to the inclusions, respectively.}}
	\label{fig:sch_inc}
\end{figure}

The problem set up and boundary conditions for the overall simple shearing motion are similar to the description given in Sec.~\ref{sec:NLE_ss}. Simulations are performed on domain sizes of $\mum{5}$ and $\mum{50}$ for the two cases i) $4$ large inclusions and ii)  $16$ small inclusions.  The volume fraction of the inclusions is fixed at $14\%$.  Fig.~\ref{fig:sch_inc} shows the location of the inclusion and mesh used for the simulations. We use the Crystal and $J_2$ plasticity models  (Eqs.~\eqref{eq:Lp_crystal} and \eqref{eq:Lp_j2}) for the plastic strain rate due to SDs  ($\bfV$ is also slightly different between the two cases). The initial ($t=0$) slip system directions in the body are assumed to be oriented at $\mdeg{30}$, $\mdeg{-30}$, and $\mdeg{0}$ w.r.t~the $x_1$ axis. To model the behavior under a quasi-static simple shear loading, we follow the algorithm in the Table \ref{tab:quasi_algo} except that we evolve the dislocation density only in the matrix since the inclusions are treated as purely elastic. The conventional plasticity results for this problem are obtained by following the algorithm outlined in Sec.~\ref{sec:new_classical_algo}. The internal boundaries at the matrix-inclusion interface are considered as unconstrained.

\begin{figure}
	\centering
	\begin{subfigure}[b]{0.495\textwidth}
		\centering
		{\includegraphics[width=.990\linewidth]{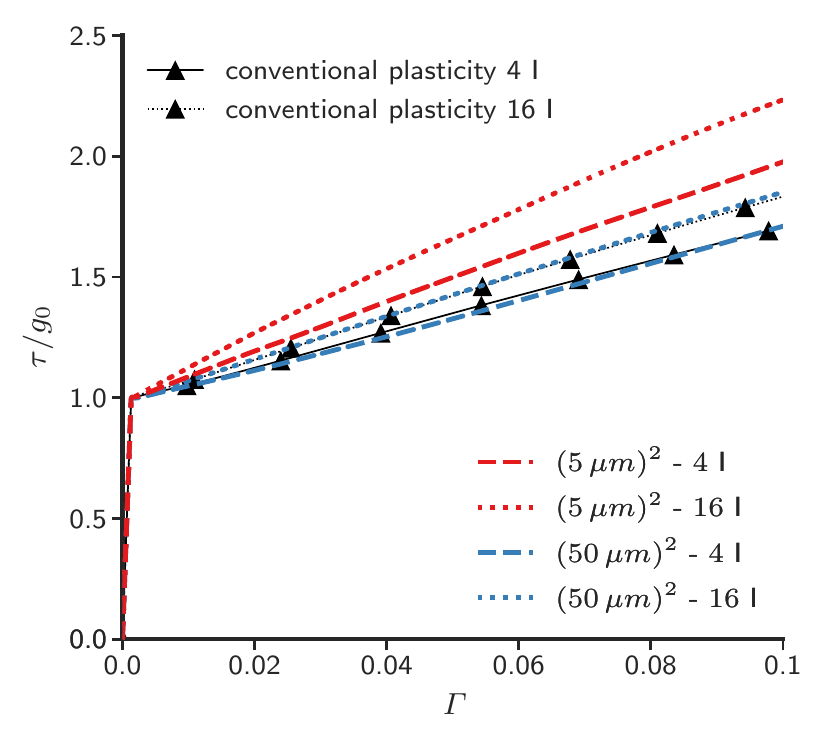}}
		\caption{}
		\label{fig:inclusion_ss_cr}
	\end{subfigure}%
	\begin{subfigure}[b]{0.495\textwidth}
		\centering
		{\includegraphics[width=.990\linewidth]{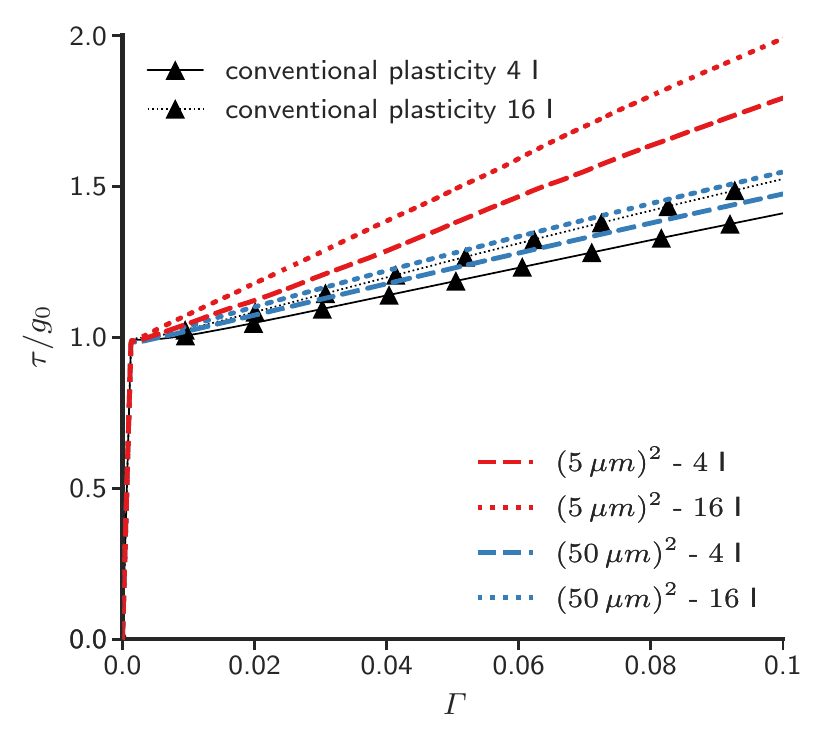}}
		\caption{}
		\label{fig:inclusion_ss_j2}
	\end{subfigure}
	\caption{Stress-strain response a) Crystal plasticity b) $J_2$ plasticity (I: Inclusions).}
	\label{fig:inclusion_ss}
\end{figure}

 Assuming $l_{inc}$ and $L$ denote the inclusion and the sample sizes respectively, the three non-dimensional geometric parameters that matter in the context of non-dimensional analysis are $(\frac{l}{L},  \frac{b}{L},  \frac{l_{inc}}{L})$. As the sample size is decreased keeping the ratio of the inclusion to sample size fixed, the first two numbers increase and it is their effect that can potentially lead to changes in the strength of the composite; here, we quantitatively establish this effect through solving the full problem, which is beyond the scope of dimensional analysis. We note that given the structure of the model, the first two non-dimensional parameters can have an effect on strength only in the presence of dislocation density and this is trigerred, in the problems considered, by the elastic and plastic contrast in material properties between matrix and inclusion. It is also clear that in conventional plasticity such a size effect cannot arise under self similar geometric scaling of the types of samples with inclusions shown in Fig.~\ref{fig:sch_inc}. Figures \ref{fig:inclusion_ss_cr} and  \ref{fig:inclusion_ss_j2} show the stress-strain response ($\tau$ vs.~$\mGamma$) of the samples for the crystal and $J_2$ plasticity MFDM models, respectively.  MFDM is able to model the dependence  of the  mechanical response of the material on the size of the reinforcing particles. For both the models, the response of the sample with $16$ small inclusions is harder than the sample with  $4$ larger inclusions keeping the volume fraction of inclusions fixed.
 
 As an algorithmic aside, we have verified that the results obtained from the novel algorithm  for conventional plasticity outlined in Sec.~\ref{sec:new_classical_algo} are consistent with those from the MFDM plasticity algorithm with $k_0=0$, $\bfV=0$ for the problem setup with $4$ inclusions, which also serves as a self-consistent verification for these algorithms.

 \begin{table}
 	\centering
 	\begin{tabular}{ c c c }
 		\myhline
 		& Ratio of $\tau - \tau_0$ on halving particle size &\\ \myhline
 		Experimental & Crystal & $J_2$\\ \myhline
 		1.414 & 1.258 & 1.237
 		\\ \hline 
 	\end{tabular}
 	\caption{Comparison of enhancement in strength with experimental data taken from \cite{ebeling1966dispersion}.}
 	\label{tab:inc_data_exp}
 \end{table}

 

 

Table \ref{tab:inc_data_exp} presents the ratios of $\tau - \tau_0$ at 10\% strain, where $\tau_0$ is the initial yield stress, when the inclusion size is halved while keeping the volume fraction fixed at $\sim14 \%$, for the crystal  and $J_2$ plasticity models. These results are compared with the experimental trend observed in \cite{ebeling1966dispersion} over a large class of data, with the conclusion, for the tests conducted here, that the mentioned ratio should be equal to the square root of the reciprocal of the factor by which the inclusion size is reduced, i.e., the strengthening ratio should be $\sqrt{2}$, for a decrease in average particle diameter by a factor of $\half$.  The maximum volume fraction of inclusions was $1 \%$ in the experiments, and the average particle diameter ranged from $0.06 - 0.18 \mu m$. A single slip system was predominantly activated in the experiments, which is in accord with
the orientation chosen by us. The data and abstracted trend in \cite[Eq.(3) and Sec.~5] {ebeling1966dispersion} suggest that the strengthening ratio considered here should be independent of the Burgers vector, the applied strain, and the volume fraction.  Our computational results are in good agreement with the experimental data.

\begin{figure}
	\centering
	\centering
	\begin{subfigure}[b]{0.495\textwidth}
		\centering
		{\includegraphics[width=.990\linewidth]{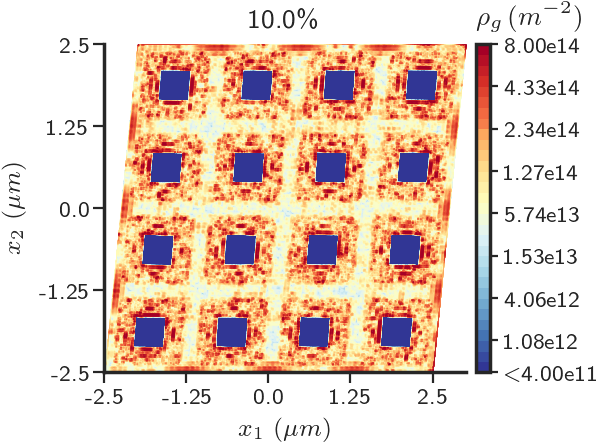}}
		\caption{}
		\label{fig:inclusion_an_j2_16_5mic}
	\end{subfigure}%
	\begin{subfigure}[b]{0.495\textwidth}
		\centering
		{\includegraphics[width=.990\linewidth]{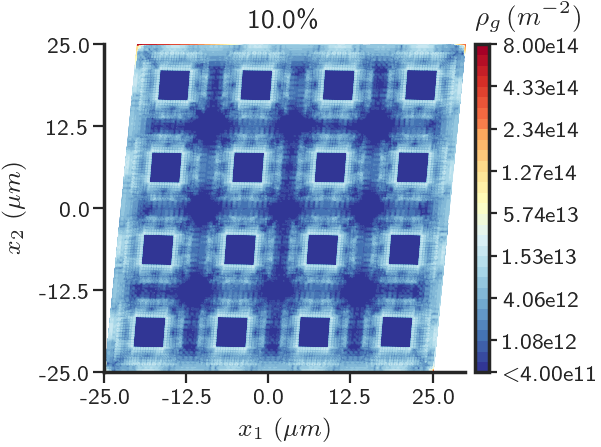}}
		\caption{}
		\label{fig:inclusion_an_j2_16_50mic}
	\end{subfigure}
		\caption{Dislocation density at $10\%$ strain when the size of inclusions is decreased keeping the volume fraction same for $J_2$ plasticity model. a) $\mum{5}$ domain size with $16$ inclusions b) $\mum{50}$ domain size with $16$ inclusions.}
	\label{fig:inclusion_an_j2}
\end{figure}

Figure \ref{fig:inclusion_an_j2} shows  the distribution of the dislocation density for different domain sizes and  inclusions for the $J_2$ plasticity models. It can be seen that for a given sample size, the average dislocation density norm in the domain is greater when the inclusions are smaller in size. Also, for the larger sample sizes $\mum{50}$, we see that the magnitude of  dislocation density is smaller as compared to the corresponding $\mum{5}$ domain sizes and therefore less hardening is observed.

\begin{figure}
		\centering
		{\includegraphics[width=.50\linewidth]{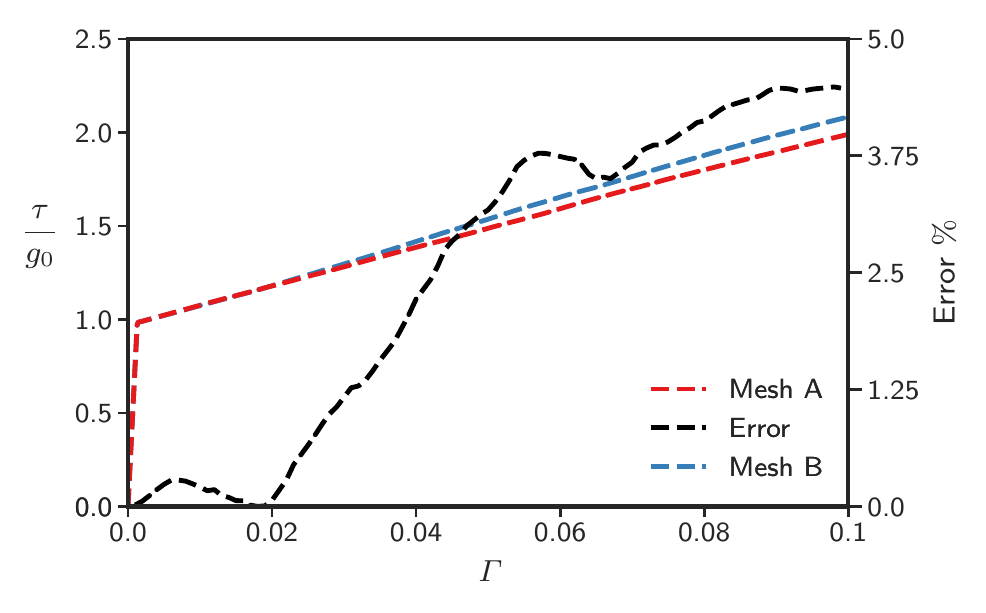}}
		\caption{Convergence in the stress-strain response for the $\mum{5}$ sample size with $16$ inclusions for the $J_2$ MFDM model. \textrm{Mesh A} refers to the non-uniform mesh shown in Fig.~\ref{fig:mesh_16}  and $\text{Mesh B}$ refers to the uniformly refined mesh with $256$ elements in each direction.}
		\label{fig:inclusion_convergence}
\end{figure}

\begin{figure}
	\centering
	\centering
	\begin{subfigure}[b]{0.495\textwidth}
		\centering
		{\includegraphics[width=.990\linewidth]{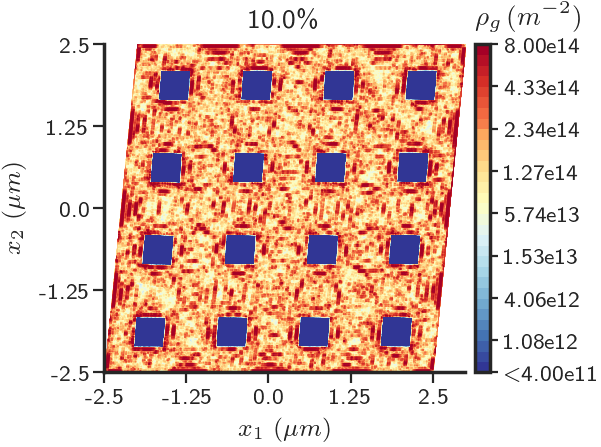}}
		\caption{}
		\label{fig:inclusion_an_j2_16_5mic_ref}
	\end{subfigure}%
	\centering
	\begin{subfigure}[b]{0.495\textwidth}
		\centering
		{\includegraphics[width=.990\linewidth]{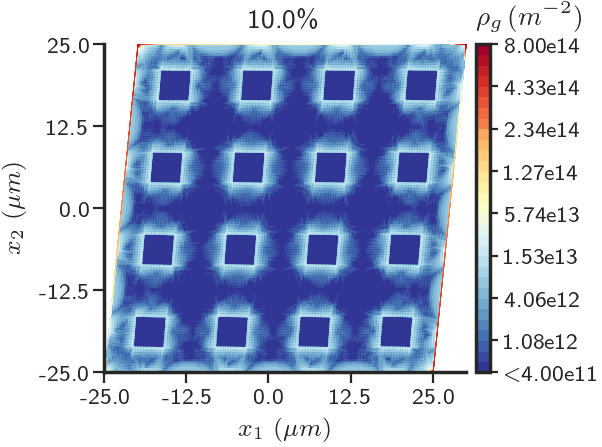}}
		\caption{}
		\label{fig:inclusion_an_j2_16_50mic_ref}
	\end{subfigure}\\
	\centering
\begin{subfigure}[b]{0.495\textwidth}
	\centering
	{\includegraphics[width=.990\linewidth]{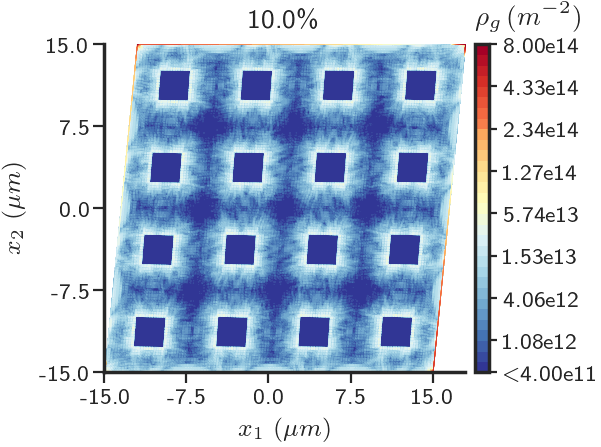}}
	\caption{}
	\label{fig:inclusion_an_j2_16_30mic_ref}
\end{subfigure}\\
	\caption{Dislocation density distribution at $10\%$ strain  for different sample sizes with $16$ inclusions for the $J_2$ MFDM model using a uniformly refined mesh with $256$ elements in each direction. a) $\mum{5}$ b) $\mum{50}$ c)$\mum{30}$.}
	\label{fig:inclusion_convergence_all}
\end{figure}

\added{ In what follows, we briefly study the  convergence of our results, choosing the problem with $16$ inclusions for the $J_2$ MFDM model.
Figure \ref{fig:inclusion_convergence} shows the convergence in the stress-strain plot for the $\mum{5}$ sample size. \textrm{Mesh A} refers to the non-uniform mesh shown in Fig.~\ref{fig:mesh_16}  and $\text{Mesh B}$ refers to a uniformly refined mesh with $256$ elements in each direction. The maximum error is $4.74\%$ at $10\%$ strain. We did not observe any visible difference in the stress-strain response corresponding to the graded and uniformly fine meshes, for the $\mum{50}$ sample size.

Figures \ref{fig:inclusion_an_j2_16_5mic_ref}, \ref{fig:inclusion_an_j2_16_50mic_ref},  and \ref{fig:inclusion_an_j2_16_30mic_ref} show the $\rho_g$ distribution at $\mGamma = 0.1$ for the $\mum{5}$, $\mum{30}$, and $\mum{50}$ sample sizes with $16$ inclusions using the uniformly refined Mesh B. The graded microstructural patterns, following the graded mesh, observed in Figure  \ref{fig:inclusion_an_j2}  are an artifact of the choice of the mesh. However, the dislocation density distribution in Figures \ref{fig:inclusion_an_j2_16_5mic} and \ref{fig:inclusion_an_j2_16_50mic} qualitatively resembles Figures \ref{fig:inclusion_an_j2_16_5mic_ref} and \ref{fig:inclusion_an_j2_16_50mic_ref}, respectively, in the regions close to the inclusions.  The dislocation density generates near the inclusion boundary for all the sample sizes. These dislocation boundary layers are clearly separated for the $\mum{50}$ sample size. With a decrease in sample and inclusion size Figure \ref{fig:inclusion_convergence_all} shows that the (rescaled) distance between the boundary layers decreases to the extent that the dislocation density is distributed through out the body for the $\mum{5}$ sample size.}

\subsubsection{Elastically stiffer inclusions}
In this section, we model the same phenomena for the case when the inclusions are harder than the matrix. The Young's modulus and Poisson's ratio for the inclusion are taken to be $220$ GPa and $0.17$,  respectively. All other material parameters and problem setup remains the same. 

\begin{figure}
	\centering
	\begin{subfigure}[b]{0.495\textwidth}
		\centering
		{\includegraphics[width=.990\linewidth]{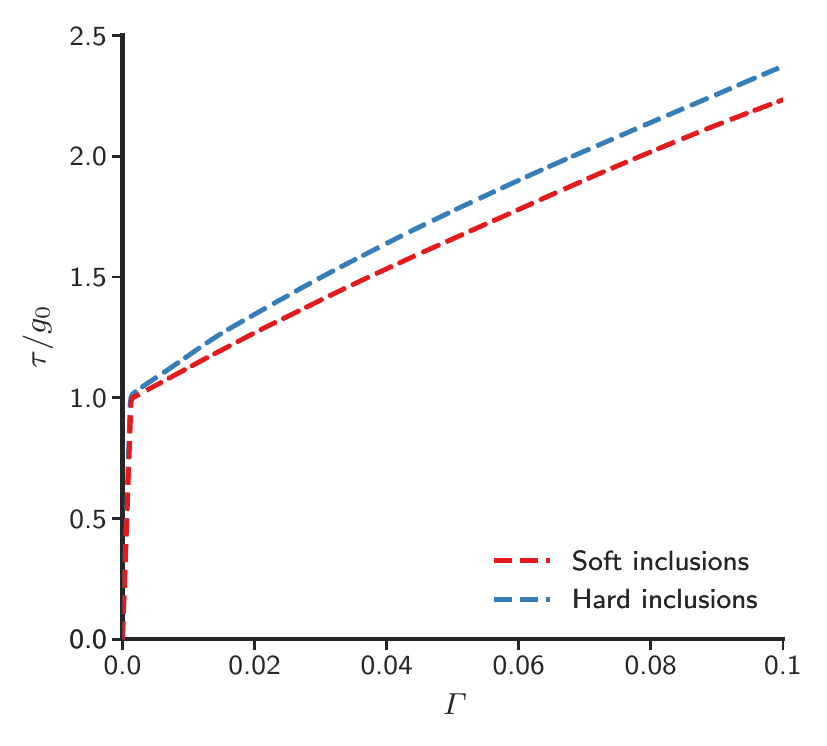}}
		\caption{}
		\label{fig:inclusion_ss_cr_s}
	\end{subfigure}%
	\begin{subfigure}[b]{0.495\textwidth}
		\centering
		{\includegraphics[width=.990\linewidth]{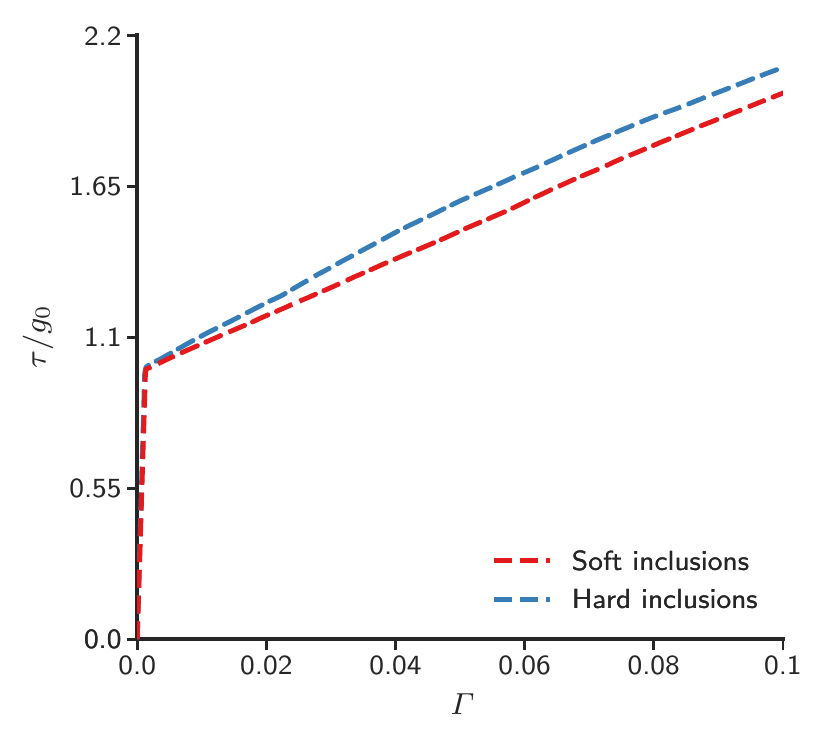}}
		\caption{}
		\label{fig:inclusion_ss_j2_s}
	\end{subfigure}
	\caption{Effect of inclusion  stiffness on the stress-strain response of the $\mum{5}$ sample with $16$ inclusions. a) Crystal plasticity b) $J_2$ plasticity. }
	\label{fig:inclusion_ss_s}
\end{figure}

Figures \ref{fig:inclusion_ss_cr_s} and  \ref{fig:inclusion_ss_j2_s} shows the stress-strain response for this case for crystal and $J_2$ plasticity models of MFDM, respectively. As expected, the material hardening is  more pronounced when inclusions are harder than the matrix as this increases the average elastic modulii of the domain. There is enhancement in the purely elastic response of the material for the case of stiffer inclusions  (that cannot be seen due to the large strain range covered by the results).

\subsection{(Stress-uncoupled) FDM with inertia}
\label{sec:dyna_fdm}
We present some results of calculations where inertia is accounted for in the numerical formulation. The full algorithm for the dynamic case appears in Table \ref{tab:dyna_algo}. The results presented here are for the case when the dislocation velocity, $\bfV_0$, below is uncoupled to the underlying stress field  i.e. ii is prescribed. The governing equations are
\begin{align}\label{eqn:strs_unc_vel}
tr(\bfL)\bfalpha+\dot{\bfalpha}-\bfalpha\bfL^T &= -curl\left(\bfalpha\times \bfV_0 \right) - \tilde{\epsilon}(|\bfalpha|)\, curl\,curl\, \bfalpha \notag\\
&= -curl\left(\bfalpha\times \bfV_0 \right) + \tilde{\epsilon}(|\bfalpha|)\, div\,grad\, \bfalpha 
\end{align}
along with \eqref{eq:stress_svk} and \eqref{eq:rate_form}. As shown subsequently, under this evolution the dislocation core changes shape as it moves and is spread over progressively smaller area, with the contraction mathematically expected for this case of a system of, instead of a scalar, wave equations due to the coupling to the velocity field, as shown in \cite{acharya2011equation}. Without the term involving the Laplacian, the dislocation density then becomes larger so as to maintain constant Burgers vector. Therefore, we assume a non-zero core energy with a non-dimensional value of  $\tilde{\epsilon} = 0.2 |\bfalpha b|^2$ to avoid large dislocation density. The form of the second term on the rhs of \eqref{eqn:strs_unc_vel} is motivated by the thermodynamic constitutive structure of the full FDM theory  \cite{acharya2011microcanonical} where the dislocation velocity has a contribution from the core energy of the form $\bfX(curl \bfalpha)^T \bfalpha$ that provides an rhs contribution to the dislocation density evolution of 
\begin{align}\label{eqn:alpha_coeff}
-\left[curl\left( \bfalpha \times \bfX ( curl \,\bfalpha)^T \bfalpha\right)\right] =  (\bfalpha \otimes \bfalpha ): div \, grad \, \bfalpha + \textrm{additional terms}
\end{align}

 When the evolution of dislocations is restricted to be in a single direction, say $x$, this term takes the form $|\bfalpha|^2 \bfalpha_{,xx}$, up to constants \cite{zhang2015single}. This along with \eqref{eqn:alpha_coeff} serves as motivation for the diffusive term in \eqref{eqn:strs_unc_vel}.

 We study the three following problems:

\begin{enumerate}
	\item A single edge dislocation moving in the $\bfe_1$ direction at constant velocity. The solution reflects similarities to the longitudinal propagation of dynamic shear bands.
	
	\item Two dislocations of opposite sign moving towards each other with prescribed velocities  and  annihilating.
	
	\item A single dislocation in a pre-strained body and moving at a speed greater than the linear elastic shear wave speed $v_s$ of the material. This gives rise to an (unsymmetric) propagating Mach cone  in the domain.
\end{enumerate}

\begin{figure}
	\centering
	{\includegraphics[width=.6\linewidth]{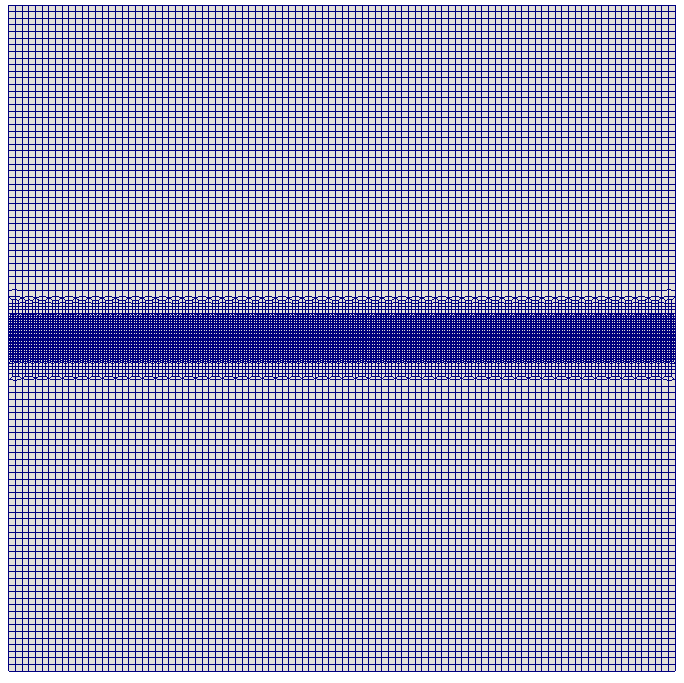}}
	\caption{Mesh with $23,000$ elements  used in Sections \ref{sec:dyna_sd} and \ref{sec:dyna_md}}.
	\label{fig:dyna_mesh}
\end{figure}

Although the dislocation velocity is uncoupled to the underlying stress field, the dislocation density evolution is coupled to the evolving deformation of the body through the gradient of  the material velocity field $\bfv$  which is a finite deformation effect.

In the formulation, $b$ is a fundamental length scale that correlates with the core size. At the atomistic scale, $b$ denotes the Burgers vector of the full dislocation in the material. At the mesoscale, it could be a length scale related to the core-size of a shear band, a measurable quantity. We use the non-uniform mesh  shown in Fig.~\ref{fig:dyna_mesh} which is highly refined in a layer where the dislocation density will evolve. Table \ref{tab:dyna_d} presents the values of the material constants used for the simulations presented in this section. The non-dimensional time $t^*$ is given as $t^*= \frac{t\,v_s}{b}$, where $t$ is the  dimensional time.

\begin{table}[H]
	\centering
	\begin{tabular}{ c c c c  }
		\myhline
		Parameter & $\dfrac{\rho v^2_s}{\mu}$ & $\dfrac{E}{\mu}$ & $\nu$ \\[.35cm] \myhline 
 Value  & $1$ & $2.6$ & $.33$ \\ \hline
	\end{tabular}
	\caption{Parameter values used for modeling dislocation evolution with inertia.}
	\label{tab:dyna_d}
\end{table}

\subsubsection{Single dislocation}
\label{sec:dyna_sd}
The problem is set up as follows: a single edge dislocation with Burgers vector $\bfb = b\bfe_1$ is  assumed to be present in a domain of  dimensions $[-20b, 20b] \times [-20b, 20b]$.  The edge dislocation is modeled by prescribing an initial dislocation density tensor $\bfalpha$ of the form
\begin{align}
\begin{split}
\alpha_{13}(x_1, x_2, t = 0) &= \varphi_0 \left(1 - tanh^2\left(\dfrac{|\bfx - \bfp|}{b}\right)\right),\\
\alpha_{ij}(x_1, x_2, t = 0) &= 0 ~ \rm{if} ~ i \neq 1 ~\rm{and}~ j \neq 3.
\label{eq:dyna_alpha13}
\end{split}
\end{align}  $\varphi_0$ is a constant chosen to give a dislocation of strength $b$ by ensuring $\int_{A_0} \alpha_{13} \, dA = b$, where $A_0$ is any area patch in the domain at $t = 0$ that encloses the dislocation core. $x_1$ and $x_2$ are the in-plane coordinates of any point $\bfx$ in the domain and $\bfp$ denotes the initial position of the center of the dislocation core. At all times, the dislocation is assumed to be moving with a constant velocity $\bfV = \frac{v_s}{2}\bfe_1$.  For the computation presented below, $\bfp$ is taken to be $\bf0$. 
The system is then evolved by following the algorithm given in Table \ref{tab:dyna_algo}.

\begin{figure}
	\centering
	\begin{subfigure}[b]{0.495\textwidth}
		\centering
		{\includegraphics[width=0.9\linewidth]{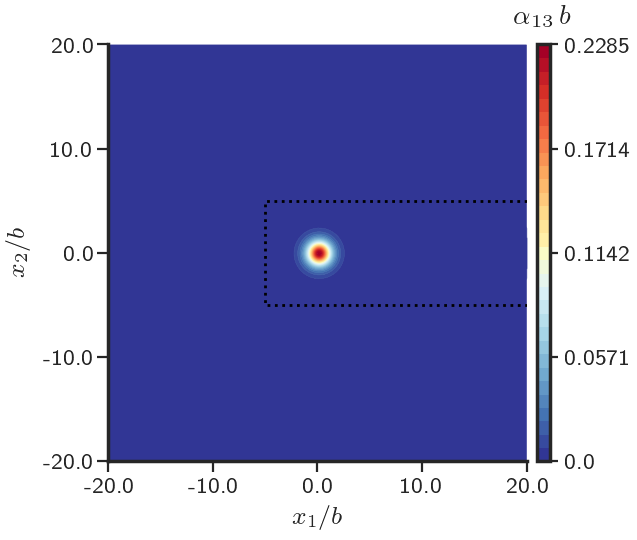}}
		\caption{}
\label{fig:dyna_sd_alpha13_t0}
	\end{subfigure}%
	\begin{subfigure}[b]{0.495\textwidth}
		\centering
		{\includegraphics[width=0.9\linewidth]{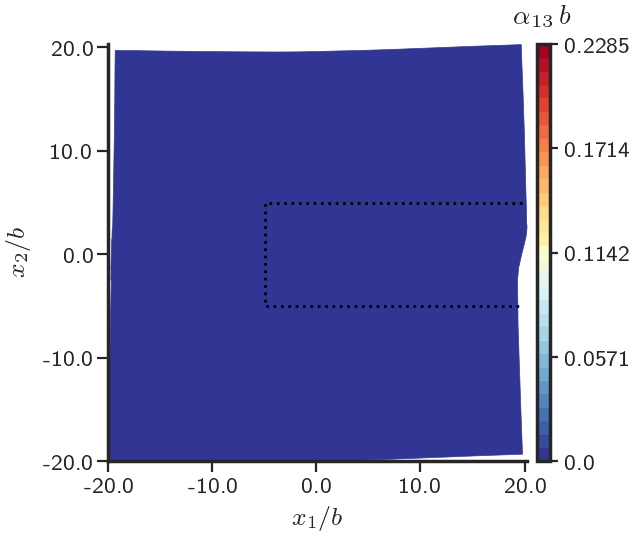}}
		\caption{}
\label{fig:dyna_sd_alpha_13_slipstep}
	\end{subfigure}
	\caption{Dislocation density distribution in the body at different times  a) $t^* = 0$ b) $t^* = 50$, i.e.~when dislocation exits the body.}
\label{fig:dyna_sd_alpha13_ttmps}
\vspace{4mm}
	\centering
	\begin{subfigure}[b]{0.33\textwidth}
		\centering
		{\includegraphics[width=0.9\linewidth]{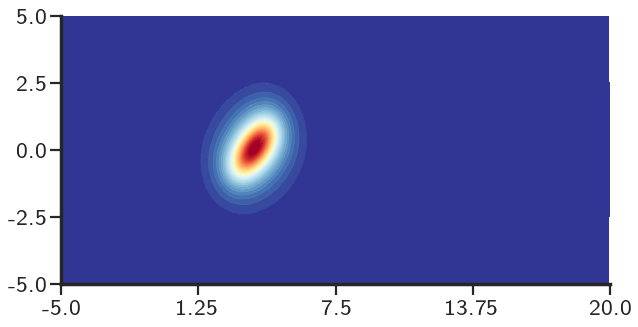}}
		\caption{}
	\end{subfigure}%
	\begin{subfigure}[b]{0.33\textwidth}
		\centering
{\includegraphics[width=0.9\linewidth]{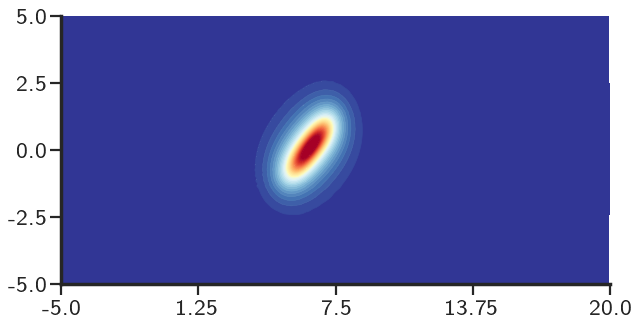}}
		\caption{}
	\end{subfigure}%
	\begin{subfigure}[b]{0.33\textwidth}
		\centering
{\includegraphics[width=0.9\linewidth]{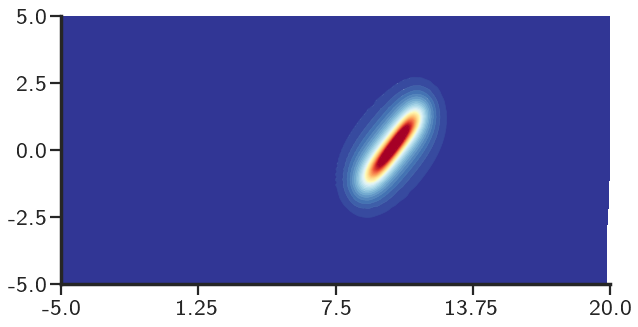}}
		\caption{}
	\end{subfigure}\\
	\begin{subfigure}[b]{0.33\textwidth}
		\centering
		{\includegraphics[width=0.9\linewidth]{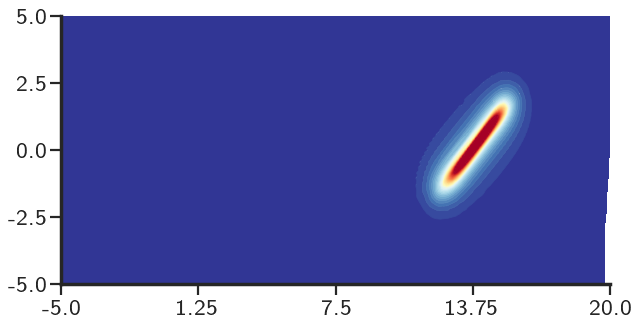}}
		\caption{}
	\end{subfigure}
	\begin{subfigure}[b]{0.33\textwidth}
		\centering
{\includegraphics[width=0.9\linewidth]{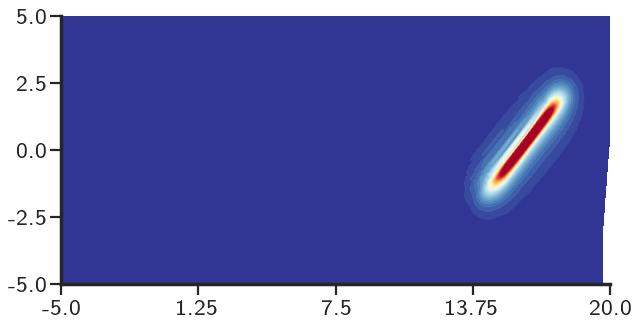}}
		\caption{}
	\end{subfigure}%
	\begin{subfigure}[b]{0.33\textwidth}
		\centering
{\includegraphics[width=0.9\linewidth]{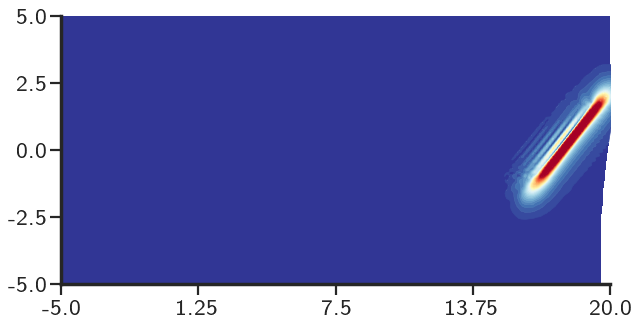}}
		\caption{}
	\end{subfigure}
	\caption{Dislocation density distribution in the body at different times $t^*$  a) $7.5$ b) $12.5$ c) $20$ d) $27.5$ e) $32.5$ f) $37.5$.}
	\label{fig:dyna_sd_alpha13_evol}
\end{figure}


Figures \ref{fig:dyna_sd_alpha13_t0}  shows the dislocation density distribution in the body at $t=0$. Figure \ref{fig:dyna_sd_alpha_13_slipstep} shows the formation of a slip step in the body upon exit of the dislocation. Figures \ref{fig:dyna_sd_alpha13_evol} show the evolution of the dislocation density in the region highlighted in the black box at various time instants.  As the evolution progresses, the dislocation density is not radially symmetric anymore, and tilts towards the right. The distribution is almost elliptical with major axis along the direction of tilting.  The tilt is towards the right because of the positive $(\bfalpha \times \bfV)_{12}$ component of the plastic strain rate generated by the motion of the dislocation.  Therefore, a negative dislocation ($\bfb = -b\bfe_1$) moving in the $-\bfe_1$ direction also tilts towards the right as the plastic strain rate component $(\bfalpha \times \bfV)_{12}$  is still positive which can be seen in Fig.~\ref{fig:dyna_md_alpha13_evol}.  Although not shown here, we have verified that a negative dislocation ($\bfb = -b\bfe_1$) moving in the $\bfe_1$ direction tilts towards the left as the plastic strain rate component $(\bfalpha\times\bfV)_{12}$ is negative for this case.

 Although the tilt in the core shape is self-consistent for the idealized, `kinematic' dislocation motion considered here, the observed core shapes, especially at equilibria, are not to be considered as physically representative of those to be obtained when full effects of stress, core energy, and non-convex generalized stacking fault energies are taken into account, e.g. \cite{zhang2015single}.


\begin{figure}
	\centering
	\begin{subfigure}[b]{0.495\textwidth}
		\centering
		{\includegraphics[width=0.9\linewidth]{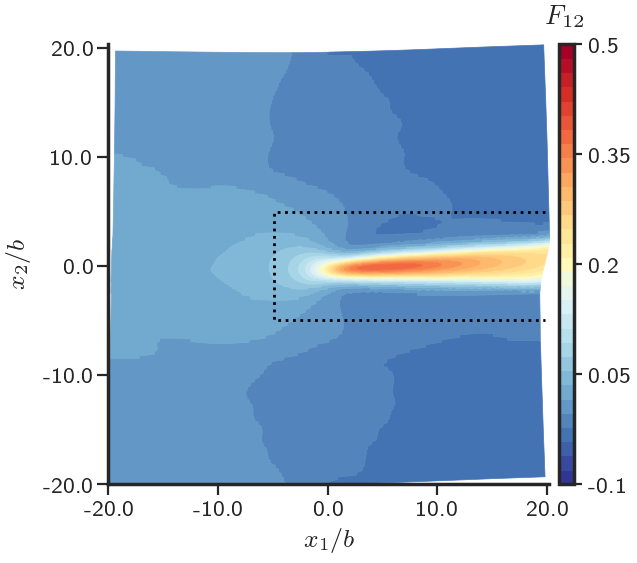}}
		\caption{}
		\label{fig:dyna_sd_F12_slipstep}
	\end{subfigure}
	\caption{$F_{12}$ component of deformation gradient at $t^* = 50$ when the dislocation exits the body.}
	\label{fig:dyna_sd_F12_ttmps}
	\vspace{4mm}
	\centering
	\begin{subfigure}[b]{0.33\textwidth}
		\centering
		{\includegraphics[width=0.9\linewidth]{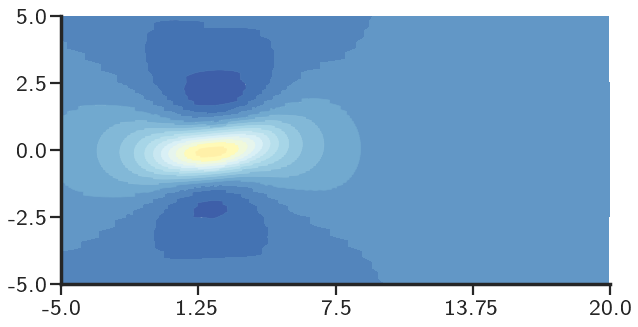}}
		\caption{}
	\end{subfigure}%
	\begin{subfigure}[b]{0.33\textwidth}
		\centering
		{\includegraphics[width=0.9\linewidth]{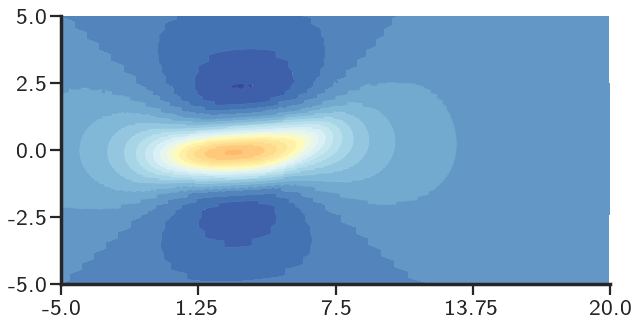}}
		\caption{}
	\end{subfigure}%
	\begin{subfigure}[b]{0.33\textwidth}
		\centering
		{\includegraphics[width=0.9\linewidth]{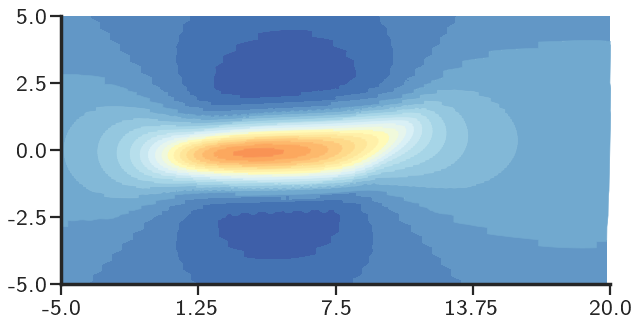}}
		\caption{}
	\end{subfigure}\\
	\begin{subfigure}[b]{0.33\textwidth}
		\centering
		{\includegraphics[width=0.9\linewidth]{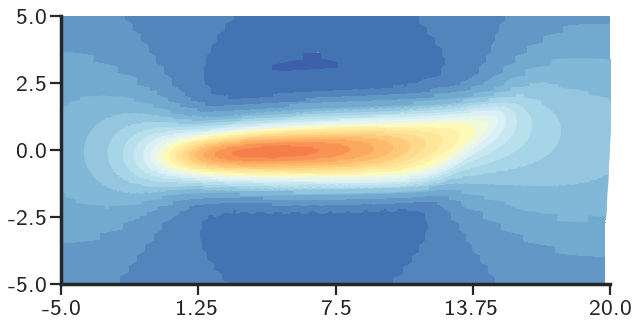}}
		\caption{}
	\end{subfigure}
	\begin{subfigure}[b]{0.33\textwidth}
		\centering
		{\includegraphics[width=0.9\linewidth]{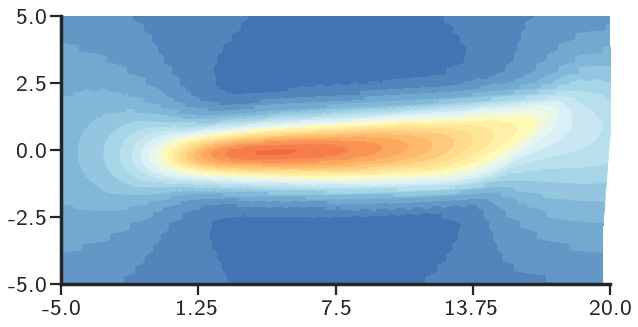}}
		\caption{}
	\end{subfigure}%
	\begin{subfigure}[b]{0.33\textwidth}
		\centering
		{\includegraphics[width=0.9\linewidth]{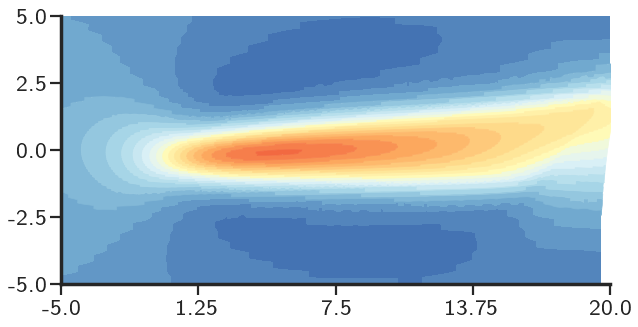}}
		\caption{}
	\end{subfigure}
	\caption{$F_{12}$ in the highlighted region at different times, $t^*$  a) $7.5$ b) $12.5$ c) $20$ d) $27.5$ e) $32.5$ f) $37.5$.}
	\label{fig:dyna_sd_def_F12_evol}
\end{figure}

Figures \ref{fig:dyna_sd_F12_ttmps} and  \ref{fig:dyna_sd_def_F12_evol} show the evolution of the $F_{12}$ component of the deformation gradient $\bfF$ as the dislocation core moves.  There is an accumulation of shear deformation in the wake of the moving dislocation. The body is permanently deformed when the dislocation exits the body with a clear region where slip  occurred, mimicking a shear band.
\subsubsection{Multiple dislocations}
\label{sec:dyna_md}
The problem is set up as follows: Two edge dislocations with Burgers vectors $\bfb^{1} = b\bfe_1$ and $\bfb^{2} = -b\bfe_1$, respectively, are  considered in a domain with dimensions $[-20b, 20b] \times [-20b, 20b]$. The edge dislocations are modeled by prescribing an initial  dislocation density tensor $\bfalpha$ of the form
\begin{align*}
\begin{split}
\alpha_{13}(x_1, x_2, t = 0) &= \varphi_0 \left( tanh^2\left(\dfrac{|\bfx - \bfp|}{b}\right)   - tanh^2\left( \dfrac{|\bfx - \bfq|}{b}\right) \right)\\
\alpha_{ij}(x_1, x_2, t = 0) &= 0 ~ \rm{if} ~ i \neq 1 ~\rm{and}~ j \neq 3.
\end{split}
\end{align*}
The value of $\varphi_0$ is same as in Eq.~\eqref{eq:dyna_alpha13} which ensures the strength of each dislocation is $b$. $\bfp$ and $\bfq$ are the initial position vectors of the two dislocations. In the computed example below,  $\bfp$ and $\bfq$ are taken as $-5b\bfe_1$  and $5b\bfe_1$, respectively. At all times, each dislocation is assumed to be moving with a prescribed velocity given by
\begin{align}
\bfV^i = \dfrac{v_s}{2.0} \bfe_2 \times \bfl^i
\end{align}
where $\bfl^i = \frac{1}{|\bfb^i|} (\bfb^i \otimes \bfe_3)^T\bfe_1$ is the line direction of the $i^{th}$ dislocation. Therefore, the positive and negative dislocations move in the $\bfe_1$ and $-\bfe_1$ directions, respectively, at the same constant speed. The system is then evolved by following the algorithm given in Table \ref{tab:dyna_algo}.


\begin{figure}
	\centering
	\begin{subfigure}[b]{0.495\textwidth}
		\centering
		{\includegraphics[width=0.9\linewidth]{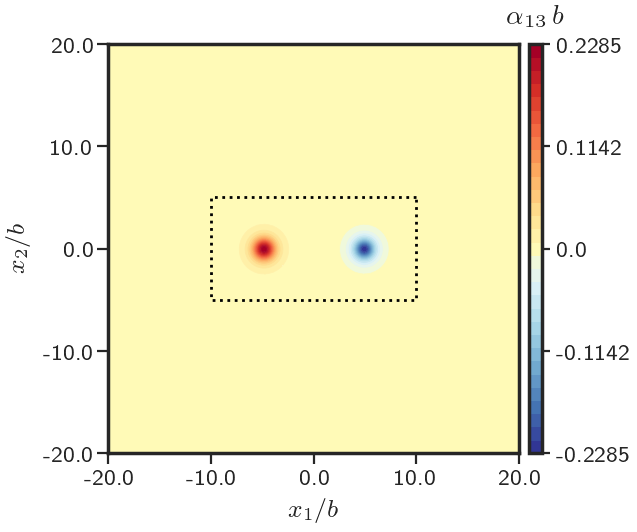}}
		\caption{}
		\label{fig:dyna_md_alpha13_t0}
	\end{subfigure}%
	\begin{subfigure}[b]{0.495\textwidth}
	\centering
	{\includegraphics[width=0.9\linewidth]{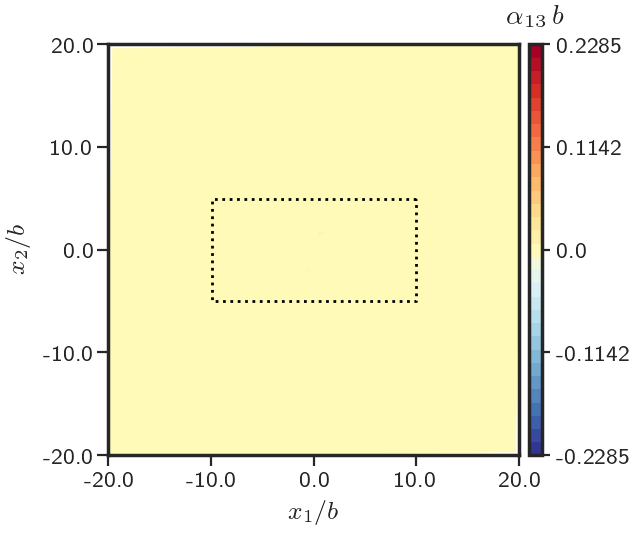}}
	\caption{}
	\label{fig:dyna_md_alpha13_tend}
\end{subfigure}\\
	\caption{Dislocation density distribution in the body at a) $t^* = 0$ b) $t^* = 15$, i.e.~after annihilation.}
	\label{fig:dyna_md_alpha13_ttmps}

	\centering
	\begin{subfigure}[b]{0.33\textwidth}
		\centering
		{\includegraphics[width=0.9\linewidth]{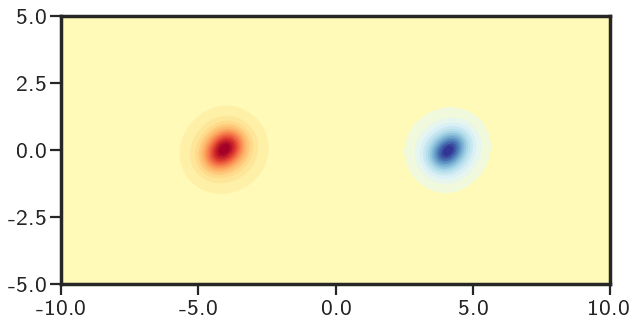}}
		\caption{}
	\end{subfigure}%
	\begin{subfigure}[b]{0.33\textwidth}
		\centering
		{\includegraphics[width=0.9\linewidth]{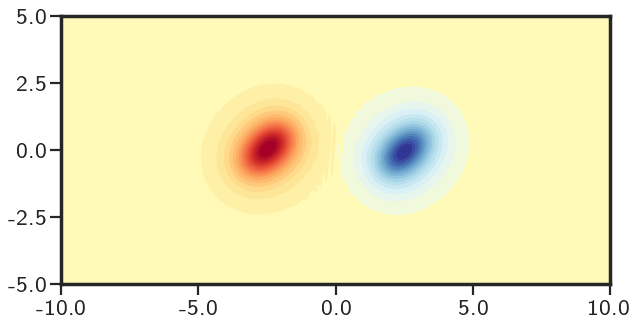}}
		\caption{}
	\end{subfigure}\\
	\begin{subfigure}[b]{0.33\textwidth}
		\centering
		{\includegraphics[width=0.9\linewidth]{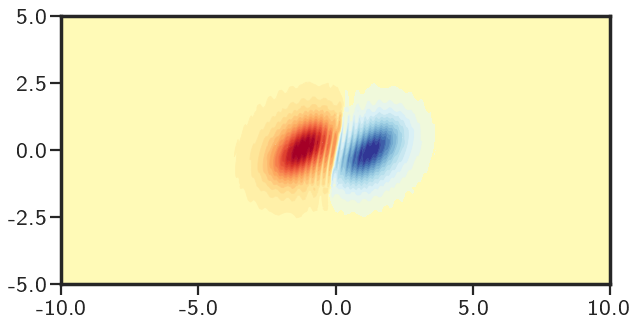}}
		\caption{}
	\end{subfigure}%
	\begin{subfigure}[b]{0.33\textwidth}
		\centering
		{\includegraphics[width=0.9\linewidth]{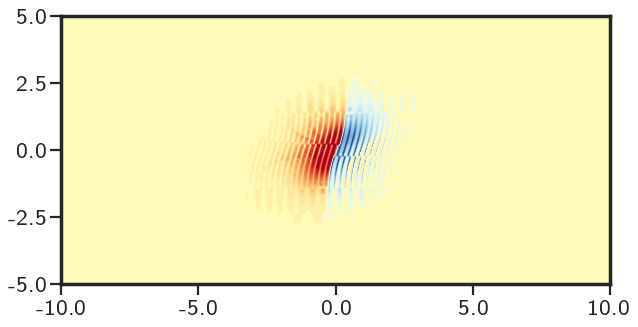}}
		\caption{}
	\end{subfigure}%

	\caption{Dislocation density distribution in the highlighted region at different times $t^*$  a) $2.5$ b) $5$ c) $7.5$ d) $10$.}
	\label{fig:dyna_md_alpha13_evol}
\end{figure}

\begin{figure}
	\centering
	\begin{subfigure}[b]{0.495\textwidth}
		\centering
		{\includegraphics[width=0.9\linewidth]{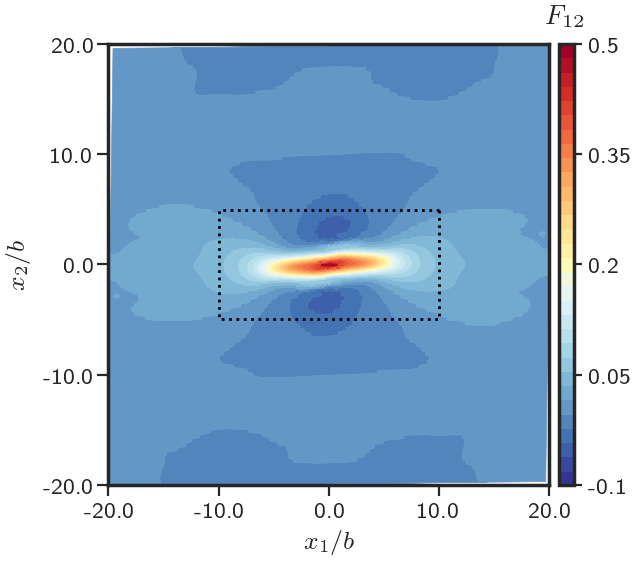}}
		\caption{}
		\label{fig:dyna_md_def_F12_t0}
	\end{subfigure}\\
	\caption{$F_{12}$ component of the deformation gradient at $t^* = 15$, i.e~after annihilation.}
	\label{fig:dyna_md_def_F12_ttmps}

	\centering
	\begin{subfigure}[b]{0.33\textwidth}
		\centering
		{\includegraphics[width=0.9\linewidth]{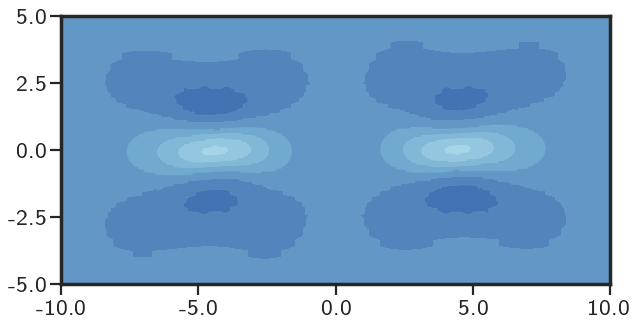}}
		\caption{}
	\end{subfigure}%
	\begin{subfigure}[b]{0.33\textwidth}
		\centering
		{\includegraphics[width=0.9\linewidth]{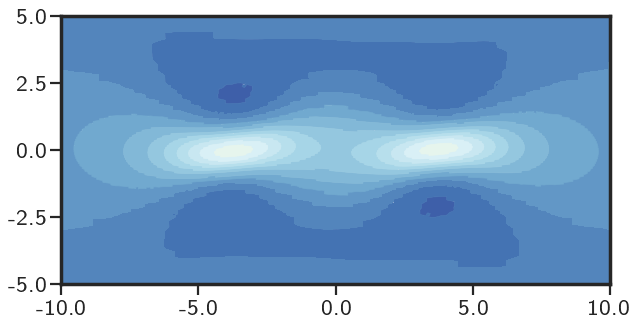}}
		\caption{}
	\end{subfigure}\\
	\begin{subfigure}[b]{0.33\textwidth}
		\centering
		{\includegraphics[width=0.9\linewidth]{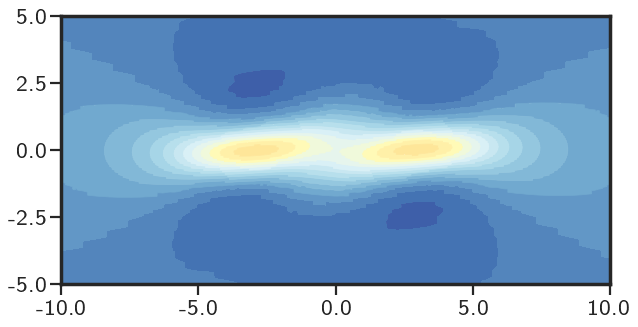}}
		\caption{}
	\end{subfigure}
	\begin{subfigure}[b]{0.33\textwidth}
		\centering
		{\includegraphics[width=0.9\linewidth]{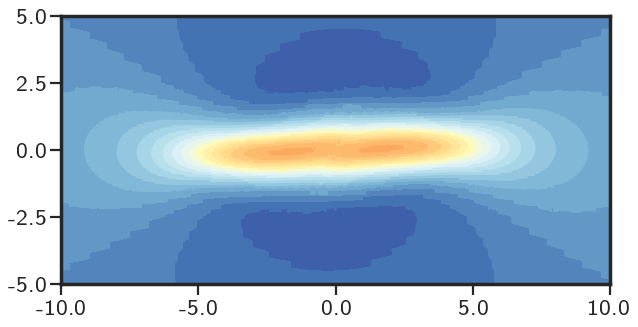}}
		\caption{}
	\end{subfigure}%
	\caption{$F_{12}$ component of the deformation gradient in the highlighted region at different times $t^*$  a) $2.5$ b) $5$ c) $7.5$ d) $10$.}
	\label{fig:dyna_md_def_F12_evol}
\end{figure}

Figure \ref{fig:dyna_md_alpha13_evol} shows the dislocation positions at various non-dimensional times, $t^*$.  It can be seen that the two dislocations of opposite sign annihilate each other leaving behind a slipped embryo along the path of the dislocations in the domain. This is demonstrated in Figure \ref{fig:dyna_md_def_F12_evol} which shows the evolution of the $F_{12}$ component of the deformation gradient in the domain at the corresponding times.

As noted in \cite{varadhan2006dislocation}, for a $1$-d small deformation case (dislocation evolution is uncoupled to stress field and material velocity gradient), when dislocation densities of opposite sign meet, a discontinuity (shock) develops, grows, and finally disappears. Here, we have presented a full finite deformation scheme that is able to qualitatively resolve the shock and model the result of the interaction, which is that the dislocations are annihilated.

\subsubsection{Mach cone}
\label{sec:dyna_fdm_mach}

\begin{figure}
	\centering
	{\includegraphics[width=.6\linewidth]{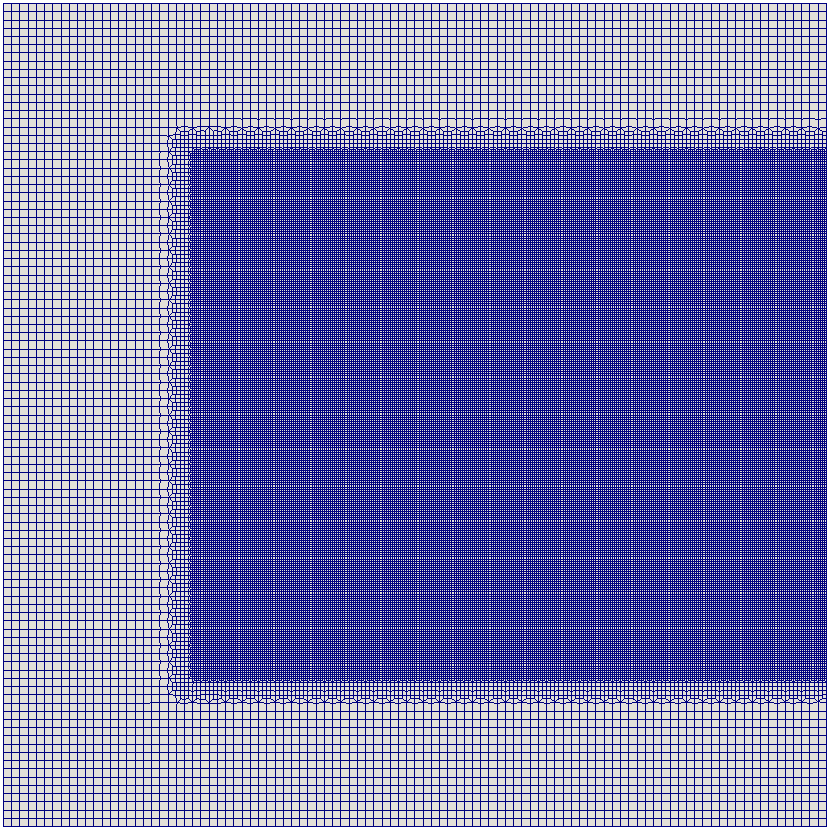}}
	\caption{Mesh with $86,000$ elements for modeling Mach cone formation. }
	\label{fig:mach_mesh}
\end{figure}

The problem is set up as follows: a single edge dislocation with Burgers vector $\bfb = b\bfe_1$ is
 assumed to be present in a square domain of dimensions $[-20b, 20b] \times [-20b, 20b]$.
 The edge dislocation is modeled by prescribing an initial dislocation density of the form given by  Eq.~\eqref{eq:dyna_alpha13}.  Figure  \ref{fig:mach_mesh} shows the non-uniform mesh with approximately $86K$ elements used to capture the accurate motion of the stress waves in the domain. 
 
 The problem is solved in $2$ steps:
 
 \begin{itemize}
 	\item In the first step, the body is elastically deformed ($\bfV = \bf0$) under a quasistatic simple shear loading until it reaches a  strain of $\mGamma =  14.8\%$.
 	
 	\item In the second step, the forces on the  boundary are held constant and system is  evolved dynamically (following the algorithm from Table \ref{tab:dyna_algo}) with dislocation velocity prescribed as $\bfV = 2\,v_s\bfe_1$.
 \end{itemize}

\begin{figure}
	\centering
	\begin{subfigure}{.495\linewidth}
		{\includegraphics[width=\linewidth]{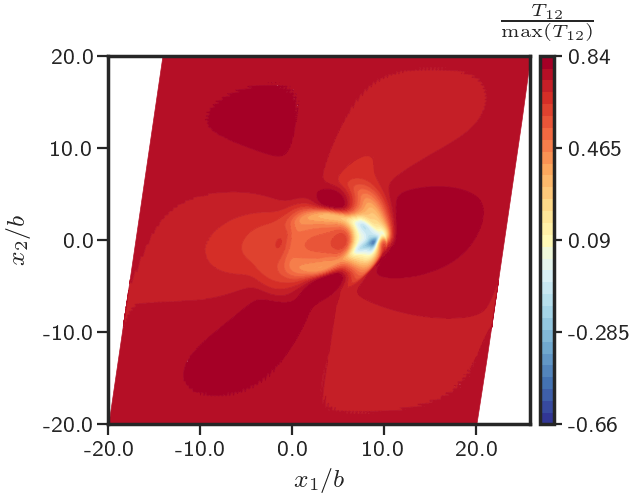}}
		\caption{}
	\end{subfigure}%
	\begin{subfigure}{.495\linewidth}
		{\includegraphics[width=\linewidth]{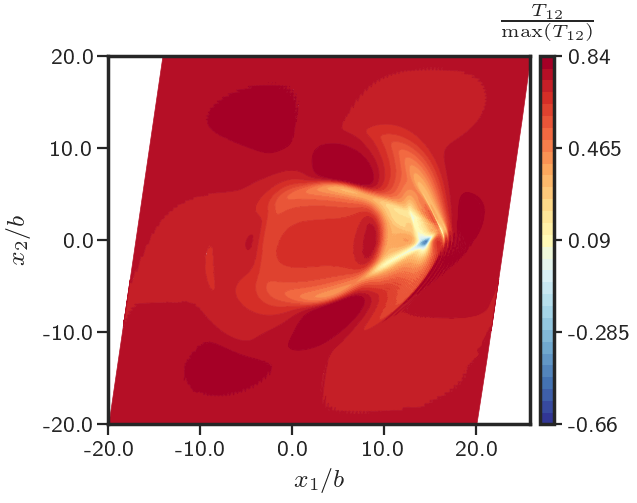}}
		\caption{}
	\end{subfigure}\\
	\begin{subfigure}{.495\linewidth}
		\centering
		{\includegraphics[width=\linewidth]{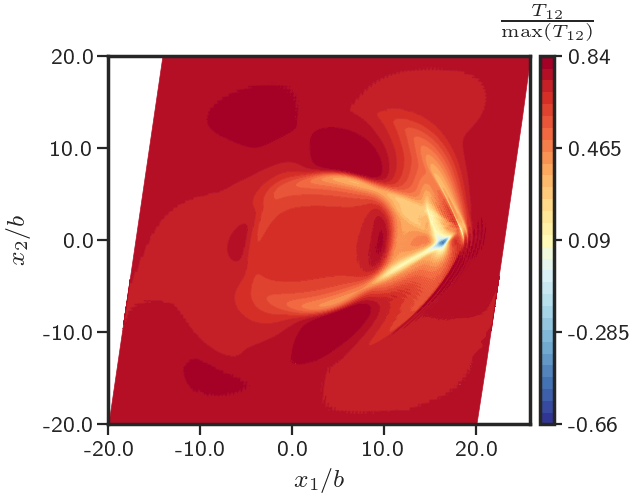}}
		\caption{}
	\end{subfigure}%
	\begin{subfigure}{.495\linewidth}
		\centering
		{\includegraphics[width=\linewidth]{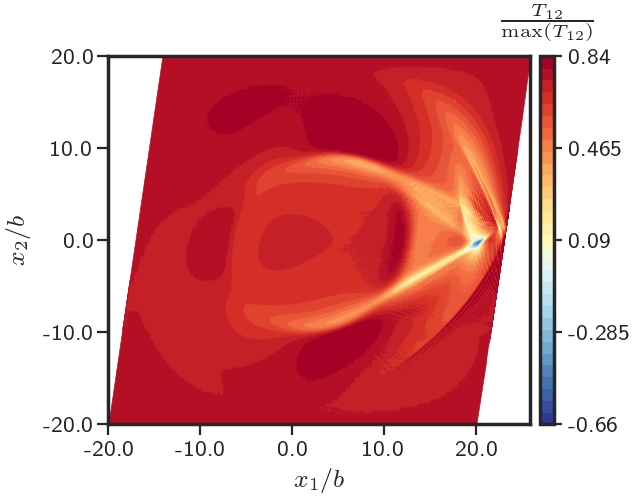}}
		\caption{}
		\label{fig:dyna_mc_stress12_evol_last}
	\end{subfigure}
	\caption{Cauchy stress (shear component) wave at different non-dimensional times in a pres-stressed body.}
	\label{fig:dyna_mc_stress12_evol}
\end{figure}

\begin{figure}
	\centering
	\begin{subfigure}[b]{0.495\textwidth}
		\centering
			{\includegraphics[width=\linewidth]{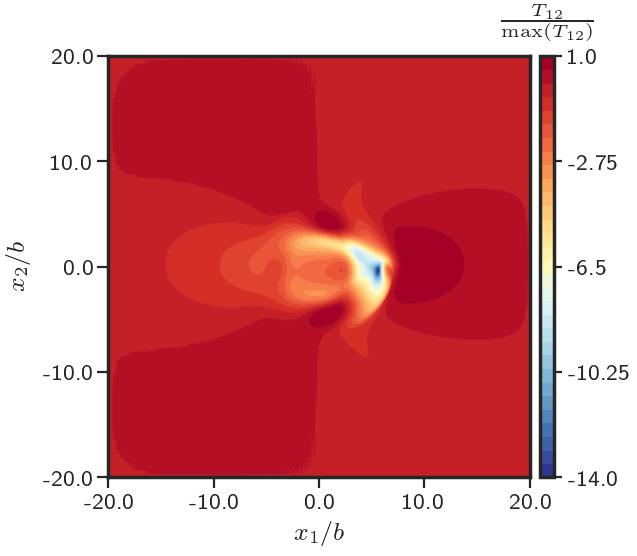}}
		\caption{}
	\end{subfigure}%
	\begin{subfigure}[b]{0.495\textwidth}
		\centering
			{\includegraphics[width=\linewidth]{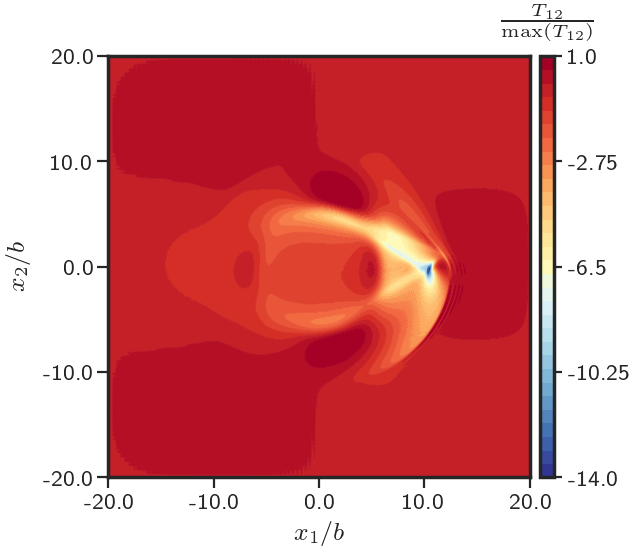}}
		\caption{}
	\end{subfigure}\\
	\begin{subfigure}[b]{0.495\textwidth}
		\centering
			{\includegraphics[width=\linewidth]{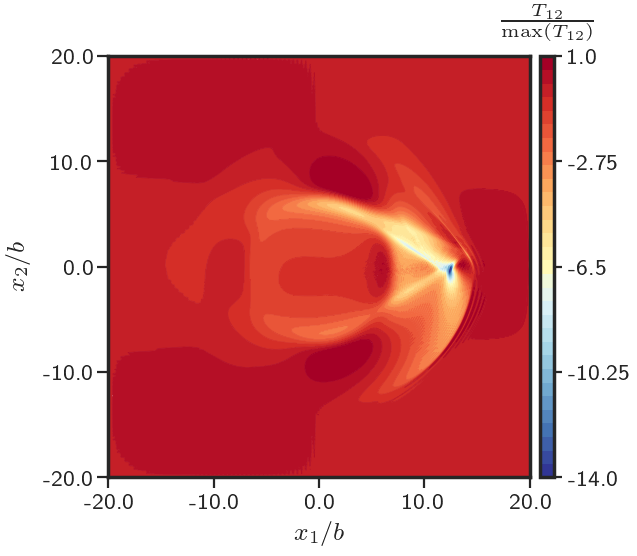}}
		\caption{}
	\end{subfigure}%
	\begin{subfigure}[b]{0.495\textwidth}
		\centering
			{\includegraphics[width=\linewidth]{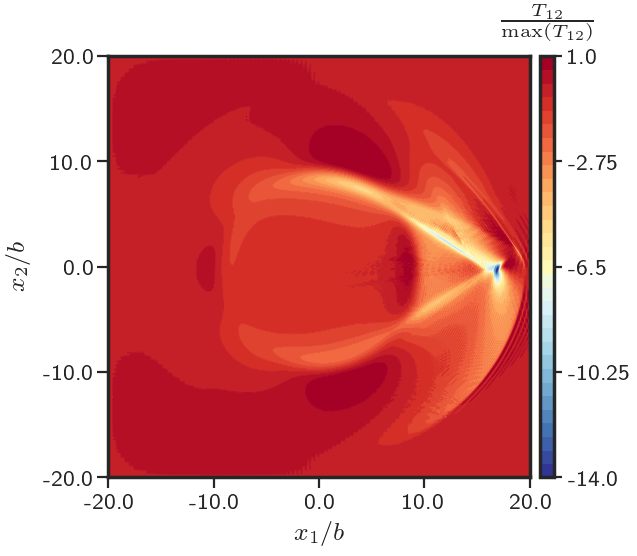}}
		\caption{}
\label{fig:dyna_mc_stress12_evold}		
	\end{subfigure}
\caption{Cauchy stress (shear component) wave at different non-dimensional times.}
\label{fig:dyna_mc_stress12_evol_np}
\end{figure}

Figure \ref{fig:dyna_mc_stress12_evol} shows the  $T_{12}$ component of the stress field $\bfT$ of the dislocation moving  at twice the linear elastic shear wave speed of the material  under an applied pre-strain of $\mGamma = 14.8\%$ at various non-dimensional times, $t^*$.  It can be seen from Fig.~\ref{fig:dyna_mc_stress12_evol_last}   that a Mach cone is formed behind the dislocation core and whose wings subtend angles that are  different between the top and the bottom regions. Such asymmetry is absent when the same phenomenon is modeled under no prestress as shown in Fig.~\ref{fig:dyna_mc_stress12_evol_np}. The nonlinearity is the main reason that leads to this asymmetric behaviour in the Mach cone. This is because  for a nonlinear elastic material at finite deformation, the difference in the stress fields  (primarily $T_{11}$), and their coupling to the prestress, between the top and the bottom of the dislocation core greatly affects the local stiffness of the system. This leads to different magnitude of the (`local') shear wave speed at points above and below the dislocation in contrast to the linear theory.  Similar observations were also made in \cite{zhang2015single} in a setting wherein  geometric nonlinearities in the total deformation and elastic constitutive equation are allowed, and the plastic distortion field is evolved with dislocation velocity coupled to the underlying stress field, but no kinematic nonlinearities in the dislocation density evolution are taken into account.


%
%
%
%


\section{Concluding Remarks}
This paper presented results of a first model of mesoscale crystal/$J_2$ plasticity of unrestricted geometric and material nonlinearities  where  static and dynamic fields of dislocation distributions can be explicitly calculated and dislocation transport, and its attendant plastic strain rate, is fundamentally accounted. Staggered numerical algorithms for the quasistatic and dynamic (with inertia) cases were devised and implemented in a finite element framework.  To our knowledge, this accomplishment stands as a first computational implementation of a partial differential equation based model of the mechanics of dislocations at finite deformations (the alternative is Molecular Dynamics, with computation of static elastic fields being the overlap of scope of the models). The model has the  following attractive features:  

\begin{enumerate}
	\item  Computation of finite deformation stress fields of arbitrary (evolving) dislocation distributions in finite bodies of arbitrary shape and elastic anisotropy under general  boundary conditions;
	
	\item Built-in kinematic ingredient to   allow modeling of longitudinal propagation of plastic wavefront as a fundamental kinematical feature of plastic flow;
	
	\item No involvement of a multiplicative decomposition of the deformation gradient, a plastic distortion tensor, or a choice of a reference configuration to describe the micromechanics of plasticity arising from dislocation motion and prediction of plastic spin with isotropic $J_2$ plasticity assumptions.
\end{enumerate}

 The developed computational tool was then used to study problems of significant scientific interest, after verification of the numerical formulation and algorithm. We showed significant differences in the stress field of a single edge dislocation obtained from finite deformation FDM and small deformation theory existing even in  large extended regions away from the dislocation core. The sharply contrasting predictions of the linear and nonlinear FDM theories for the stress field of a spatially homogeneous dislocation distribution was demonstrated.

We demonstrated size effect with crystal and $J_2$ plasticity MFDM under simple shear loading in a rate-dependent setting while using the simplest possible isotropic model of work hardening. Through the motion of localized dislocation cores, the built-in kinematic ingredient in MFDM that potentially allows for the modeling of longitudinal propagation of shear bands without the involvement of any ad-hoc failure criteria was shown.  In the  stress-uncoupled scenario, we showed the  formation of a slip step upon motion and subsequent exit of a dislocation core from the body, as well as the annihilation of two edge dislocations of opposite sign moving towards each other. We also explored the role of finite deformation leading to differences in the (propagating) Mach cone wings for the case of a single dislocation moving  at twice the linear elastic shear wave speed of the material,  with and without pre-strain in the body. The model has also been succesfully used to predict stressed dislocation pattern formation, including dipolar dislocation walls under simple shear loading, and unloaded stressed metastable dislocation microstructures \cite{arora_acharya_ijss}.

\subsubsection*{Acknowledgements}
This research was funded by the Army Research Office grant number ARO-W911NF-15-1-0239. This work also used the Extreme Science and Engineering Discovery Environment (XSEDE) \cite{xsede1}, through generous XRAC grants of supercomputing resources, which is supported by National Science Foundation grant number ACI-1548562.  We gratefully acknowledge the Pittsburgh Supercomputing Center (PSC), and Prof.~Jorge Vin\~als and the Minnesota Supercomputing Institute (URL: http://www.msi.umn.edu) for providing computing resources that contributed to the research results reported within this paper.

\begin{appendices}
	
	\section{(Mesoscale) Field Dislocation Mechanics (M)FDM}\label{app:MFDM}
	Significant portions of this section are common with \cite{arora_acharya_ijss}, a paper developed concurrently with this work.  We include this material here for the sake of being self-contained and since the theory being discussed is quite recent.
	
	Field Dislocation Mechanics (FDM) was developed in \cite{acharya2001model, acharya2003driving, acharya2004constitutive} building on the pioneering works of Kr\"oner \cite{kroner1981continuum}, Willis \cite{willis1967second},  Mura \cite{mura1963continuous}, and Fox \cite{fox1966continuum}.  The theory utilizes a tensorial description of dislocation density \cite{nye1953, bilby1955continuous}, which is related to special gradients of the (inverse) elastic distortion field.  The governing equations of FDM at finite deformation are presented below:
	\begin{subequations}
		\begin{align}
		&~\mathring{\bfalpha}\equiv tr(\bfL)\,\bfalpha+\dot{\bfalpha}-\bfalpha\bfL^T = -curl\left(\bfalpha\times \bfV \right)
		\label{eq:fdm_alpha}\\[1mm]
		&~\bfW = \bfchi+grad\bff ; \quad\bfF^{e} := \bfW^{-1} \nonumber\\[1.25mm]
		&\left.\begin{aligned}
		& curl\bfW = curl{\bfchi} = -\bfalpha\\
		&div{\bfchi} = \bf0
		\label{eq:fdm_chi}
		\end{aligned}~~~~\qquad\qquad\right\}\\[1.25mm]
		&~div\left(grad\dot{\bff}\right) = div\left(\bfalpha\times \bfV - \dot{\bfchi}-\bfchi\bfL\right)\label{eq:fdm_fevol}\\
		&~div\,[\bfT(\bfW)]  = \begin{cases}
		\bf 0 & \text{quasistatic} \\
		\rho\dot{\bfv} & \text{dynamic}. \\
		\end{cases}
		\label{eq:fdm_eqb}
		\end{align}
		\label{eq:fdm}
	\end{subequations}
	
	Here, $\bfF^e$ is the elastic distortion tensor, ${\bfchi}$ is the incompatible part of $\bfW$, $\bff$ is the plastic position vector \cite{roy2006size}, $grad\bff$ represents the compatible part of $\bfW$, $\bfalpha$ is the dislocation density tensor, $\bfv$ represents the material velocity field, $\bfL=grad\bfv$ is the velocity gradient, and $\bfT$ is the (symmetric) Cauchy stress tensor. The dislocation velocity, $\bfV$, at any point is the instantaneous velocity of the dislocation complex at that point relative to the material; at the microscopic scale, the dislocation complex at most points consists of single segment with well-defined line direction and Burgers vector. At the same scale, the mathematical model assigns a single velocity to a dislocation junction, allowing for a systematic definition of a thermodynamic driving force on a dislocation complex that consistently reduces to well-accepted notions when the complex is a single segment, and which does not preclude dissociation of a junction on evolution.
	
	The statement of dislocation density evolution \eqref{eq:fdm_alpha} is derived from the fact that the rate of change of Burgers vector content of any arbitrary area patch has to be equal to the flux of dislocation lines into the area patch carrying with them their corresponding Burgers vectors. Equation \eqref{eq:fdm_chi} is the fundamental statement of elastic incompatibility and relates the dislocation density field to the incompatible part of the inverse elastic distortion field $\bfW$. It can be derived by considering the closure failure of the image of every closed loop in the current configuration on mapping by $\bfW$. Equation \eqref{eq:fdm_fevol} gives the evolution equation for the compatible part of the inverse elastic distortion field. It can be shown to be related to the permanent deformation that arises due to dislocation motion \cite{acharya2004constitutive}. The field $grad \bff$ can also be viewed as the gradient of the inverse deformation for purely elastic deformations. Equation \eqref{eq:fdm_eqb} is the balance of linear momentum (in the absence of body forces). Balance of mass is assumed to hold in standard form, and balance of angular momentum is satisfied by adopting a symmetric stress tensor.

	Equation \eqref{eq:fdm} is augmented with constitutive equations for the dislocation velocity $\bfV$ and the stress $\bfT$ in terms of $\bfW$ and $\bfalpha$ \cite{acharya2004constitutive, zhang2015single} to obtain a closed system. It can also be succinctly reformulated as
	\begin{equation}\label{eqn:HJ}
	\begin{split}
	\dot{\bfW} & = -\bfW \bfL - (curl \bfW) \times \bfV\\
	~div\,[\bfT(\bfW)]  &= \begin{cases}
	\bf 0 & \text{quasistatic} \\
	\rho\dot{\bfv} & \text{dynamic}, \\
	\end{cases}
	\end{split}
	\end{equation} 
	but since the \emph{system} of Hamilton-Jacobi equations in \eqref{eqn:HJ}$_1$ is somewhat daunting, we work with \eqref{eq:fdm} instead, using a Stokes-Helmholtz decomposition of the field $\bfW$ and the evolution equation for $\bfalpha$ in the form of a conservation law.

	FDM \cite{acharya2001model, acharya2003driving, acharya2004constitutive} is a model for the representation of dislocation mechanics at a scale where individual dislocations are resolved. In order to develop a model of plasticity that is applicable to mesoscopic scales, a space-time averaging filter is applied to microscopic FDM  \cite{acharya2006size, acharya2011microcanonical, babic1997average} and the resulting averaged model is called Mesoscale Field Dislocation Mechanics (MFDM).  For any microscopic field $m$, the weighted, space-time running average field $\overline{m}$ is given as
	\begin{align*}
	\overline{m}(\bfx, t) := \dfrac{1}{\int_{B(\bfx)} \int_{I(t)} w(\bfx-\bfx', t-t')d\bfx' dt'  } {\int_{\mLambda} \int_{\mOmega} w(\bfx-\bfx', t-t') \,m(\bfx',t') d\bfx' dt'},
	\end{align*} where $\mOmega$ is the body and $\mLambda$ is a sufficiently large interval of time. $B(\bfx)$ is a bounded region within the body around the point $\bfx$ with linear dimension of the spatial resolution of the model to be developed, and $I(t)$ is a  bounded interval contained in  $\mLambda$. The weighting function $w$ is non-dimensional and assumed to be  smooth in the variables $\bfx, \bfx', t, t'$. For fixed $\bfx$ and $t$, $w$ is only non-zero in $B(\bfx) \times I(t)$ when viewed as a function of $\bfx'$ and $t'$. 
	
	MFDM is obtained by applying this space-time averaging filter to the FDM equations \eqref{eq:fdm} with the assumption that all averages of products are equal to the product of averages except for $\overline{\bfalpha \times \bfV}$. The governing equations of MFDM \cite{acharya2004constitutive,roy2005finite, roy2006size, acharya2011microcanonical, arora_acharya_ijss} at finite deformation (without body forces) can be written as 
\begin{subequations}
\begin{align}
&~\mathring{\overline\bfalpha}\equiv tr(\overline\bfL)\,\overline\bfalpha+\dot{\overline\bfalpha}-\overline\bfalpha\overline\bfL^T = -curl\left(\overline\bfalpha\times \overline\bfV + \bfL^p\right)\label{eq:mfdm_a_app}\\[1mm]
&~\overline\bfW = \overline\bfchi+grad\overline\bff \nonumber\\[1.25mm]
&\left.\begin{aligned}
&curl{\overline{\bfW}} = curl{\overline\bfchi} = -\overline\bfalpha\\
&div{\overline\bfchi} = \bf0
\label{eq:mfdm_chi_app} 
\end{aligned}\right\}\\[1.25mm]
&~div\left(grad\dot{\overline\bff}\right) = div\left(\overline\bfalpha\times \overline\bfV + \bfL^p - \dot{\overline\bfchi}-\overline\bfchi\overline\bfL\right)\label{eq:mfdm_fevol_app}\\
&~div\,[\overline\bfT(\overline\bfW)]  = \begin{cases}
\bf 0 & \text{quasistatic} \\
\overline\rho\,\dot{\overline\bfv} & \text{dynamic}, \\
\end{cases}
\label{eq:mfdm_f_app}
\end{align}
\label{eq:mfdm_app}
\end{subequations}
where $\bfL^p$ is defined as
\begin{align}\label{eqn:Lp}
\bfL^p(\bfx,t) := \overline{(\bfalpha - \overline{\bfalpha}(\bfx,t)) \times \bfV}(\bfx,t) = \overline{\bfalpha \times \bfV}(\bfx,t) - \overline{\bfalpha}(\bfx,t) \times \overline{\bfV}(\bfx,t).
\end{align}

The barred quantities in \eqref{eq:mfdm_app} are simply the weighted, space-time, running averages of their corresponding microscopic fields. The field $\overline{\bfalpha}$ is  the Excess Dislocation Density (ED). The microscopic density of Statistical Dislocations (SD)  at any point is defined as the difference between the microscopic dislocation density $\bfalpha$ and its  averaged field $\overline{\bfalpha}$:
\begin{equation*}
\bfbeta(\bfx,\bfx',t,t') = \bfalpha (\bfx',t') - \overline{\bfalpha}(\bfx,t),
\end{equation*}
which implies
\begin{align}
\label{eqn:tot_gnd_ssd}
\begin{split}
\rho_t &= \sqrt{\rho_g^2 + \rho_s^2}\\
\rho_t(\bfx, t) := \sqrt{ \overline{ \left( \dfrac{|\bfalpha  |}{b}\right)^2} (\bfx,t)} \ \ ; \  \rho_g(\bfx, t) &:=    \dfrac{|\overline{\bfalpha}(\bfx,t)|}{b} \ \  ; \ \ \rho_s(\bfx, t) := \sqrt{ \overline{ \left( \frac{ |\bfbeta|}{b} \right)^2 }(\bfx, t) },
\end{split}
\end{align}
with $b$ the magnitude of the Burgers vector of a dislocation in the material, $\rho_t$ the \emph{total dislocation density}, $\rho_g$ the magnitude of ED (commonly referred to as the geometrically necessary dislocation density), and $\rho_s$ is, up to a scaling constant, the root-mean-squared SD. We refer to $\rho_s$ as the scalar statistical dislocation density (\emph{ssd}).   It is important to note that spatially unresolved dislocation loops below the scale of resolution of the averaged model do not contribute to the ED ($\overline\bfalpha$) on space time averaging of the microscopic dislocation density, due to sign cancellation. Thus, the magnitude of the ED is an inadequate approximation of the total dislocation density. Similarly, a consideration of `symmetric' expansion of unresolved dislocation loops shows that the plastic strain rate produced by SD, $\bfL^p$ \eqref{eqn:Lp}, is not accounted for in $\overline{\bfalpha} \times \overline{\bfV}$, and thus the latter is not a good approximation of the total averaged plastic strain rate $\overline{\bfalpha \times \bfV}$.

In MFDM, closure assumptions are made for the field $\bfL^p$ and the evolution of $\rho_s$, as is standard in most, if not all, averaged versions of nonlinear microscopic models, whether of real-space or kinetic theory type. As such, these closure assumptions can be improved as necessary (and increasingly larger systems of such a hierarchy of nonlinear pde can be formally written down for MFDM). In this paper, we adopt simple and familiar closure statements from (almost) classical crystal   and $J_2$ plasticity theories and present the finite element formulation for the model.  Following the works of Kocks, Mecking, and co-workers \cite{mecking1981kinetics, estrin1984unified} we describe the evolution of $\rho_s$ through a statement, instead, of evolution of material strength $g$ described by \eqref{eq:softening}; $\bfL^p$ is defined by \eqref{eq:Lp_crystal} (or \eqref{eq:Lp_j2}) following standard assumptions of crystal/$J_2$ plasticity theory and thermodynamics. A significant part of the tensorial structure of \eqref{eq:Lp_crystal}  and \eqref{eq:Lp_j2}  can be justified by  elementary averaging considerations of dislocation motion on a family of slip planes under the action of their Peach-K\"{o}ehler driving force \cite{acharya2012elementary}. 

\emph{Below, and in system \eqref{eq:mfdm} as well as the rest of the paper, we drop the overhead bars for convenience in referring to averaged quantities}. 

As shown in \cite{acharya2015dislocation}, \eqref{eq:mfdm_a} and \eqref{eq:mfdm_chi} imply
\begin{equation}
\dot{\bfW} + \bfW\bfL = \bfalpha \times \bfV + \bfL^p
\label{eq:vel_grad}
\end{equation} up to the gradient of a vector field, which is re-written as
\begin{equation*}
\bfL = \dot{\bfF^{e}} {\bfF^{e-1}} + \bfF^{e}(\bfalpha\times\bfV+\bfL^p),
\end{equation*}
where $\bfF^e := \bfW^{-1}$. This can be interpreted as the decomposition of the velocity gradient into an elastic part, given by $\dot{\bfF^{e}} {\bfF^{e-1}}$, and a plastic part given by  $\bfF^{e}(\bfalpha\times\bfV+\bfL^p)$. The plastic part is defined by the motion of dislocations, both resolved and unresolved, on the current configuration and \emph{no notion of any pre-assigned reference configuration is needed}. Of  significance is also the fact that \emph{MFDM involves no notion of a plastic distortion tensor and yet produces (large) permanent deformation}.

\section{Calculation of $\parderiv{\bfT}{\bfF^e}$}
\subsection{Saint-Venant-Kirchhoff material}
\label{app:dTdFe_svk}
For a Saint-Venant-Kirchhoff material whose stress response is given by Eq.~\eqref{eq:stress_svk},  the partial derivative of $\bfT$ w.r.t.~$\bfF^e$ can be obtained as
\begin{align*}
T_{ij} &= F^e_{ik} \mathbb{C}_{klrs}E_{rs}F^{eT}_{lj}\\
\frac{\partial T_{ij}}{\partial F^e_{mn}} &= \parderiv{F^e_{ik}}{F^e_{mn}}\mathbb{C}_{klrs}E_{rs}F^e_{jl} + F^e_{ik}\mathbb{C}_{klrs}\parderiv{E_{rs}}{F^e_{mn}}F^e_{jl}  +  F^e_{ik}\mathbb{C}_{klrs}E_{rs}\parderiv{F^e_{jl}}{F^e_{mn}} \\ 
E^e_{rs} &= \frac{1}{2}\left(F^{eT}_{rp}F^e_{ps} - \delta_{rs}\right) \implies \parderiv{E^e_{rs}}{F^e_{mn}} = \dfrac{1}{2} ( \delta_{rn} F^e_{ms} + F^e_{mr}\delta_{sn}) \\
\frac{\partial T_{ij}}{\partial F^e_{mn}} &= \delta_{im}\mathbb{C}_{nlrs}E_{rs}F^e_{jl} +  \dfrac{1}{2} F^e_{ik}\mathbb{C}_{klrs}\left[F^e_{ms}\delta_{rn}+F^e_{mr}\delta_{ns}\right]F^e_{jl}  +  F^e_{ik}\mathbb{C}_{knrs}E_{rs}\delta_{jm}.
\end{align*}

\subsection{Neo-Hookean material}
\label{app:dTdFe_nh}
For the Neo-Hookean Material whose stress is given by Eq.~\eqref{eq:stress_nh},  the partial derivative of $\bfT$ w.r.t.~$\bfF^e$ can be obtained as
\begin{align*}
T_{ij} &= \mu ( F^e_{ik} F^{eT}_{kj} - \delta_{ij})\\
\parderiv{T_{ij}}{F^e_{mn}} &= \mu \left( \parderiv{F^e_{ik}}{F^e_{mn}} F^e_{jk} + F^e_{ik} \parderiv{F^e_{jk}}{F^e_{mn}}  \right)\\
&= \mu\left( \delta_{im} F^e_{jn} + F^e_{in} \delta_{jm} \right)
\end{align*}

\section{Calculation of $\parderiv{\bfT}{(grad\bfz)}$}
\subsection{Saint-Venant-Kirchhoff material}
\label{app:dTdH_svk}

At small deformation, writing $\bfF^e = \bfI + \bfU^e$, the linearised stress for Saint-Venant-Kirchhoff material is then given as
\begin{align*}
T_{ij} &= F^e_{ik} \mathbb{C}_{klrs}E_{rs}F^{eT}_{lj}\\
T_{ij} &= \dfrac{1}{2} ( \delta_{ik} +  U^e_{ik}) \mathbb{C}_{klrs}\left[(\delta_{ar} + U^e_{ar}) (\delta_{as} + U^e_{as}) - \delta_{rs} \right](\delta_{jl} + U^{eT}_{lj})\\
& \approx  \dfrac{1}{2} \mathbb{C}_{klrs}(U^e_{rs} + U^e_{sr})\\
& \approx  \mathbb{C}_{klrs}U^e_{rs}
\end{align*}
wherein the symmetry in the last two indices of $\bbC$ has been used.
Writing $\bfH = grad\bfz$,  $\parderiv{\bfT}{\bfH}$ can then be calculated as
\begin{align*}
\parderiv{T_{ij}}{H_{ab}} &= \parderiv{T_{ij}}{U^e_{pq}}\parderiv{U^e_{pq}}{H_{ab}}\\
&=\bbC_{klrs} \delta_{pr} \delta_{qs} \delta_{pa} \delta_{qb}\\
&=\bbC_{klab}.
\end{align*}

\subsection{Neo-Hookean material}
\label{app:dTdH_nh}
For Neo-Hookean material whose stress response is given by \eqref{eq:stress_nh}, the linearised stress can be calculated as:
\begin{align*}
\bfT &= \mu( \bfF^e\bfF^{eT} - \bfI )\\
T_{ij}&= \mu (F^{eT}_{ip} F^e_{pj}  - \delta_{ij})\\
&=\mu ( (\delta_{pi} + U^{e}_{pi})( \delta_{pj} + U^e_{pj})  - \delta_{ij})\\
&\approx \mu( U^e_{ij} + U^e_{ji} )
\end{align*}

Writing $\bfH= grad\bfz$,  $\parderiv{\bfT}{\bfH}$ can then be calculated as
\begin{align*}
\parderiv{T_{ij}}{H_{ab}} &= \parderiv{T_{ij}}{U^e_{pq}}\parderiv{U^e_{pq}}{H_{ab}}\\
&=\mu (\delta_{ip}\delta_{jq} +\delta_{jp} \delta_{iq}) \delta_{pa} \delta_{qb}\\
&=\mu  (\delta_{ia}\delta_{jb} +\delta_{ja} \delta_{ib})
\end{align*}

\end{appendices}

\cleardoublepage

\bibliographystyle{alpha}
\bibliography{gen_bib}

\end{document}